\documentclass[reprint,amsmath,amssymb,aps,superscriptaddress,floatfix,prmaterials,longbibliography]{revtex4-2}

\usepackage{graphicx}
\usepackage{dcolumn}
\usepackage{bm}
\usepackage{color}
\usepackage[caption=false]{subfig}
\usepackage{booktabs}
\usepackage{braket}

\newcommand*\UNS{UNb$_6$Sn$_6$}
\newcommand*\TNS{ThNb$_6$Sn$_6$}

\begin{document}

\title{Magnetic order and physical properties of the Kagome metal \UNS}

\author{Z.~W.~Riedel}
\affiliation{Los Alamos National Laboratory, Los Alamos, New Mexico 87545, USA}

\author{W.~Simeth}
\affiliation{Los Alamos National Laboratory, Los Alamos, New Mexico 87545, USA}

\author{C.~S.~Kengle}
\affiliation{Los Alamos National Laboratory, Los Alamos, New Mexico 87545, USA}

\author{S.~M.~Thomas}
\affiliation{Los Alamos National Laboratory, Los Alamos, New Mexico 87545, USA}

\author{J.~D.~Thompson}
\affiliation{Los Alamos National Laboratory, Los Alamos, New Mexico 87545, USA}

\author{A.~O.~Scheie}
\affiliation{Los Alamos National Laboratory, Los Alamos, New Mexico 87545, USA}

\author{F.~Ronning}
\affiliation{Los Alamos National Laboratory, Los Alamos, New Mexico 87545, USA}

\author{C.~Lane}
\affiliation{Los Alamos National Laboratory, Los Alamos, New Mexico 87545, USA}

\author{Jian-Xin Zhu}
\affiliation{Los Alamos National Laboratory, Los Alamos, New Mexico 87545, USA}

\author{P.~F.~S.~Rosa}
\affiliation{Los Alamos National Laboratory, Los Alamos, New Mexico 87545, USA}

\author{E.~D.~Bauer}
\affiliation{Los Alamos National Laboratory, Los Alamos, New Mexico 87545, USA}

\begin{abstract}
The $RM_6X_6$ family of materials ($R$~=~rare-earth, $M$~=~transition metal, $X$~=~Ga, Si, Ge, Sn) produces an array of emergent phenomena, such as charge density waves, intrinsic Hall effects, and complex magnetic order, due to its Kagome net of transition metal atoms, its local-moment magnetic anisotropies, and its extensive chemical tunability. Here, we report a new ``166" material containing both an actinide (uranium) and a 4$d$ transition metal (niobium) to investigate the properties of a 5$f$-4$d$ electron 166 system. \UNS\ crystallizes in the hexagonal $P$6/$mmm$ space group with a small degree of disorder due to shifts in the size of the CoSn-like cages along the $c$ axis. Upon cooling at zero magnetic field, the material undergoes two magnetic phase transitions at $T_\mathrm{2}$~=~46~K and $T_\mathrm{N}$~=~43~K. The low-temperature, zero-field phase is an antiferromagnet with ordered uranium moments and a {\textbf{k}}=(0,0,1/2) propagation vector determined by neutron diffraction. Remarkably, with a magnetic field applied along the $c$ axis, five additional magnetic transitions occur, evidenced by magnetization and resistivity data, before the moment saturates at 2.62~${\mu}_{\mathrm{B}}$/U at 2~K and $\ge$13.6~T. In two magnetic phase regions, the Hall resistivity of \UNS\ significantly deviates from the magnetization, suggesting that the phases have a large Berry curvature or a change in the Fermi surface. The unknown magnetic ordering of the field-dependent phases of \UNS\ demonstrates the complexity of the 5$f$-4$d$ 166 system and encourages further study of its properties.
\end{abstract}

\maketitle

\section{Introduction}
A Kagome lattice of atoms consists of corner-sharing triangles that create geometric frustration, often leading to emergent phenomena. Using a tight-binding model with $s$ orbitals, the electronic band structure of a pristine, isolated Kagome lattice contains a flat band, a van Hove singularity at the M point, and a Dirac point at the K point \cite{kiesel2013unconventional,yin2022topological}. In a 3D layered crystal structure, these band features may interact with nearby layers and create novel intertwined ground states. For example, if the flat band sits near the Fermi level, the density of states increase can enhance Kondo energy scales of $f$-electron layers. An ideal Kagome system, therefore, is highly tunable to enable band structure engineering.

The chemical diversity of the $RM_{6}X_{6}$ materials family ($R$~=~rare-earth, $M$~=~transition metal, $X$~=~Ga, Si, Ge, Sn) provides significant flexibility. The ``166" compounds have often been studied for their structural complexity \cite{venturini2006filling,fredrickson2008origins} and rich magnetic phase diagrams \cite{clatterbuck1999magnetic,kimura2006high,riberolles2024new,ghimire2020competing,li2024highfield,victa2025high}. The simplest 166 structure type contains two flat Kagome layers of transition metal atoms and has $P$6/$mmm$ space group symmetry. Other structure types include disordered derivatives of the hexagonal cell as well as orthorhombic and monoclinic distortions with slightly buckled Kagome layers and occasional incommensurate superstructures \cite{malaman1997magnetic,waerenborgh2005crystal,venturini2006filling,fredrickson2008origins,tan2023abundant,feng2024catalogue}. Within the 166 family, emergent properties such as charge density waves \cite{arachchige2022charge,tuniz2023dynamics,pokharel2023frustrated,kim2023infrared,hu2024phonon,yi2024tuning,ortiz2024stability}, large intrinsic anomalous Hall effects \cite{asaba2020anomalous,dhakal2021anisotropically}, Chern topological magnetism \cite{yin2020quantum}, and possible quantum criticality \cite{guo2023triangular} are reported. 

Despite the extensive number of reported 166 materials, only a small portion contain actinides or 4$d$ transition metals. 
Since the radial distribution of the 5$f$ wavefunctions of the actinides is larger than that of the 4$f$ electrons in the lanthanides, they are more likely to hybridize with the Kagome band features and to produce complex, anisotropic magnetic interactions.
The actinide-containing materials UCo$_6$Ge$_6$ \cite{buchholz1981intermetallische} and UFe$_6$Ge$_6$ \cite{gonccalves1994ufe6ge6} (or U$_{x}$Fe$_3$Ge$_3$, $x\leq0.6$ \cite{waerenborgh2005crystal}) crystallize in the ordered HfFe$_6$Ge$_6$-type and disordered Y$_{0.5}$Co$_3$Ge$_3$-type structures, respectively. 
UFe$_6$Ge$_6$ orders antiferromagnetically at 322~K before transitioning to ferromagnetic order at 230~K \cite{gonccalves1994ufe6ge6}.

Recently, UV$_6$Sn$_6$ expanded the actinide 166 family. UV$_6$Sn$_6$ has multiple uranium-driven magnetic phases, a complex crystal structure, and abnormal field-dependent properties \cite{thomas2025uv6sn6,patino2025incom}.
Regarding 4$d$ transition metal systems, only niobium-containing 166 compounds are reported \cite{oshchapovsky2010tbnb6sn6,yue2012syntheses,ortiz2024stability,xiao2025kagome}.
Similar to the vanadium Kagome lattice in the $R$V$_6$Sn$_6$ family \cite{pokharel2021electronic,zhang2022electronic,zeng2024magnetic}, the niobium lattice does not order magnetically in the $R$Nb$_6$Sn$_6$ family \cite{ortiz2024stability,xiao2025kagome}.
Extending the niobium series to the actinides may, therefore, allow us to study the interplay of 5$f$ electron interactions and 4$d$ Kagome features without additional magnetic contributions from the transition metal. Here, we characterize the complex magnetic phase space and physical properties of the compound \UNS.

\section{Materials and methods}
\subsection{Synthesis}
Single crystals of \UNS\ were grown by first arc melting pieces of depleted uranium and niobium (Alfa Aesar, 99.95\%) together in a 1:6 molar ratio on a water-cooled copper hearth under an argon atmosphere with a zirconium getter. The resultant button was cut into small pieces and combined with tin shot (Thermo Scientific, 99.999\%) in an alumina crucible for a total U:Nb:Sn molar ratio of 1:6:100. The crucible was sealed in a quartz tube under vacuum with an additional inverted alumina crucible separated by a frit disc. The sample was heated to 1150$^{\circ}$C at 150$^{\circ}$C/h, held there for 18~h, then slow cooled to 800$^{\circ}$C at 2$^{\circ}$C/h. At 800$^{\circ}$C, the tube was inverted and centrifuged to remove the excess flux. Additional flux was removed with dilute HCl etching. The resulting crystals were hexagonal plates around 40~$\mu$m thick and 1~mm wide. The \UNS\ plate thickness was roughly doubled by performing an oscillating temperature profile upon cooling: after the max 1150$^{\circ}$C/18~h hold, the sample was cooled to 850$^{\circ}$C (20$^{\circ}$C/h), then heated to 1050$^{\circ}$C (75$^{\circ}$C/h), then cooled to 850$^{\circ}$C (10$^{\circ}$C/h), then heated to 1150$^{\circ}$C (75$^{\circ}$C/h), and finally cooled to 800$^{\circ}$C (2$^{\circ}$C/h). Thin, hexagonal single crystals of \TNS\ were also grown with the same synthesis procedure.

\subsection{Characterization}
Single crystal x-ray diffraction data were collected using a Bruker D8 Venture equipped with Mo-K$\alpha$ radiation. The crystal structures of \UNS\ and \TNS\ were solved using the \textsc{SHELXT} intrinsic phasing algorithm \cite{sheldrick2015shelxt}, and subsequent refinements and analysis utilized \textsc{SHELXL} \cite{sheldrick2008short,sheldrick2015crystal}, \textsc{Platon} \cite{spek2003single}, and \textsc{Olex2} \cite{dolomanov2009olex2}. Crystal structure images were generated with \textsc{VESTA} \cite{momma2011vesta}. 

Magnetic susceptibility and magnetization measurements of \UNS\ with $H{\parallel}c$ were performed in a Quantum Design Physical Property Measurement System (PPMS), using the vibrating sample magnetometer option. Samples were mounted with GE 7031 Varnish. Data were collected while sweeping the field at 25~Oe/s. Magnetic susceptibility data for \UNS\ and \TNS\ with $H{\perp}c$ were collected with the vibrating sample mode of a Quantum Design Magnetic Property Measurement System (MPMS) equipped with a SQUID magnetometer. 

Heat capacity data were collected on thin plates of \UNS\ and \TNS\ using a PPMS, utilizing a quasiadiabatic thermal relaxation method. Significant sample-to-stage coupling issues were present for the thin crystals during our initial measurements, so to improve the coupling below 60~K, we placed a sapphire plate (Meller Optics) on top of the samples, then subtracted the additional sapphire contribution. 

Longitudinal resistivity and transverse  resistivity measurements were carried out simultaneously in a PPMS using standard four-probe configurations with an ac resistance bridge (Lake Shore, model 372) and a Stanford Research Systems lock-in amplifier (SR860) with the magnetic field applied along the $c$ axis. To account for voltage lead misalignment, resistance data at positive and negative applied magnetic fields were collected and were averaged with the formula ($R_{H>0}$~-~$R_{H<0}$)/2 for the transverse signal, taking care to average data with the same heating and applied-field histories. Similarly, misalignment of the leads in the longitudinal signal was accounted for by using the average ($R_{H>0}$~+~$R_{H<0}$)/2. Resistance data were collected at the same rate as the magnetization field sweeps (25~Oe/s).

Neutron powder diffraction was carried out on the high-resolution powder diffractometer for thermal neutrons (HRPT) at PSI \cite{fischer2000high}. For the experiments, 3.223~g of powder were loaded into a vanadium can under a helium atmosphere. Diffraction data were taken using neutrons of wavelength 2.45~\AA\ at temperatures 1.5~K (foreground) and 100~K (background).
The magnetic structure was determined from the magnetic scattering intensity, obtained by subtracting the background from the foreground data set. The thermal change of lattice parameters leads to a shift of structural Bragg peaks to larger scattering angles at lower temperatures and results in over-subtractions at angles $2\theta>50^{\circ}$. Magnetic structure refinements were therefore restricted to scattering angles $2\theta<50^{\circ}$, where the subtraction yields strictly nonnegative intensity. Group theoretical symmetry analysis was done with \textsc{FullProf} \cite{rodriguez1993recent} assuming $P$6/$mmm$ space group symmetry.

\section{Results and Discussion}
\subsection{Crystal Structure}
\begin{figure}
    \centering
    \includegraphics[width=0.98\columnwidth]{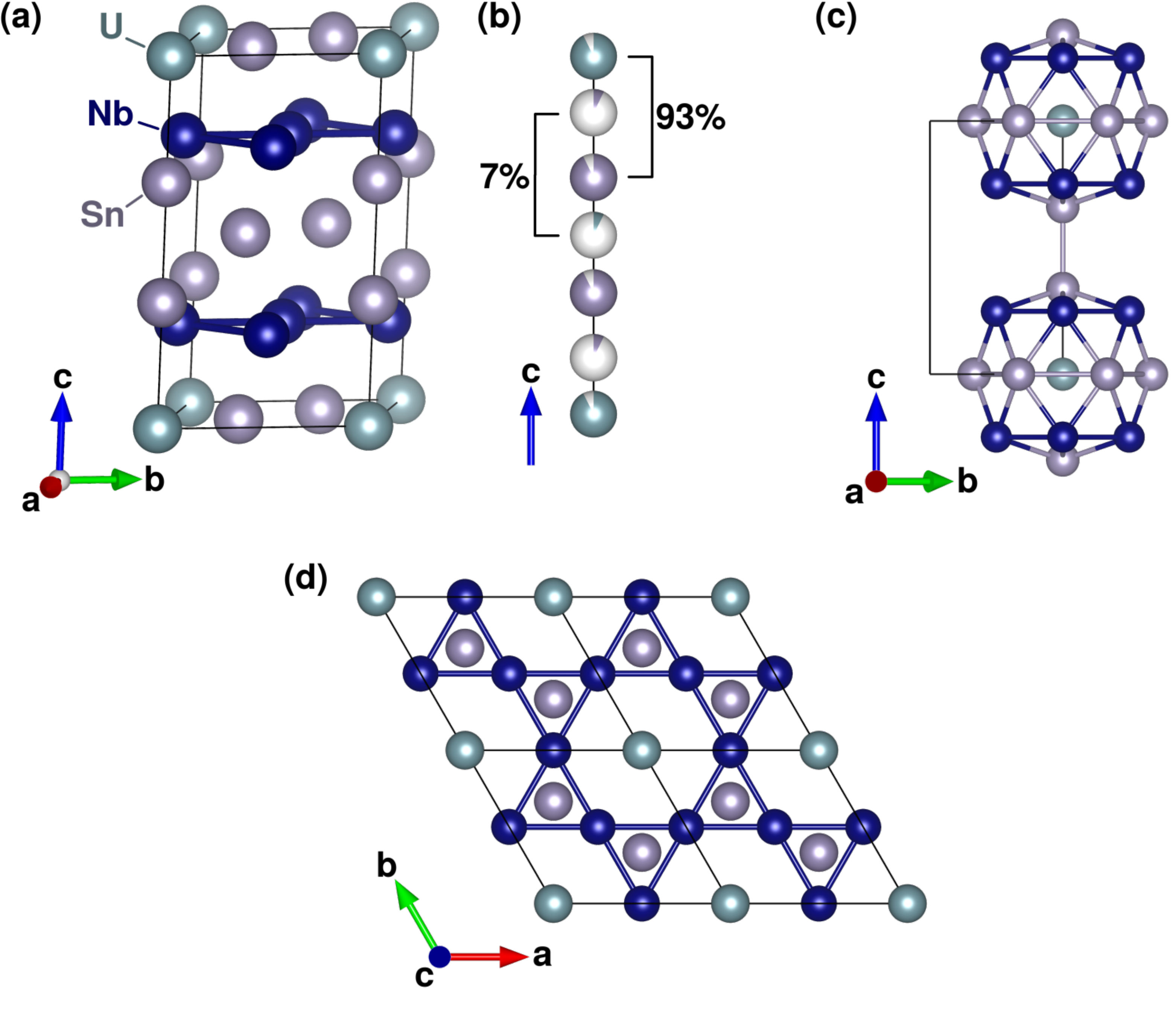}
    \caption{(a) The \UNS\ unit cell is shown without disorder. (b) Disorder model of the U and Sn sites along the $c$ axis. A higher occupancy (93\%) U/Sn pair and a lower occupancy (7\%) U/Sn pair are marked. (c) A Sn--Sn ``dumbbell" connects the stuffed CoSn-like cages. U/Sn disorder contracts the dumbbell while expanding the cages along $c$. (d) The Nb Kagome and Sn hexagonal lattices within the $ab$ plane.}
    \label{fig:structure}
\end{figure}

The single crystal x-ray data for \UNS\ was fit to a disordered, SmMn$_6$Sn$_6$-type structure \cite{malaman1997magnetic} with $P$6/$mmm$ space group symmetry and lattice parameters of $a$~=~5.7570(3)~\AA\ and $c$~=~9.5061(7)~\AA\ (Fig.~\ref{fig:structure}). 
The SmMn$_6$Sn$_6$-type structure is an intermediate between the fully ordered HfFe$_6$Ge$_6$ type and the half-unit-cell Y$_{0.5}$Co$_3$Ge$_3$ type.
With an ordered structure model, significant residual electron density was present along the $c$ axis. The residual electron density could not be accounted for with distortions producing rhombohedral, orthorhombic, or monoclinic symmetry, as in other 166 materials \cite{venturini2001crystallographic,venturini2006filling,pokharel2023frustrated}. 
The SmMn$_6$Sn$_6$ structure type was previously also observed for 
\textit{R}Cr$_6$Ge$_6$ (\textit{R}=Y, Gd--Lu) \cite{schobinger1997atomic,schobinger1997ferrimagnetism,konyk2021phase,romaka2022interaction,romaka2024structure,yang2024crystal}, \textit{R}V$_6$Sn$_6$ (\textit{R}=Sm--Gd, Dy--Tm, Lu) \cite{romaka2011peculiarities,romaka2019lu,huang2023anisotropic}, YbMn$_6$Sn$_6$ \cite{xia2006ybmn6sn6}, Yb$_{1-x}$Fe$_6$Sn$_6$ \cite{mazet2002study}, LuFe$_6$Sn$_6$ \cite{schobinger1998fe}, and \textit{R}Nb$_6$Sn$_6$ (\textit{R}=Ce--Sm) \cite{ortiz2024stability}.

The disordered structure splits the uranium and tin sites along the $c$ axis into corresponding higher occupancy (``primary") and lower occupancy (``secondary") positions.
Both primary sites have the same occupancy, and the sum of the primary sites' occupancy with that of the secondary sites is 100\%. Therefore, the 166 stoichiometry is maintained. In our case, unconstrained occupancies converged to within 1\% of the constrained occupancy values, so we maintained the constraint on the 166 stoichiometry. The resulting occupancies for \UNS\ were 92.7(3)\% occupancy of the primary uranium (U1A) and tin (Sn1A) sites and 7.3(3)\% occupancy of the secondary uranium (U1B) and tin (Sn1B) sites (Fig.~\ref{fig:structure}b).

In our final, constrained model, the primary uranium and tin sites are separated by 3.22~\AA, and the secondary sites are separated by 3.25~\AA. The Sn--Sn distances along the $c$ axis are 3.07~\AA\ for the primary sites and 3.00~\AA\ for the secondary. When we view the structure of \UNS\ as a stuffed CoSn phase \cite{fredrickson2008origins}, Nb-Sn cages along the $c$ axis surround a central U atom (Fig.~\ref{fig:structure}c). Therefore, the Sn--Sn distance decrease for the secondary sites can be interpreted as a slight contraction of the ``dumbbells" connecting the large Nb-Sn cages paired with a corresponding expansion of the cages along $c$. 
This cage expansion is distinct from the case of the hexagonal-orthorhombic inter-growth structures of other stuffed CoSn phases \cite{fredrickson2008origins}, where the $R$--$X$ bond distances are the same for the $R$ atoms at the $z=0$ and $z=0.5$ unit cell coordinates. 
The flat niobium Kagome layers (Fig.~\ref{fig:structure}d) are separated by 4.72~\AA\ between unit cells and 4.78~\AA\ within the unit cell.
\TNS\ also crystallizes in the disordered $P$6/$mmm$ structure [$a$~=~5.7925(4)~\AA, $c$~=~9.541(1)~\AA] with a primary/secondary site splitting of 94.0(2)/6.0(2)\%. 

As noted below, no evidence for a charge density wave (CDW) transition, which occurs in LuNb$_6$Sn$_6$ \cite{ortiz2024stability} and ScV$_6$Sn$_6$ \cite{meier2023tiny}, was found for \UNS. This supports the ``rattling" origin of the CDW, where relatively small Sc or Lu atoms allow space for Sn movement in the 166 family \cite{meier2023tiny}. Uranium is notably larger than lutetium, thus restricting the ability of the tin atoms to rattle along $c$. For ScV$_6$Sn$_6$ and LuNb$_6$Sn$_6$, the Sn1-Sn1 distance is $\sim$3.2--3.3~\AA, and the (Sc/Lu)1-Sn1 distance is $\sim$2.9--3.1~\AA, whereas for \UNS, the Sn1-Sn1 distance is smaller (3.00-3.07~\AA) and the U1-Sn1 distance is larger (3.22-3.25~\AA), thus expanding the uranium-centered cages and restricting Sn1 movement. Similarly, though the Sn1 displacement parameter along $c$ ($U_{\mathrm{33}}$) is larger than that along $a$ ($U_{\mathrm{11}}$) for \UNS, the values do not differ dramatically as in LuNb$_6$Sn$_6$ \cite{ortiz2024stability} or ScV$_6$Sn$_6$ \cite{meier2023tiny}. 
The supplemental material \cite{supplemental} contains the refined atomic positions and anisotropic atomic displacement parameters for \UNS\ and \TNS. 

\subsection{Magnetic Properties}
\begin{figure}
    \centering
    \includegraphics[width=0.9\columnwidth]{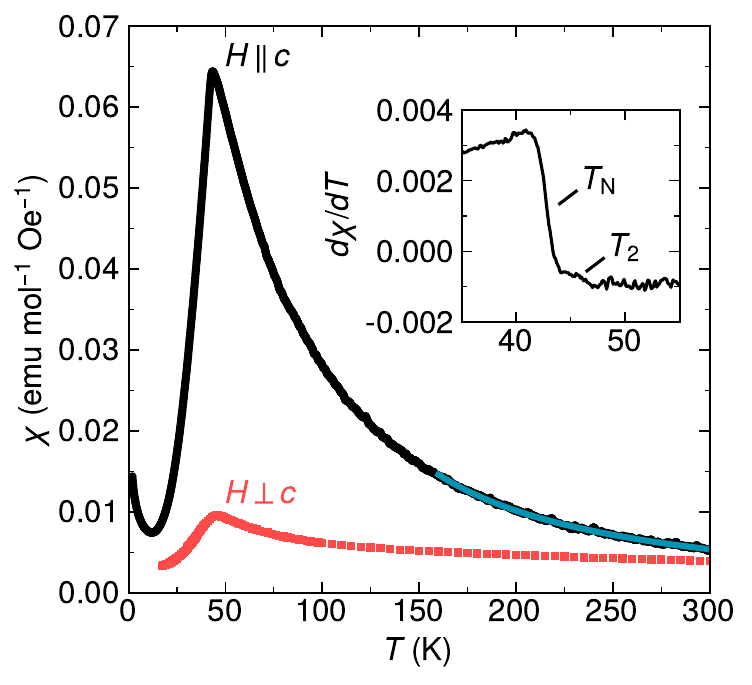}
    \caption{Anisotropic magnetic susceptibility, $\chi(T)$, of \UNS\ at 0.5~T for $H{\parallel}c$ and $H{\perp}c$. The  solid teal line shows the modified Curie-Weiss-law fit. Inset: $d\chi/dT$ vs. $T$ for $H{\parallel}c$ showing the features corresponding to $T_{\mathrm{N}}$ and $T_{\mathrm{2}}$.}
    \label{fig:chi}
\end{figure}

The magnetic susceptibility of \UNS\ with a 0.5~T field applied parallel and perpendicular to the $c$ axis is shown in Fig.~\ref{fig:chi}. To improve the $H{\perp}c$ data signal, several smaller crystals were measured along with the larger crystal used to collect the $H||c$ data. The additional crystals contained a small Nb$_3$Sn impurity ($\sim$7~vol\%), evidenced by a superconducting transition at 17.5~K (not shown). 
A kink in the magnetic susceptibility indicates long-range antiferromagnetic ordering at 43.5~K for $H{\parallel}c$. An additional transition appears in the derivative of the susceptibility (d$\chi$/d$T$) at 46~K for $H{\parallel}c$. For $H{\perp}c$, a peak appears at 46~K.
While \UNS\ exhibits long range magnetic order, \TNS\ is a Pauli paramagnet with no magnetic transitions (see Fig.~S3 in \cite{supplemental}). Therefore, we conclude that only the uranium atoms order magnetically in \UNS.
The $H{\parallel}c$ inverse magnetic susceptibility ($\chi^{-1}$) above 160~K was fit to a modified Curie-Weiss law (Eq.~\ref{eq:mCW}):
\begin{equation} \label{eq:mCW}
    (\chi - \chi_{\mathrm{0}}) ^{-1} = \frac{T-\theta_{\mathrm{CW}}}{C},
\end{equation}
yielding an effective magnetic moment of $\mu_\mathrm{eff}$=3.69(4)~${\mu}_{\mathrm{B}}$/U. Since the free ion values are similar for a 5$f^2$ (3.58~$\mu_{\mathrm{B}}$) or 5$f^3$ (3.62~$\mu_{\mathrm{B}}$) valence configuration of uranium, we cannot distinguish them here. The Curie-Weiss parameter ($\theta_\mathrm{CW}$) is 55(2)~K, and the temperature-independent contribution ($\chi_\mathrm{0}$) is \mbox{-1.6(1)$\times$10$^{-3}$~emu~mol$^{-1}$~Oe$^{-1}$}. The positive Curie-Weiss parameter suggests the presence of ferromagnetic interactions within the material despite the long-range antiferromagnetic ordering, but our model does not account for contributions from excited crystal field level mixing, limiting its interpretability. 
Therefore, the $H{\parallel}c$ and $H{\perp}c$ magnetic susceptibility were also fit to several U$^{3+}$ and U$^{4+}$ crystal field models using \textsc{PyCrystalField} \cite{scheie2021pycrystalfield}. Details are in the supplemental material \cite{supplemental}. 
The U$^{3+}$ models provided a qualitatively better fit and consistently led to a ground state doublet dominated by $|\pm\frac{7}{2}\rangle$, but the fits are underconstrained and should be treated cautiously.
That said, the relative abundance of trivalent $R$ elements in the $R$Nb$_6$Sn$_6$ and $R$V$_6$Sn$_6$ families [excluding nonmagnetic exceptions of ThV$_6$Sn$_6$ \cite{xiao2024preparation} and (Ti,Zr,Hf)V$_6$Sn$_6$ \cite{he2024quantum} presumed to contain tetravalent $R$ atoms] and the assertion that UV$_6$Sn$_6$ is better modeled with U$^{3+}$ in magnetic susceptibility \cite{thomas2025uv6sn6} and neutron data \cite{patino2025incom} suggest that the uranium atoms have more 5$f^3$ character in \UNS. 

\begin{figure*}
    \centering
    \includegraphics[width=1.75\columnwidth]{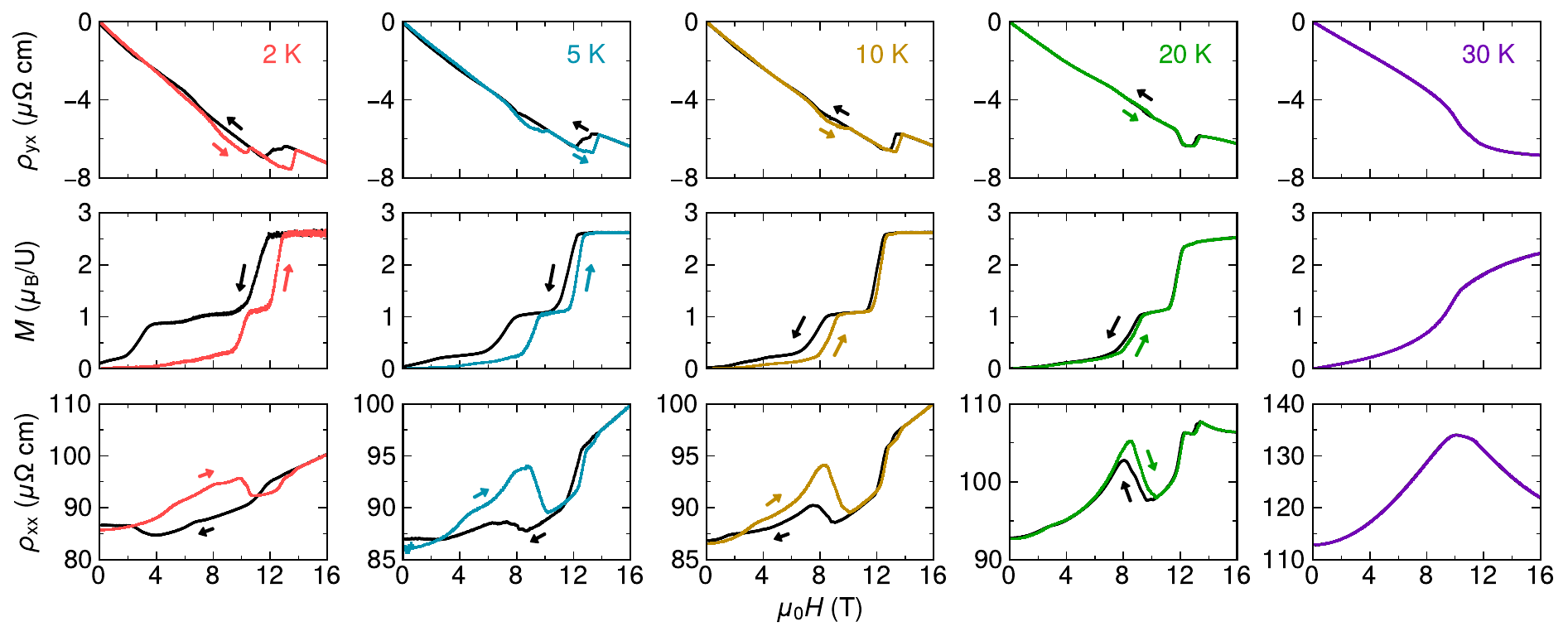}
    \caption{The transverse (Hall) and longitudinal resistivity of \UNS\ up to 16~T are presented with the corresponding magnetization curves for $H{\parallel}c$. Increasing (colorful) and decreasing (black) magnetic field sweeps are indicated by arrows. The 30~K curves have no detectable hysteresis.}
    \label{fig:Hall_M_Rxx}
\end{figure*}

The field-dependent magnetization of \UNS\ contains multiple metamagnetic transitions with $H{\parallel}c$ (Fig.~\ref{fig:Hall_M_Rxx}). Above 13.6~T, the magnetization at 2~K saturates at 2.62~${\mu}_{\mathrm{B}}$/U, below the free-ion values ($g_J \cdot J~{\mu}_{\mathrm{B}}$) of 3.20 ${\mu}_{\mathrm{B}}$/U for 5$f^2$ and 3.27~${\mu}_{\mathrm{B}}$/U for 5$f^3$, likely due to crystal-field splitting of the $J$ multiplet. 
Notably, at 2~K, 5~K, and 10~K, the material has a remnant magnetization (Fig.~S4 \cite{supplemental}). 
No remnant magnetization was observed at 20~K or 30~K. The origin of this behavior is unclear, though it may stem from the site disorder of the crystal structure producing a second uranium magnetic site or from a small moment modulation along the $c$ axis direction. 

\begin{figure}
    \centering
    \includegraphics[width=\columnwidth]{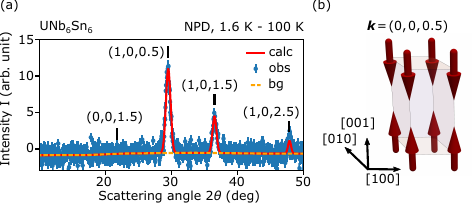}
    \caption{(a) Magnetic neutron powder diffraction (NPD) data. Diffraction intensity (blue data points) was recorded with neutrons of wavelength $\lambda=2.45$~\AA\ and was temperature subtracted ($T=1.6-100$~K). The observed intensity (obs) was fit with an A-type antiferromagnetic structure with wave vector $\bm{k}=(0,0,1/2)$ and moments pointing along the $c$-axis (red line, calc). For the refinement, a background dependent on scattering angle was considered (orange dashed line, bg). (b) Spin texture as considered for the refinement in (a).}
    \label{fig:npd}
\end{figure}

We performed zero-field, powder neutron diffraction measurements to determine the antiferromagnetic phase ordering at low temperature. The magnetic diffraction intensity is presented in Fig.~\ref{fig:npd} and indicates A-type antiferromagnetic order (ferromagnetic (001) layers that are antiferromagnetically coupled along [001]) with moments along [001]. Magnetic Bragg peaks in the pattern can be well indexed with a conventional hexagonal unit cell of parameters $a=5.76$~\AA\ and $c=9.50$~\AA\ and a magnetic propagation vector $\bm{k}=(0,0,1/2)$. The presence of scattering intensity at momentum transfer $\bm{Q}=(1,0,0.5)$ and the absence of a peak at $\bm{Q}=(0,0,1.5)$ suggest antiferromagnetism with moments pointing along the $c$ axis \cite{boothroyd2020principles}. To quantify this, two irreducible representations were considered: $\Gamma_3$, which has one basis vector pointing along the $c$ axis, and $\Gamma_9$, which has two basis vectors in the hexagonal basal plane. The fit quality parameters $\chi^2=1.2$ and $R=5.5$ for $\Gamma_3$ and $\chi^2=1.8$ and $R=450$ for $\Gamma_9$ indicate relatively good agreement for $\Gamma_3$.

Knowing the zero-field magnetic structure contains anti-aligned (001) ferromagnetic planes, the magnetization plateaus can be revisited.
With increasing magnetic field, a prominent plateau emerges at a magnetization between 40-44\% of the saturation value at 2~K.
With decreasing field, two prominent plateaus appear at 38-40\% and 32-34\% of the 2~K saturation magnetization.
The increasing field plateau is in magnetic phase P6, as defined in Fig.~\ref{fig:magpd_rhoH}a.
The P6 plateau is near (2/5)$M_\mathrm{sat}$, suggesting a propagation vector of $\textbf{k}$=(0,0,7/10). One possible stacking arrangement of the ferromagnetic planes with this propagation vector is [$\uparrow\uparrow\downarrow\uparrow\uparrow\downarrow\uparrow\uparrow\uparrow\downarrow$]. 
On the field down sweep, the prominent plateau near (1/3)$M_\mathrm{sat}$ indicates that there is magnetic phase separation or that \UNS\ is locked into a higher magnetization phase not present on the up sweep with a propagation vector of $\textbf{k}$=(0,0,2/3) and [$\uparrow\downarrow\uparrow$] stacking.
The remnant magnetization at 2~K on the field down sweep is 4.0\% of the saturation magnetization (0.105~$\mu_\mathrm{B}$/U), suggesting one uncompensated ferromagnetic plane of uranium atoms is present every 25 planes. This would give a wave vector of \textbf{k}=(0,0,13/25), very near \textbf{k}=(0,0,1/2). Continuing the field down sweep past 0~T reveals a coercivity field of \mbox{-0.48~T}. Further neutron measurements and/or resonant elastic x-ray scattering are needed to confirm the proposed magnetic propagation vectors in the applied-field phases. 

The step-like behavior of the magnetization resembles that of ``Devil's staircase" materials where increasing field progressively aligns the magnetic moments, e.g. CeSb \cite{rossat1977phase,kuroda2020devil}. Such materials can be described by an axial next-nearest neighbor Ising (ANNNI) model \cite{selke1988annni} with three exchange interactions: a ferromagnetic in-plane interaction ($J_0$), an antiferromagnetic nearest-neighbor interaction along $c$ ($J_1$), and a competing ferromagnetic next-nearest-neighbor interaction along $c$ ($J_2$). 

\subsection{Heat Capacity}
\begin{figure}
    \centering
    \includegraphics[width=0.9\columnwidth]{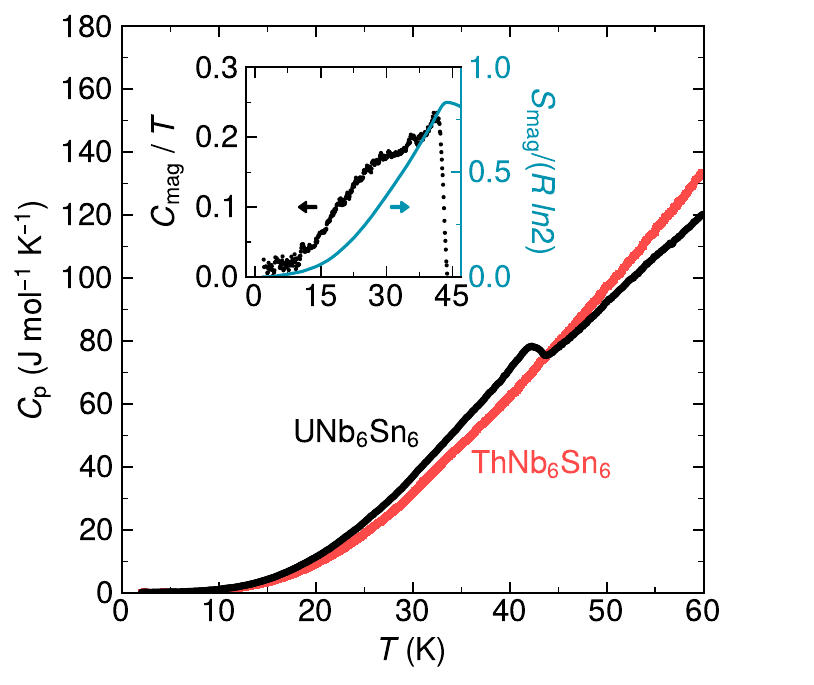}
    \caption{Zero-field heat capacity, $C_p(T)$, of \UNS\ (black) and \TNS\ (red) showing the phase transition at $T_\mathrm{N}=43$~K. Inset: Using \TNS\ crystals to estimate the lattice contribution gives an entropy recovery of close to $Rln$2 at the transition.}
    \label{fig:heat_capacity}
\end{figure}

The zero-field heat capacity, $C_p(T)$, of \UNS\ is displayed in Fig.~\ref{fig:heat_capacity}, which shows a broad peak. The midpoint of the peak's rise is at 43~K, 0.5~K below the antiferromagnetic transition peak observed in the 0.5~T magnetic susceptibility (Fig.~\ref{fig:chi}), but no discernible feature is observed at $T_\mathrm{2}=46$~K. No features corresponding to magnetic order were found in the heat capacity of \TNS, supporting the conclusion that only the uranium atoms order magnetically in \UNS. The heat capacity of \TNS\ was treated as a lattice contribution and subtracted from the \UNS\ heat capacity to obtain the 5$f$ contribution to the heat capacity. The heat capacity of \TNS, though, crosses that of \UNS\ just above the antiferromagnetic transition, producing a nonphysical negative magnetic heat capacity. The crossing may be due to mass uncertainty or the aforementioned data collection complications for the extremely thin crystals. Despite this complication, the $5f$ contribution to the entropy, defined as \mbox{$S$=$\int$[( C$_\mathrm{UNb_6Sn_6}$-C$_\mathrm{ThNb_6Sn_6}$)/$T$]~$dT$}, shows that \UNS\ recovers roughly $Rln$2 entropy at the antiferromagnetic transition, suggesting a uranium ground state 5$f$ doublet in the crystalline electric field. A linear fit to $C_\mathrm{p}$/$T$=${\beta}T^2$+${\gamma}$ at low temperatures provides values for the Sommerfeld coefficient ($\gamma$) and Debye temperature ($\theta_D$, extracted from $\beta$), as shown in Fig.~S6 \cite{supplemental}. The resulting coefficients are $\gamma$=26(2)~mJ~mol$^{-1}$~K$^{-2}$ and $\beta$=0.54(5)~mJ~mol$^{-1}$~K$^{-4}$ ($\theta_D$=360(11)~K). Other 166 compounds have similar Sommerfeld coefficients on the order of tens of mJ~mol$^{-1}$~K$^{-2}$ \cite{ishii2013ycr6ge6,pokharel2021electronic,pokharel2022highly,lyu2024anomalous,xiao2024preparation}, including UV$_6$Sn$_6$ (15-40~mJ~mol$^{-1}$~K$^{-2}$) \cite{thomas2025uv6sn6,patino2025incom}.

\subsection{Longitudinal Resistivity} \label{sec:rhoxx}
\begin{figure}
    \centering
    \includegraphics[width=0.8\columnwidth]{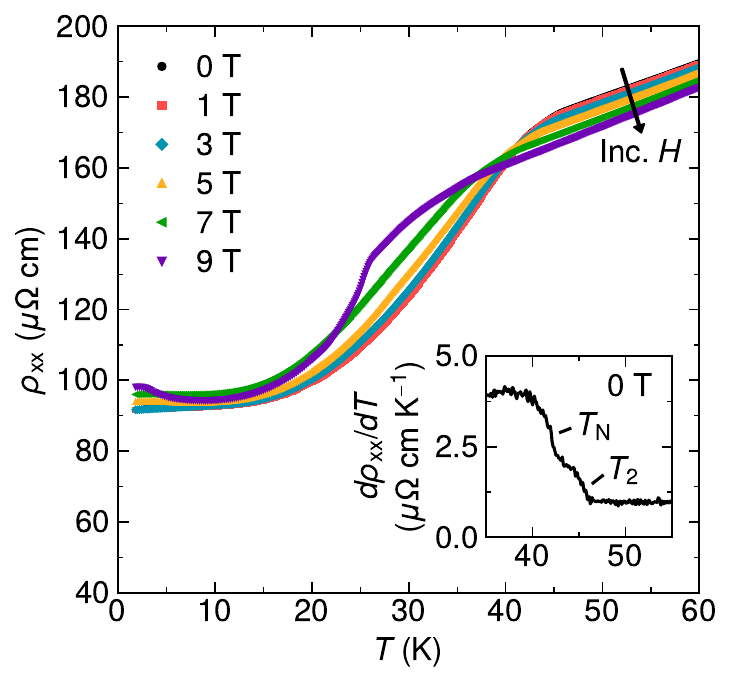}
    \caption{The longitudinal resistivity, $\rho_\mathrm{xx}(T)$, of \UNS. Inset: Zero-field $d\rho_\mathrm{xx}/dT$ vs. $T$ showing features corresponding to $T_{\mathrm{N}}$ and  $T_{\mathrm{2}}$.}
    \label{fig:rhoT}
\end{figure}

The longitudinal resistivity, $\rho_\mathrm{xx}(H,T)$, of \UNS\ in magnetic fields up to 9 T ($H{\parallel}c$) also contains several features corresponding to magnetic transitions. Fig.~\ref{fig:rhoT} shows constant-field data collected on heating where two slope changes, most prominent in the 9~T data, correspond to the magnetic transitions $T_{\mathrm{2}}$ and $T_{\mathrm{N}}$. At lower magnetic fields, the shifts are more pronounced in $d{\rho_{\mathrm{xx}}}/dT$ (Fig.~\ref{fig:rhoT} inset). At zero field, $T_\mathrm{N}$~=~42.5~K and $T_\mathrm{2}$~=~46~K, which agree well with the magnetic susceptibility data (Fig.~\ref{fig:chi} inset). 
The zero-field curve up to 300~K (Fig.~S7 in \cite{supplemental}) shows no indication of a CDW transition, in contrast to data for ScV$_6$Sn$_6$ \cite{arachchige2022charge} and LuNb$_6$Sn$_6$ \cite{ortiz2024stability}. 

\begin{figure*}
    \centering
    \subfloat{\includegraphics[width=1\columnwidth]{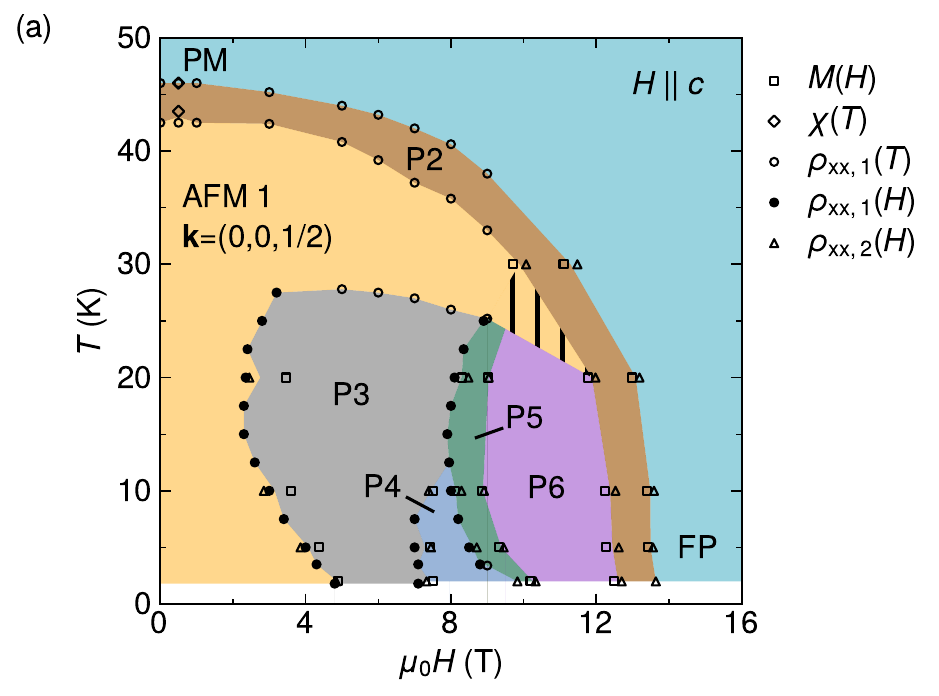}}
    \hspace{2em}
    \subfloat{\includegraphics[width=0.824\columnwidth]{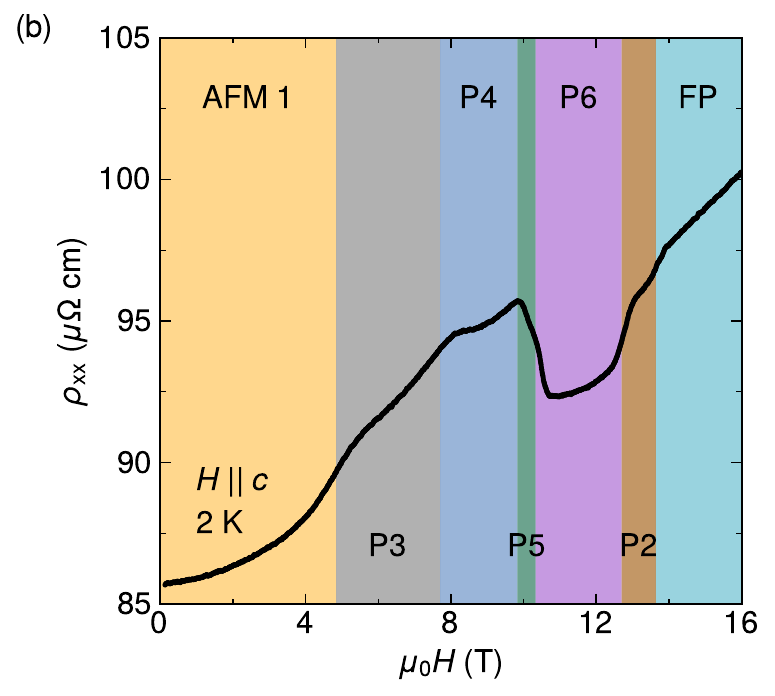}}
    \caption{(a) The magnetic field-temperature ($H$-$T$) phase diagram of \UNS\ with $H{\parallel}c$ contains six magnetically ordered phases (AFM~1, P2--P6). Boundaries are drawn using averaged field values for transitions observed in multiple characterization methods. The magnetic phase of the striped region is unclear and tentatively assigned to AFM~1. Constant-temperature data points are from increasing field after zero-field cooling, and constant-field data points are from increasing temperature after zero-field cooling. (b) $\rho_\mathrm{xx,2}(H)$ at 2~K with increasing field shows the various phase boundaries (AFM~1, P2--P6, and FP) defined by peaks in the derivative, $d\rho_\mathrm{xx}/dH$.}
    \label{fig:magpd_rhoH}
\end{figure*}

Magnetic field sweeps of the longitudinal resistivity also delineate the magnetic phase transitions of \UNS\ with $H{\parallel}c$. Several representative curves are shown in Fig.~\ref{fig:Hall_M_Rxx} (with additional curves shown in Fig.~S8 in \cite{supplemental}). Below 16~T and 12.5~K, six magnetic phases are observed. Starting with the zero-field antiferromagnetic phase (AFM~1), with increasing field, resistivity features indicate magnetic transitions of AFM~1 to ``P3" to ``P4" to ``P5" to ``P6". After P6, \UNS\ transitions to ``P2", the phase previously identified with $T_\mathrm{2}$ in $\rho_\mathrm{xx}$($T$) and $\chi$($T$). 
P2 may be an incommensurate antiferromagnetic phase reminiscent of the ordering in GdV$_6$Sn$_6$ \cite{porter2023incommensurate} or in UV$_6$Sn$_6$ \cite{patino2025incom}. Above 13.6~T, the longitudinal resistivity slope remains constant at 2~K, indicating a transition to the field-polarized (``FP") state also observed in the magnetization. 
The sequence of magnetic phase transitions in \UNS\ at 2~K is shown in Fig.~\ref{fig:magpd_rhoH}b. 

In the magnetization data, the P2 to FP transition was difficult to observe at 2~K because of increased noise at the highest fields. However, the P2-FP transition was found in the less noisy magnetization curves at higher temperatures, and a moving average of the 2~K data contains a feature similar to the P2-FP transition features in the other magnetization data (see Fig.~S5 in \cite{supplemental}). For the resistivity curves collected to 16~T at 2~K and 5~K, there is significant hysteresis at every applied field, except in the FP region. Because of the misalignment of the previously discussed up sweep and down sweep magnetization plateaus, it is generally difficult to assign hysteresis to specific magnetic phase transitions, but for the clearest up and down sweep pair of transitions, between P6 and P2, the hysteresis significantly decreases when the temperature increases to 10~K. At 30~K, no hysteresis is discernible.

A magnetic field-temperature ($H$-$T$) phase diagram of \UNS\ for $H{\parallel}c$ is constructed from features in $\chi(T)$, $M(H)$, and $\rho_\mathrm{xx}(H,T)$ (Fig.~\ref{fig:magpd_rhoH}a). To build the phase diagram, only data collected with increasing field or temperature after initial zero-field cooling (ZFC) was used. Different single crystals were used for the longitudinal field sweeps up to 9~T ($\rho_\mathrm{xx,1}$) and the field sweeps up to 16~T ($\rho_\mathrm{xx,2}$). The different measurements and samples show good agreement, with the largest deviation being at the AFM~1 to P3 boundary where the transition is subtle in the magnetization curves. The magnetic phase present in the striped region in Fig.~\ref{fig:magpd_rhoH}a is unclear based on the available data. Since the phase diagram is restricted to two free parameters, temperature and field, the Gibbs phase rule limits the phase boundaries to meet at a triple point, assuming only first order transitions. However, second order transitions could allow for a quadruple point \cite{laughlin2019magnetic} where P3, P5, P6, and AFM~1 meet. We cannot make a definitive statement in this case, so we tentatively assign the region to AFM~1 based on the perceived curvature of the phase boundaries. It may, however, be an extension of the P5 and/or P6 regions. 

\subsection{Hall Effect}
Compared with the longitudinal resistivity, the Hall data (transverse resistivity, $\rho_\mathrm{yx}(H)$) of \UNS\ does not as closely follow the magnetization features. 
Most notably, the sharp rises in the magnetization when \UNS\ transitions from P4 to P5 and from P6 to P2 occur in ranges where the Hall resistivity is relatively flat (see comparisons in Fig.~S10 in \cite{supplemental}).
Often, the transverse resistivity can be modeled with a linear dependence on the applied field (ordinary Hall component) and a linear dependence on the magnetization (anomalous Hall component). The anomalous Hall component may also be broken into an intrinsic component determined by the material's Berry curvature and into an extrinsic component proportional to $\rho_{\mathrm{xx}}$ or $\rho_{\mathrm{xx}}^{2}$. The extrinsic component accounts for behavior such as skew and/or side-jump scattering \cite{nagaosa2010anomalous}. For many $d$/$f$-electron materials (e.g., Mn$_3$Sn \cite{nakatsuji2015large}, Co$_3$Sn$_2$S$_2$ \cite{wang2018large}, and $R$Mn$_6$Sn$_6$ compounds \cite{asaba2020anomalous,dhakal2021anisotropically,zeng2022large,jones2024origin}), the intrinsic Hall contributions dominate the transport behavior when $\rho_\mathrm{xx}$ is of order 10$^0$-10$^2$~$\mu\Omega$~cm \cite{nagaosa2010anomalous}, as is the case for \UNS. 

\begin{figure}
    \centering
    \includegraphics[width=0.9\columnwidth]{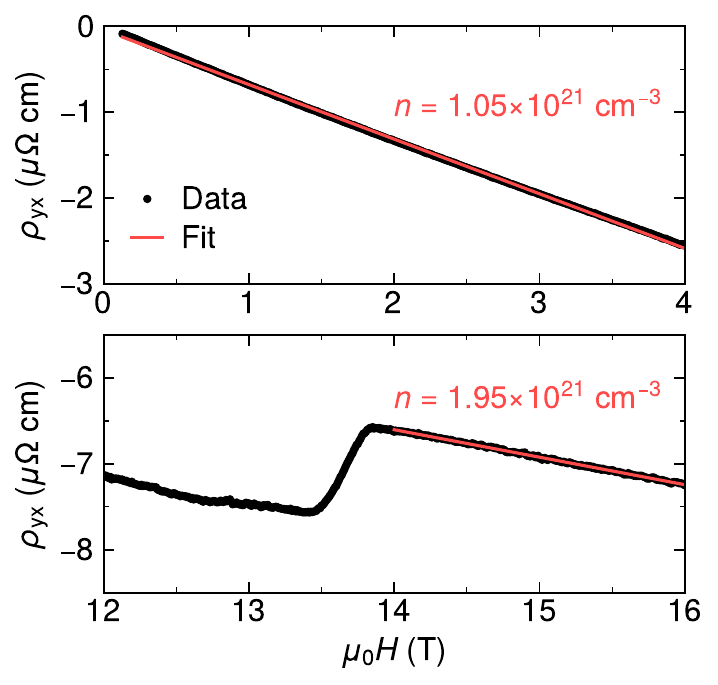}
    \caption{The carrier concentration of the AFM~1 and field-polarized (FP) magnetic phases at 2~K are extracted from the ordinary Hall coefficient assuming single-band behavior. The fits are to the magnetic field ranges 0--4~T for AFM~1 (ordinary and anomalous terms) and 14--16~T for FP (ordinary term only). The slope is roughly halved in the FP phase, leading to a carrier concentration change of almost a factor of 2.}
    \label{fig:carrier}
\end{figure}

At 2~K, the resistivity in the FP phase is fit well with an ordinary Hall term, yielding a Hall coefficient ($R_\mathrm{H}$) of -3.20$\times$10$^{-3}$~cm$^{3}$/C for the range 14--16~T. Electrons are the dominant charge carriers. Assuming a single-carrier model, the relationship $R_\mathrm{H}$~=~($ne$)$^{-1}$, where $n$ is the carrier concentration and $e$ is the elementary charge, was used to determine a carrier concentration of 1.95$\times$10$^{21}$~cm$^{-3}$ for FP (Fig.~\ref{fig:carrier}). In the paramagnetic state at 60~K, the transverse resistivity is also fit well with a single-carrier model (Fig.~S9 in \cite{supplemental}), yielding \mbox{$R_\mathrm{H}$=-4.49$\times$10$^{-3}$~cm$^{3}$/C} and $n=$1.39$\times$10$^{21}$~cm$^{-3}$ for 0-16~T.

A simple model of the transverse resistivity (ZFC, increasing field), including an ordinary and anomalous term, i.e., 
\begin{equation} \label{eq:Hall}
   \rho_\mathrm{yx} = R_H (\mu_0 H) + \beta (M), 
\end{equation}
reasonably describes the low-field phases AFM~1, P3, and P4. An example fit in the AFM~1 region is shown in Fig.~\ref{fig:carrier}. But deviations occur in higher-field phases, especially for P5 and P2, where $\rho_\mathrm{yx}$ and $M$ differ significantly. Introducing a term proportional to $\rho^2_{\mathrm{xx}} M$ does not improve the fit. Alternative fitting methods, including fitting the Hall resistivity to Eq.~\ref{eq:Hall} piece-wise in the magnetic phases, are presented in the supplemental material \cite{supplemental}; these fits lead to the same conclusion that significant differences exist between the Hall resistivity and magnetization in the P5 and P2 phases. Deviations from a simple anomalous term proportional to magnetization could signal that the \UNS\ Hall resistivity is dominated by intrinsic effects in those regions. On the other hand, these deviations could be due to data collection procedures, sample variability, or misalignment of the \UNS\ crystals, but the same sweep rates were used in the $M$($H$), $\rho_\mathrm{yx}$($H$), and $\rho_\mathrm{xx}$($H$) measurements, and the $M$($H$) and $\rho_\mathrm{xx}$($H$) curves contain reasonably similar transition points despite being collected on different \UNS\ crystals (see the magnetic phase diagram in Fig.~\ref{fig:magpd_rhoH}a for transition points and Fig.~S10 in \cite{supplemental} for feature comparisons). Still, in the P2 region, an alternative to Berry curvature explanations for the large deviation is a significant change in the Fermi surface and/or multiband behavior. Along those lines, the Hall resistivity slope in the P2 region is similar to that of the FP region, previously noted in Fig.~\ref{fig:carrier} to be roughly half that observed for AFM~1, and the similarity of the P2 and FP magnetization values suggests similar magnetic structures. The P2 and FP regions, though, cannot be fit to the same ordinary and anomalous Hall terms without ignoring the transverse resistivity jump at the transition between them.

\begin{figure}
    \centering
    \includegraphics[width=0.9\columnwidth]{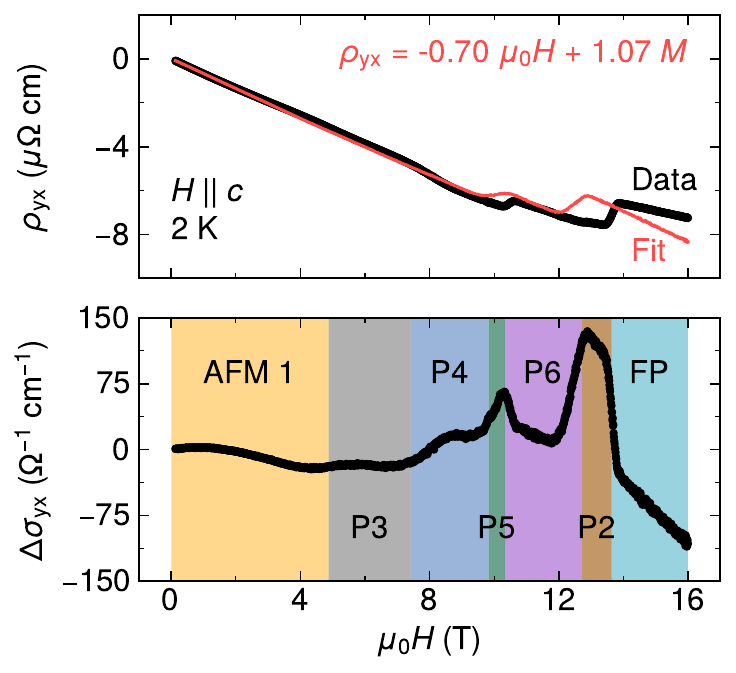}
\caption{The transverse resistivity at 2~K of \UNS\ is fit up to 16~T with linear terms proportional to applied magnetic field and magnetization. The data and fit are converted to conductivity, and the difference curve shows pronounced changes in the P2 and P5 magnetic phase regions, defined as in Fig.~\ref{fig:magpd_rhoH}a.}
    \label{fig:Hall_fit}
\end{figure}

To further highlight the regions where the fit to Eq.~\ref{eq:Hall} poorly accounts for the transverse resistivity, the resistivity data and fit are converted to conductivity via Eq.~\ref{eq:conductivity}:
\begin{equation} \label{eq:conductivity}
    \sigma_{\mathrm{yx}} = \frac{-\rho_{\mathrm{yx}}}{\rho^2_{\mathrm{xx}} + \rho^2_{\mathrm{yx}}}
\end{equation}
The difference between the fit and the data (${\Delta}{\sigma}_{\mathrm{yx}} = {\sigma}_{\mathrm{Data}} - {\sigma}_{\mathrm{Fit}}$) is shown in the lower panel of Fig.~\ref{fig:Hall_fit} with regions shaded based on the phase boundaries of the magnetic phase diagram in Fig.~\ref{fig:magpd_rhoH}a. The largest conductivity differences occur within the P2 and P5 phases. (The slope change in the FP region may be accounted for by the change in carrier concentration as discussed above.) Similar fit deviations in other materials indicate large Berry curvature in noncollinear spin textures and topologically nontrivial phases \cite{chen2014anomalous,thomas2016hall,asaba2020anomalous,chen2021large,shang2021anomalous}, though the large number of magnetic transitions in \UNS, the aforementioned potential Fermi surface changes, and the unknown magnetic structures of those phases should qualify a similar conclusion for \UNS\ with the currently available data. Assuming the conductivity difference does represent an intrinsic contribution to the Hall response, the maximum of the difference curve is 134.5~${\Omega}^{-1}$~cm$^{-1}$ in the P2 region. A two-dimensional electronic structure can demonstrate a Hall effect quantized in units of $e^2/h$ \cite{von2017quantum}, so the difference maximum is converted to the value 0.3301~$e^2$/$hc$, where $e$, $h$, and $c$ are the elementary charge, Planck constant, and $c$ lattice parameter, respectively. Further, since \UNS\ contains two Kagome layers per unit cell, the value reduces to 0.1650~$e^2$/$h$ per layer. This value is lower than that of GdMn$_6$Sn$_6$ (0.27~$e^2$/$h$ per Kagome layer) \cite{asaba2020anomalous}.

\section{Band structure}
\begin{figure} 
    \hspace{0.55em}
    \subfloat{%
      \includegraphics[width=0.8\columnwidth]{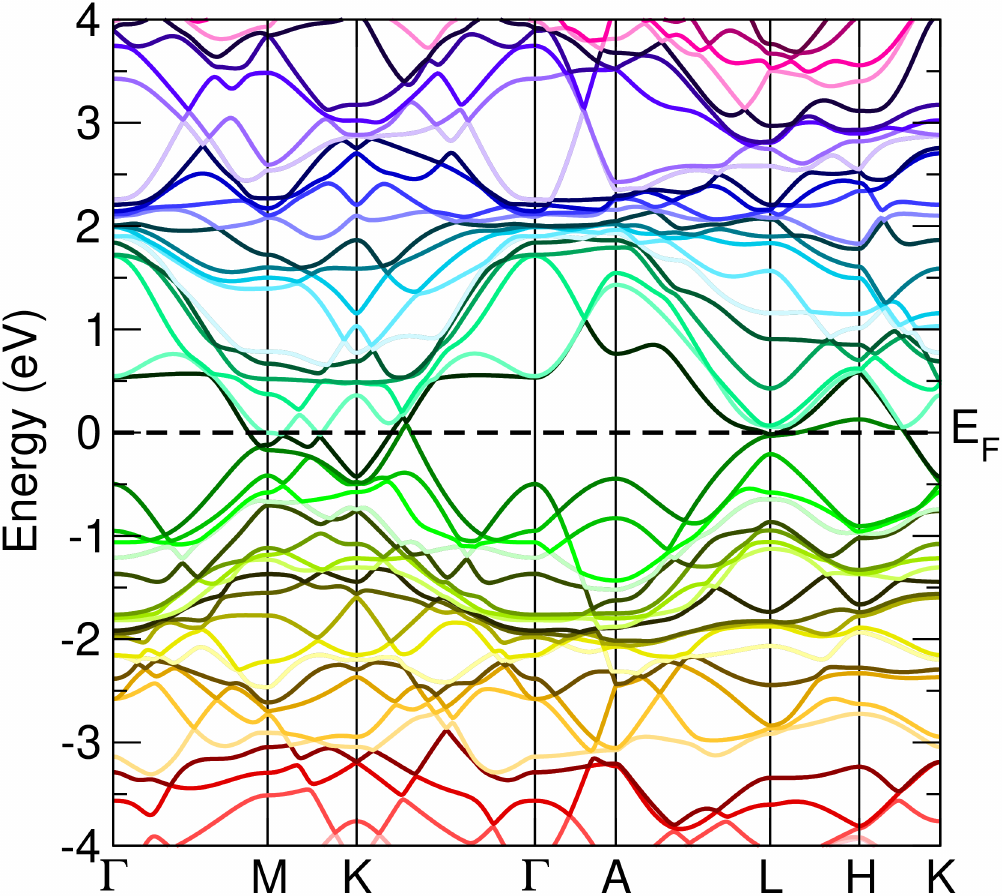}%
    }

    \subfloat{%
      \includegraphics[width=0.75\columnwidth]{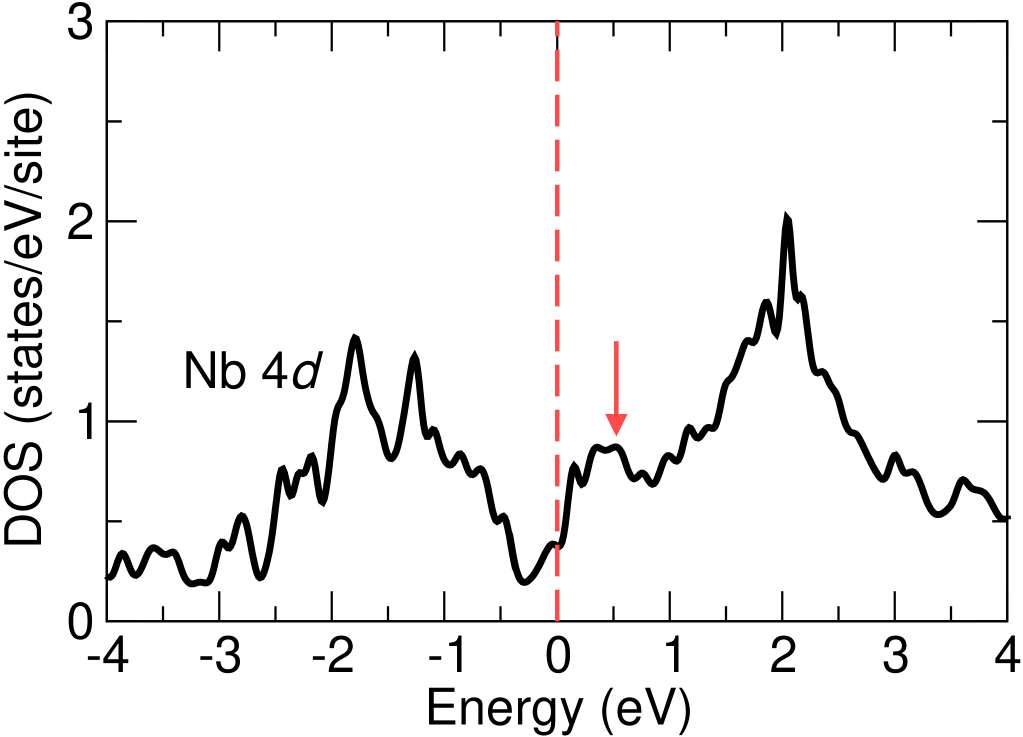}%
    }
    \caption{The electronic band structure of \UNS\ and the corresponding Nb 4$d$ electron density of states (DOS) are calculated assuming localized uranium 5$f$ electrons. The DOS arrow points to the feature associated with the Kagome layer flat band.}
    \label{fig:dft}
\end{figure}

Fig.~\ref{fig:dft} shows the calculated electronic band structure and density of states (DOS) of \UNS\ along various high-symmetry momentum paths in the Brillouin zone. Assuming trivalent uranium, the uranium 5$f$ electrons are treated as core states. The crystal structure is also assumed to be fully ordered. A very narrow band at 0.5~eV above the Fermi energy ($E_\mathrm{F}$) originates from the destructive interference of the niobium 4$d$ electrons on the Kagome lattice. This narrow band is clearly seen in the $\Gamma$ plane of the Brillouin zone, while it is obscured by multiple intersecting bands and avoided crossings in the $A$ plane. Interestingly, in both the band structure and the density of states, the niobium band appears broader than that of the corresponding vanadium 3$d$ states in UV$_6$Sn$_6$ \cite{thomas2025uv6sn6}. Since 4$d$ orbitals are more delocalized, they hybridize more readily with neighboring tin atoms and facilitate strong interlayer coupling, thereby weakening the interference effect and broadening the band. This suggests that 4$d$ and 5$d$ transition metals may intrinsically warp the Kagome flat band more than the more localized $d$ orbitals of 3$d$ elements. Additionally, several van Hove singularities are observed at or near $E_\mathrm{F}$ for momentum points $M$ and $L$, along with multiple possible Dirac band crossings along $M$--$K$ and $K$--$\Gamma$. The van Hove singularity at $L$ distinguishes the band structure of \UNS\ from that of the nonmagnetic LuNb$_6$Sn$_6$ \cite{ortiz2024stability}, where the same band is roughly 0.25-0.5~eV above $E_\mathrm{F}$. Similarly, for UV$_6$Sn$_6$, a van Hove singularity near $E_\mathrm{F}$ at $L$ is evidenced by band structure calculations and ARPES data \cite{thomas2025uv6sn6}.

\section{Conclusions}
The Kagome metal \UNS\ demonstrates a complex interplay of crystallographic disorder and competing magnetic exchange interactions, leading to a cascade of uranium-driven, field-induced transitions when a magnetic field is applied along the $c$ axis. The magnetic susceptibility of \UNS\ suggests a 5$f^2$ or 5$f^3$ electronic configuration for uranium and nonmagnetic niobium, and the zero-field heat capacity indicates a ground state doublet. At zero magnetic field, \UNS\ orders magnetically below $T_\mathrm{2}=46$ K in a narrow temperature-field region (P2), likely in an incommensurate or long-period commensurate phase, before entering the AFM~1 phase below $T_\mathrm{N}=42.5-43$~K. AFM~1 is an A-type antiferromagnetic phase with a (0,0,1/2) propagation vector. At 2~K with $H{\parallel}c$, \UNS\ undergoes six magnetic phase transitions, culminating in a field-polarized state above 13.6~T. At 2~K, the magnetization and resistivity are strongly hysteretic. The transverse resistivity suggests that two of the magnetic phases (P2, P5) may have a large intrinsic Berry curvature contribution or have significant Fermi surface changes relative to neighboring magnetic phases. The unknown magnetic phases of \UNS\ offer an exciting opportunity to probe emergent behavior in Kagome materials.

\begin{acknowledgments}
Work at Los Alamos National Laboratory was performed under the auspices of the U.S. Department of
Energy, Office of Basic Energy Sciences, Division of Materials Science and Engineering. Z.W.R., C.S.K., A.O.S., and W.S. acknowledge support from the Laboratory Directed Research and Development program. C.S.K. gratefully acknowledges the support of the U.S. Department of Energy through the G. T. Seaborg Institute.
This work is partly based on experiments performed at the Swiss spallation neutron source SINQ, Paul Scherrer Institute, Villigen, Switzerland.
The authors thank D. Sheptyakov for support with the neutron powder diffraction experiments.
The electronic structure calculations were supported in part by the Center for Integrated Nanotechnologies, a DOE BES user facility, in partnership with the LANL Institutional Computing Program for computational resources. Additional computations were performed at the National Energy Research Scientific Computing Center (NERSC), a U.S. Department of Energy Office of Science User Facility located at Lawrence Berkeley National Laboratory, operated under Contract No. DE-AC02-05CH11231 using NERSC award ERCAP0028014.
\end{acknowledgments}

\bibliography{UNb6Sn6.bib}

\begin{thebibliography}{82}%
\makeatletter
\providecommand \@ifxundefined [1]{%
 \@ifx{#1\undefined}
}%
\providecommand \@ifnum [1]{%
 \ifnum #1\expandafter \@firstoftwo
 \else \expandafter \@secondoftwo
 \fi
}%
\providecommand \@ifx [1]{%
 \ifx #1\expandafter \@firstoftwo
 \else \expandafter \@secondoftwo
 \fi
}%
\providecommand \natexlab [1]{#1}%
\providecommand \enquote  [1]{``#1''}%
\providecommand \bibnamefont  [1]{#1}%
\providecommand \bibfnamefont [1]{#1}%
\providecommand \citenamefont [1]{#1}%
\providecommand \href@noop [0]{\@secondoftwo}%
\providecommand \href [0]{\begingroup \@sanitize@url \@href}%
\providecommand \@href[1]{\@@startlink{#1}\@@href}%
\providecommand \@@href[1]{\endgroup#1\@@endlink}%
\providecommand \@sanitize@url [0]{\catcode `\\12\catcode `\$12\catcode
  `\&12\catcode `\#12\catcode `\^12\catcode `\_12\catcode `\%12\relax}%
\providecommand \@@startlink[1]{}%
\providecommand \@@endlink[0]{}%
\providecommand \url  [0]{\begingroup\@sanitize@url \@url }%
\providecommand \@url [1]{\endgroup\@href {#1}{\urlprefix }}%
\providecommand \urlprefix  [0]{URL }%
\providecommand \Eprint [0]{\href }%
\providecommand \doibase [0]{https://doi.org/}%
\providecommand \selectlanguage [0]{\@gobble}%
\providecommand \bibinfo  [0]{\@secondoftwo}%
\providecommand \bibfield  [0]{\@secondoftwo}%
\providecommand \translation [1]{[#1]}%
\providecommand \BibitemOpen [0]{}%
\providecommand \bibitemStop [0]{}%
\providecommand \bibitemNoStop [0]{.\EOS\space}%
\providecommand \EOS [0]{\spacefactor3000\relax}%
\providecommand \BibitemShut  [1]{\csname bibitem#1\endcsname}%
\let\auto@bib@innerbib\@empty
\bibitem [{\citenamefont {Kiesel}\ \emph {et~al.}(2013)\citenamefont {Kiesel},
  \citenamefont {Platt},\ and\ \citenamefont
  {Thomale}}]{kiesel2013unconventional}%
  \BibitemOpen
  \bibfield  {author} {\bibinfo {author} {\bibfnamefont {M.~L.}\ \bibnamefont
  {Kiesel}}, \bibinfo {author} {\bibfnamefont {C.}~\bibnamefont {Platt}},\ and\
  \bibinfo {author} {\bibfnamefont {R.}~\bibnamefont {Thomale}},\ }\bibfield
  {title} {\bibinfo {title} {Unconventional {Fermi} surface instabilities in
  the kagome {Hubbard} model},\ }\href@noop {} {\bibfield  {journal} {\bibinfo
  {journal} {Physical Review Letters}\ }\textbf {\bibinfo {volume} {110}},\
  \bibinfo {pages} {126405} (\bibinfo {year} {2013})}\BibitemShut {NoStop}%
\bibitem [{\citenamefont {Yin}\ \emph {et~al.}(2022)\citenamefont {Yin},
  \citenamefont {Lian},\ and\ \citenamefont {Hasan}}]{yin2022topological}%
  \BibitemOpen
  \bibfield  {author} {\bibinfo {author} {\bibfnamefont {J.-X.}\ \bibnamefont
  {Yin}}, \bibinfo {author} {\bibfnamefont {B.}~\bibnamefont {Lian}},\ and\
  \bibinfo {author} {\bibfnamefont {M.~Z.}\ \bibnamefont {Hasan}},\ }\bibfield
  {title} {\bibinfo {title} {Topological kagome magnets and superconductors},\
  }\href@noop {} {\bibfield  {journal} {\bibinfo  {journal} {Nature}\ }\textbf
  {\bibinfo {volume} {612}},\ \bibinfo {pages} {647} (\bibinfo {year}
  {2022})}\BibitemShut {NoStop}%
\bibitem [{\citenamefont {Venturini}(2006)}]{venturini2006filling}%
  \BibitemOpen
  \bibfield  {author} {\bibinfo {author} {\bibfnamefont {G.}~\bibnamefont
  {Venturini}},\ }\bibfield  {title} {\bibinfo {title} {Filling the {CoSn}
  host-cell: {The} {HfFe$_6$Ge$_6$}-type and the related structures},\
  }\href@noop {} {\bibfield  {journal} {\bibinfo  {journal} {Zeitschrift
  f{\"u}r Kristallographie-Crystalline Materials}\ }\textbf {\bibinfo {volume}
  {221}},\ \bibinfo {pages} {511} (\bibinfo {year} {2006})}\BibitemShut
  {NoStop}%
\bibitem [{\citenamefont {Fredrickson}\ \emph {et~al.}(2008)\citenamefont
  {Fredrickson}, \citenamefont {Lidin}, \citenamefont {Venturini},
  \citenamefont {Malaman},\ and\ \citenamefont
  {Christensen}}]{fredrickson2008origins}%
  \BibitemOpen
  \bibfield  {author} {\bibinfo {author} {\bibfnamefont {D.~C.}\ \bibnamefont
  {Fredrickson}}, \bibinfo {author} {\bibfnamefont {S.}~\bibnamefont {Lidin}},
  \bibinfo {author} {\bibfnamefont {G.}~\bibnamefont {Venturini}}, \bibinfo
  {author} {\bibfnamefont {B.}~\bibnamefont {Malaman}},\ and\ \bibinfo {author}
  {\bibfnamefont {J.}~\bibnamefont {Christensen}},\ }\bibfield  {title}
  {\bibinfo {title} {Origins of superstructure ordering and incommensurability
  in stuffed {CoSn}-type phases},\ }\href@noop {} {\bibfield  {journal}
  {\bibinfo  {journal} {Journal of the American Chemical Society}\ }\textbf
  {\bibinfo {volume} {130}},\ \bibinfo {pages} {8195} (\bibinfo {year}
  {2008})}\BibitemShut {NoStop}%
\bibitem [{\citenamefont {Clatterbuck}\ and\ \citenamefont
  {Gschneidner~Jr}(1999)}]{clatterbuck1999magnetic}%
  \BibitemOpen
  \bibfield  {author} {\bibinfo {author} {\bibfnamefont {D.~M.}\ \bibnamefont
  {Clatterbuck}}\ and\ \bibinfo {author} {\bibfnamefont {K.~A.}\ \bibnamefont
  {Gschneidner~Jr}},\ }\bibfield  {title} {\bibinfo {title} {Magnetic
  properties of {RMn$_6$Sn$_6$ (R=Tb, Ho, Er, Tm, Lu)} single crystals},\
  }\href@noop {} {\bibfield  {journal} {\bibinfo  {journal} {Journal of
  Magnetism and Magnetic Materials}\ }\textbf {\bibinfo {volume} {207}},\
  \bibinfo {pages} {78} (\bibinfo {year} {1999})}\BibitemShut {NoStop}%
\bibitem [{\citenamefont {Kimura}\ \emph {et~al.}(2006)\citenamefont {Kimura},
  \citenamefont {Matsuo}, \citenamefont {Yoshii}, \citenamefont {Kindo},
  \citenamefont {Zhang}, \citenamefont {Br{\"u}ck}, \citenamefont {Buschow},
  \citenamefont {De~Boer}, \citenamefont {Lef{\`e}vre},\ and\ \citenamefont
  {Venturini}}]{kimura2006high}%
  \BibitemOpen
  \bibfield  {author} {\bibinfo {author} {\bibfnamefont {S.}~\bibnamefont
  {Kimura}}, \bibinfo {author} {\bibfnamefont {A.}~\bibnamefont {Matsuo}},
  \bibinfo {author} {\bibfnamefont {S.}~\bibnamefont {Yoshii}}, \bibinfo
  {author} {\bibfnamefont {K.}~\bibnamefont {Kindo}}, \bibinfo {author}
  {\bibfnamefont {L.}~\bibnamefont {Zhang}}, \bibinfo {author} {\bibfnamefont
  {E.}~\bibnamefont {Br{\"u}ck}}, \bibinfo {author} {\bibfnamefont {K.~H.~J.}\
  \bibnamefont {Buschow}}, \bibinfo {author} {\bibfnamefont {F.~R.}\
  \bibnamefont {De~Boer}}, \bibinfo {author} {\bibfnamefont {C.}~\bibnamefont
  {Lef{\`e}vre}},\ and\ \bibinfo {author} {\bibfnamefont {G.}~\bibnamefont
  {Venturini}},\ }\bibfield  {title} {\bibinfo {title} {High-field
  magnetization of {RMn$_6$Sn$_6$} compounds with {R=Gd, Tb, Dy and Ho}},\
  }\href@noop {} {\bibfield  {journal} {\bibinfo  {journal} {Journal of Alloys
  and Compounds}\ }\textbf {\bibinfo {volume} {408}},\ \bibinfo {pages} {169}
  (\bibinfo {year} {2006})}\BibitemShut {NoStop}%
\bibitem [{\citenamefont {Riberolles}\ \emph {et~al.}(2024)\citenamefont
  {Riberolles}, \citenamefont {Han}, \citenamefont {Slade}, \citenamefont
  {Wilde}, \citenamefont {Sapkota}, \citenamefont {Tian}, \citenamefont
  {Zhang}, \citenamefont {Abernathy}, \citenamefont {Sanjeewa}, \citenamefont
  {Bud’ko}, \citenamefont {Canfield}, \citenamefont {McQueeney},\ and\
  \citenamefont {Ueland}}]{riberolles2024new}%
  \BibitemOpen
  \bibfield  {author} {\bibinfo {author} {\bibfnamefont {S.~X.~M.}\
  \bibnamefont {Riberolles}}, \bibinfo {author} {\bibfnamefont
  {T.}~\bibnamefont {Han}}, \bibinfo {author} {\bibfnamefont {T.~J.}\
  \bibnamefont {Slade}}, \bibinfo {author} {\bibfnamefont {J.~M.}\ \bibnamefont
  {Wilde}}, \bibinfo {author} {\bibfnamefont {A.}~\bibnamefont {Sapkota}},
  \bibinfo {author} {\bibfnamefont {W.}~\bibnamefont {Tian}}, \bibinfo {author}
  {\bibfnamefont {Q.}~\bibnamefont {Zhang}}, \bibinfo {author} {\bibfnamefont
  {D.~L.}\ \bibnamefont {Abernathy}}, \bibinfo {author} {\bibfnamefont {L.~D.}\
  \bibnamefont {Sanjeewa}}, \bibinfo {author} {\bibfnamefont {S.~L.}\
  \bibnamefont {Bud’ko}}, \bibinfo {author} {\bibfnamefont {P.~C.}\
  \bibnamefont {Canfield}}, \bibinfo {author} {\bibfnamefont {R.~J.}\
  \bibnamefont {McQueeney}},\ and\ \bibinfo {author} {\bibfnamefont {B.~G.}\
  \bibnamefont {Ueland}},\ }\bibfield  {title} {\bibinfo {title} {New insight
  into tuning magnetic phases of {$R$Mn$_6$Sn$_6$} kagome metals},\ }\href@noop
  {} {\bibfield  {journal} {\bibinfo  {journal} {npj Quantum Materials}\
  }\textbf {\bibinfo {volume} {9}},\ \bibinfo {pages} {42} (\bibinfo {year}
  {2024})}\BibitemShut {NoStop}%
\bibitem [{\citenamefont {Ghimire}\ \emph {et~al.}(2020)\citenamefont
  {Ghimire}, \citenamefont {Dally}, \citenamefont {Poudel}, \citenamefont
  {Jones}, \citenamefont {Michel}, \citenamefont {Magar}, \citenamefont
  {Bleuel}, \citenamefont {McGuire}, \citenamefont {Jiang}, \citenamefont
  {Mitchell}, \citenamefont {Lynn},\ and\ \citenamefont
  {Mazin}}]{ghimire2020competing}%
  \BibitemOpen
  \bibfield  {author} {\bibinfo {author} {\bibfnamefont {N.~J.}\ \bibnamefont
  {Ghimire}}, \bibinfo {author} {\bibfnamefont {R.~L.}\ \bibnamefont {Dally}},
  \bibinfo {author} {\bibfnamefont {L.}~\bibnamefont {Poudel}}, \bibinfo
  {author} {\bibfnamefont {D.~C.}\ \bibnamefont {Jones}}, \bibinfo {author}
  {\bibfnamefont {D.}~\bibnamefont {Michel}}, \bibinfo {author} {\bibfnamefont
  {N.~T.}\ \bibnamefont {Magar}}, \bibinfo {author} {\bibfnamefont
  {M.}~\bibnamefont {Bleuel}}, \bibinfo {author} {\bibfnamefont {M.~A.}\
  \bibnamefont {McGuire}}, \bibinfo {author} {\bibfnamefont {J.~S.}\
  \bibnamefont {Jiang}}, \bibinfo {author} {\bibfnamefont {J.~F.}\ \bibnamefont
  {Mitchell}}, \bibinfo {author} {\bibfnamefont {J.~W.}\ \bibnamefont {Lynn}},\
  and\ \bibinfo {author} {\bibfnamefont {I.~I.}\ \bibnamefont {Mazin}},\
  }\bibfield  {title} {\bibinfo {title} {Competing magnetic phases and
  fluctuation-driven scalar spin chirality in the kagome metal
  {YMn$_6$Sn$_6$}},\ }\href@noop {} {\bibfield  {journal} {\bibinfo  {journal}
  {Science Advances}\ }\textbf {\bibinfo {volume} {6}},\ \bibinfo {pages}
  {eabe2680} (\bibinfo {year} {2020})}\BibitemShut {NoStop}%
\bibitem [{\citenamefont {Li}\ \emph {et~al.}(2024)\citenamefont {Li},
  \citenamefont {Victa-Trevisan},\ and\ \citenamefont
  {McQueeney}}]{li2024highfield}%
  \BibitemOpen
  \bibfield  {author} {\bibinfo {author} {\bibfnamefont {F.}~\bibnamefont
  {Li}}, \bibinfo {author} {\bibfnamefont {T.}~\bibnamefont {Victa-Trevisan}},\
  and\ \bibinfo {author} {\bibfnamefont {R.~J.}\ \bibnamefont {McQueeney}},\
  }\href {https://arxiv.org/abs/2409.04273} {\bibinfo {title} {High-field
  magnetic phase diagrams of the {$R$Mn$_6$Sn$_6$} kagome metals}} (\bibinfo
  {year} {2024}),\ \Eprint {https://arxiv.org/abs/2409.04273} {arXiv:2409.04273
  [cond-mat.str-el]} \BibitemShut {NoStop}%
\bibitem [{\citenamefont {L}\ \emph {et~al.}(2025)\citenamefont {L},
  \citenamefont {Trevisan},\ and\ \citenamefont {McQueeney}}]{victa2025high}%
  \BibitemOpen
  \bibfield  {author} {\bibinfo {author} {\bibfnamefont {N.}~\bibnamefont {L}},
  \bibinfo {author} {\bibfnamefont {T.~V.}\ \bibnamefont {Trevisan}},\ and\
  \bibinfo {author} {\bibfnamefont {R.~J.}\ \bibnamefont {McQueeney}},\
  }\bibfield  {title} {\bibinfo {title} {High-field magnetic phase diagrams of
  the {$R$Mn$_6$Sn$_6$ ($R$= Gd--Tm)} kagome metals},\ }\href@noop {}
  {\bibfield  {journal} {\bibinfo  {journal} {Physical Review B}\ }\textbf
  {\bibinfo {volume} {111}},\ \bibinfo {pages} {054410} (\bibinfo {year}
  {2025})}\BibitemShut {NoStop}%
\bibitem [{\citenamefont {Malaman}\ \emph {et~al.}(1997)\citenamefont
  {Malaman}, \citenamefont {Venturini}, \citenamefont {El~Idrissi},\ and\
  \citenamefont {Ressouche}}]{malaman1997magnetic}%
  \BibitemOpen
  \bibfield  {author} {\bibinfo {author} {\bibfnamefont {B.}~\bibnamefont
  {Malaman}}, \bibinfo {author} {\bibfnamefont {G.}~\bibnamefont {Venturini}},
  \bibinfo {author} {\bibfnamefont {B.~C.}\ \bibnamefont {El~Idrissi}},\ and\
  \bibinfo {author} {\bibfnamefont {E.}~\bibnamefont {Ressouche}},\ }\bibfield
  {title} {\bibinfo {title} {Magnetic properties of {NdMn$_6$Sn$_6$} and
  {SmMn$_6$Sn$_6$} compounds from susceptibility measurements and neutron
  diffraction study},\ }\href@noop {} {\bibfield  {journal} {\bibinfo
  {journal} {Journal of Alloys and Compounds}\ }\textbf {\bibinfo {volume}
  {252}},\ \bibinfo {pages} {41} (\bibinfo {year} {1997})}\BibitemShut
  {NoStop}%
\bibitem [{\citenamefont {Waerenborgh}\ \emph {et~al.}(2005)\citenamefont
  {Waerenborgh}, \citenamefont {Pereira}, \citenamefont {Gon{\c{c}}alves},\
  and\ \citenamefont {No{\"e}l}}]{waerenborgh2005crystal}%
  \BibitemOpen
  \bibfield  {author} {\bibinfo {author} {\bibfnamefont {J.~C.}\ \bibnamefont
  {Waerenborgh}}, \bibinfo {author} {\bibfnamefont {L.~C.~J.}\ \bibnamefont
  {Pereira}}, \bibinfo {author} {\bibfnamefont {A.~P.}\ \bibnamefont
  {Gon{\c{c}}alves}},\ and\ \bibinfo {author} {\bibfnamefont {H.}~\bibnamefont
  {No{\"e}l}},\ }\bibfield  {title} {\bibinfo {title} {Crystal structure,
  $^{57}${Fe} {M{\"o}ssbauer} spectroscopy and magnetization of
  {U$_x$Fe$_6$Sn$6$} (0${\leq}x{\leq}$0.6)},\ }\href@noop {} {\bibfield
  {journal} {\bibinfo  {journal} {Intermetallics}\ }\textbf {\bibinfo {volume}
  {13}},\ \bibinfo {pages} {490} (\bibinfo {year} {2005})}\BibitemShut
  {NoStop}%
\bibitem [{\citenamefont {Tan}\ and\ \citenamefont
  {Yan}(2023)}]{tan2023abundant}%
  \BibitemOpen
  \bibfield  {author} {\bibinfo {author} {\bibfnamefont {H.}~\bibnamefont
  {Tan}}\ and\ \bibinfo {author} {\bibfnamefont {B.}~\bibnamefont {Yan}},\
  }\bibfield  {title} {\bibinfo {title} {Abundant lattice instability in kagome
  metal {ScV$_6$Sn$_6$}},\ }\href@noop {} {\bibfield  {journal} {\bibinfo
  {journal} {Physical Review Letters}\ }\textbf {\bibinfo {volume} {130}},\
  \bibinfo {pages} {266402} (\bibinfo {year} {2023})}\BibitemShut {NoStop}%
\bibitem [{\citenamefont {Feng}\ \emph {et~al.}(2024)\citenamefont {Feng},
  \citenamefont {Jiang}, \citenamefont {Hu}, \citenamefont
  {C{\u{a}}lug{\u{a}}ru}, \citenamefont {Regnault}, \citenamefont {Vergniory},
  \citenamefont {Felser}, \citenamefont {Blanco-Canosa},\ and\ \citenamefont
  {Bernevig}}]{feng2024catalogue}%
  \BibitemOpen
  \bibfield  {author} {\bibinfo {author} {\bibfnamefont {X.}~\bibnamefont
  {Feng}}, \bibinfo {author} {\bibfnamefont {Y.}~\bibnamefont {Jiang}},
  \bibinfo {author} {\bibfnamefont {H.}~\bibnamefont {Hu}}, \bibinfo {author}
  {\bibfnamefont {D.}~\bibnamefont {C{\u{a}}lug{\u{a}}ru}}, \bibinfo {author}
  {\bibfnamefont {N.}~\bibnamefont {Regnault}}, \bibinfo {author}
  {\bibfnamefont {M.}~\bibnamefont {Vergniory}}, \bibinfo {author}
  {\bibfnamefont {C.}~\bibnamefont {Felser}}, \bibinfo {author} {\bibfnamefont
  {S.}~\bibnamefont {Blanco-Canosa}},\ and\ \bibinfo {author} {\bibfnamefont
  {B.~A.}\ \bibnamefont {Bernevig}},\ }\href {https://arxiv.org/abs/2409.13078}
  {\bibinfo {title} {Catalogue of phonon instabilities in symmetry group 191
  kagome {MT$_6$Z$_6$} materials}} (\bibinfo {year} {2024}),\ \Eprint
  {https://arxiv.org/abs/2409.13078} {arXiv:2409.13078 [cond-mat.mtrl-sci]}
  \BibitemShut {NoStop}%
\bibitem [{\citenamefont {Arachchige}\ \emph {et~al.}(2022)\citenamefont
  {Arachchige}, \citenamefont {Meier}, \citenamefont {Marshall}, \citenamefont
  {Matsuoka}, \citenamefont {Xue}, \citenamefont {McGuire}, \citenamefont
  {Hermann}, \citenamefont {Cao},\ and\ \citenamefont
  {Mandrus}}]{arachchige2022charge}%
  \BibitemOpen
  \bibfield  {author} {\bibinfo {author} {\bibfnamefont {H.~W.~S.}\
  \bibnamefont {Arachchige}}, \bibinfo {author} {\bibfnamefont {W.~R.}\
  \bibnamefont {Meier}}, \bibinfo {author} {\bibfnamefont {M.}~\bibnamefont
  {Marshall}}, \bibinfo {author} {\bibfnamefont {T.}~\bibnamefont {Matsuoka}},
  \bibinfo {author} {\bibfnamefont {R.}~\bibnamefont {Xue}}, \bibinfo {author}
  {\bibfnamefont {M.~A.}\ \bibnamefont {McGuire}}, \bibinfo {author}
  {\bibfnamefont {R.~P.}\ \bibnamefont {Hermann}}, \bibinfo {author}
  {\bibfnamefont {H.}~\bibnamefont {Cao}},\ and\ \bibinfo {author}
  {\bibfnamefont {D.}~\bibnamefont {Mandrus}},\ }\bibfield  {title} {\bibinfo
  {title} {Charge density wave in kagome lattice intermetallic
  {ScV$_6$Sn$_6$}},\ }\href@noop {} {\bibfield  {journal} {\bibinfo  {journal}
  {Physical Review Letters}\ }\textbf {\bibinfo {volume} {129}},\ \bibinfo
  {pages} {216402} (\bibinfo {year} {2022})}\BibitemShut {NoStop}%
\bibitem [{\citenamefont {Tuniz}\ \emph {et~al.}(2023)\citenamefont {Tuniz},
  \citenamefont {Consiglio}, \citenamefont {Puntel}, \citenamefont {Bigi},
  \citenamefont {Enzner}, \citenamefont {Pokharel}, \citenamefont {Orgiani},
  \citenamefont {Bronsch}, \citenamefont {Parmigiani}, \citenamefont
  {Polewczyk} \emph {et~al.}}]{tuniz2023dynamics}%
  \BibitemOpen
  \bibfield  {author} {\bibinfo {author} {\bibfnamefont {M.}~\bibnamefont
  {Tuniz}}, \bibinfo {author} {\bibfnamefont {A.}~\bibnamefont {Consiglio}},
  \bibinfo {author} {\bibfnamefont {D.}~\bibnamefont {Puntel}}, \bibinfo
  {author} {\bibfnamefont {C.}~\bibnamefont {Bigi}}, \bibinfo {author}
  {\bibfnamefont {S.}~\bibnamefont {Enzner}}, \bibinfo {author} {\bibfnamefont
  {G.}~\bibnamefont {Pokharel}}, \bibinfo {author} {\bibfnamefont
  {P.}~\bibnamefont {Orgiani}}, \bibinfo {author} {\bibfnamefont
  {W.}~\bibnamefont {Bronsch}}, \bibinfo {author} {\bibfnamefont
  {F.}~\bibnamefont {Parmigiani}}, \bibinfo {author} {\bibfnamefont
  {V.}~\bibnamefont {Polewczyk}}, \emph {et~al.},\ }\bibfield  {title}
  {\bibinfo {title} {Dynamics and resilience of the unconventional charge
  density wave in {ScV$_6$Sn$_6$} bilayer kagome metal},\ }\href@noop {}
  {\bibfield  {journal} {\bibinfo  {journal} {Communications Materials}\
  }\textbf {\bibinfo {volume} {4}},\ \bibinfo {pages} {103} (\bibinfo {year}
  {2023})}\BibitemShut {NoStop}%
\bibitem [{\citenamefont {Pokharel}\ \emph {et~al.}(2023)\citenamefont
  {Pokharel}, \citenamefont {Ortiz}, \citenamefont {Kautzsch}, \citenamefont
  {Gomez~Alvarado}, \citenamefont {Mallayya}, \citenamefont {Wu}, \citenamefont
  {Kim}, \citenamefont {Ruff}, \citenamefont {Sarker},\ and\ \citenamefont
  {Wilson}}]{pokharel2023frustrated}%
  \BibitemOpen
  \bibfield  {author} {\bibinfo {author} {\bibfnamefont {G.}~\bibnamefont
  {Pokharel}}, \bibinfo {author} {\bibfnamefont {B.~R.}\ \bibnamefont {Ortiz}},
  \bibinfo {author} {\bibfnamefont {L.}~\bibnamefont {Kautzsch}}, \bibinfo
  {author} {\bibfnamefont {S.~J.}\ \bibnamefont {Gomez~Alvarado}}, \bibinfo
  {author} {\bibfnamefont {K.}~\bibnamefont {Mallayya}}, \bibinfo {author}
  {\bibfnamefont {G.}~\bibnamefont {Wu}}, \bibinfo {author} {\bibfnamefont
  {E.-A.}\ \bibnamefont {Kim}}, \bibinfo {author} {\bibfnamefont {J.~P.~C.}\
  \bibnamefont {Ruff}}, \bibinfo {author} {\bibfnamefont {S.}~\bibnamefont
  {Sarker}},\ and\ \bibinfo {author} {\bibfnamefont {S.~D.}\ \bibnamefont
  {Wilson}},\ }\bibfield  {title} {\bibinfo {title} {Frustrated charge order
  and cooperative distortions in {ScV$_6$Sn$_6$}},\ }\href@noop {} {\bibfield
  {journal} {\bibinfo  {journal} {Physical Review Materials}\ }\textbf
  {\bibinfo {volume} {7}},\ \bibinfo {pages} {104201} (\bibinfo {year}
  {2023})}\BibitemShut {NoStop}%
\bibitem [{\citenamefont {Kim}\ \emph {et~al.}(2023)\citenamefont {Kim},
  \citenamefont {Liu}, \citenamefont {Wang}, \citenamefont {Nam}, \citenamefont
  {Pokharel}, \citenamefont {Wilson}, \citenamefont {Cho},\ and\ \citenamefont
  {Moon}}]{kim2023infrared}%
  \BibitemOpen
  \bibfield  {author} {\bibinfo {author} {\bibfnamefont {D.~W.}\ \bibnamefont
  {Kim}}, \bibinfo {author} {\bibfnamefont {S.}~\bibnamefont {Liu}}, \bibinfo
  {author} {\bibfnamefont {C.}~\bibnamefont {Wang}}, \bibinfo {author}
  {\bibfnamefont {H.}~\bibnamefont {Nam}}, \bibinfo {author} {\bibfnamefont
  {G.}~\bibnamefont {Pokharel}}, \bibinfo {author} {\bibfnamefont {S.~D.}\
  \bibnamefont {Wilson}}, \bibinfo {author} {\bibfnamefont {J.-H.}\
  \bibnamefont {Cho}},\ and\ \bibinfo {author} {\bibfnamefont {S.~J.}\
  \bibnamefont {Moon}},\ }\bibfield  {title} {\bibinfo {title} {Infrared probe
  of the charge density wave gap in {ScV$_6$Sn$_6$}},\ }\href@noop {}
  {\bibfield  {journal} {\bibinfo  {journal} {Physical Review B}\ }\textbf
  {\bibinfo {volume} {108}},\ \bibinfo {pages} {205118} (\bibinfo {year}
  {2023})}\BibitemShut {NoStop}%
\bibitem [{\citenamefont {Hu}\ \emph {et~al.}(2024)\citenamefont {Hu},
  \citenamefont {Ma}, \citenamefont {Li}, \citenamefont {Jiang}, \citenamefont
  {Gawryluk}, \citenamefont {Hu}, \citenamefont {Teyssier}, \citenamefont
  {Multian}, \citenamefont {Yin}, \citenamefont {Xu} \emph
  {et~al.}}]{hu2024phonon}%
  \BibitemOpen
  \bibfield  {author} {\bibinfo {author} {\bibfnamefont {Y.}~\bibnamefont
  {Hu}}, \bibinfo {author} {\bibfnamefont {J.}~\bibnamefont {Ma}}, \bibinfo
  {author} {\bibfnamefont {Y.}~\bibnamefont {Li}}, \bibinfo {author}
  {\bibfnamefont {Y.}~\bibnamefont {Jiang}}, \bibinfo {author} {\bibfnamefont
  {D.~J.}\ \bibnamefont {Gawryluk}}, \bibinfo {author} {\bibfnamefont
  {T.}~\bibnamefont {Hu}}, \bibinfo {author} {\bibfnamefont {J.}~\bibnamefont
  {Teyssier}}, \bibinfo {author} {\bibfnamefont {V.}~\bibnamefont {Multian}},
  \bibinfo {author} {\bibfnamefont {Z.}~\bibnamefont {Yin}}, \bibinfo {author}
  {\bibfnamefont {S.}~\bibnamefont {Xu}}, \emph {et~al.},\ }\bibfield  {title}
  {\bibinfo {title} {Phonon promoted charge density wave in topological kagome
  metal {ScV$_6$Sn$_6$}},\ }\href@noop {} {\bibfield  {journal} {\bibinfo
  {journal} {Nature Communications}\ }\textbf {\bibinfo {volume} {15}},\
  \bibinfo {pages} {1658} (\bibinfo {year} {2024})}\BibitemShut {NoStop}%
\bibitem [{\citenamefont {Yi}\ \emph {et~al.}(2024)\citenamefont {Yi},
  \citenamefont {Feng}, \citenamefont {Kumar}, \citenamefont {Felser},\ and\
  \citenamefont {Shekhar}}]{yi2024tuning}%
  \BibitemOpen
  \bibfield  {author} {\bibinfo {author} {\bibfnamefont {C.}~\bibnamefont
  {Yi}}, \bibinfo {author} {\bibfnamefont {X.}~\bibnamefont {Feng}}, \bibinfo
  {author} {\bibfnamefont {N.}~\bibnamefont {Kumar}}, \bibinfo {author}
  {\bibfnamefont {C.}~\bibnamefont {Felser}},\ and\ \bibinfo {author}
  {\bibfnamefont {C.}~\bibnamefont {Shekhar}},\ }\bibfield  {title} {\bibinfo
  {title} {Tuning charge density wave of kagome metal {ScV$_6$Sn$_6$}},\
  }\href@noop {} {\bibfield  {journal} {\bibinfo  {journal} {New Journal of
  Physics}\ }\textbf {\bibinfo {volume} {26}},\ \bibinfo {pages} {052001}
  (\bibinfo {year} {2024})}\BibitemShut {NoStop}%
\bibitem [{\citenamefont {Ortiz}\ \emph {et~al.}(2024)\citenamefont {Ortiz},
  \citenamefont {Meier}, \citenamefont {Pokharel}, \citenamefont {Chamorro},
  \citenamefont {Yang}, \citenamefont {Mozaffari}, \citenamefont {Thaler},
  \citenamefont {Gomez~Alvarado}, \citenamefont {Zhang}, \citenamefont {Parker}
  \emph {et~al.}}]{ortiz2024stability}%
  \BibitemOpen
  \bibfield  {author} {\bibinfo {author} {\bibfnamefont {B.~R.}\ \bibnamefont
  {Ortiz}}, \bibinfo {author} {\bibfnamefont {W.~R.}\ \bibnamefont {Meier}},
  \bibinfo {author} {\bibfnamefont {G.}~\bibnamefont {Pokharel}}, \bibinfo
  {author} {\bibfnamefont {J.}~\bibnamefont {Chamorro}}, \bibinfo {author}
  {\bibfnamefont {F.}~\bibnamefont {Yang}}, \bibinfo {author} {\bibfnamefont
  {S.}~\bibnamefont {Mozaffari}}, \bibinfo {author} {\bibfnamefont
  {A.}~\bibnamefont {Thaler}}, \bibinfo {author} {\bibfnamefont {S.~J.}\
  \bibnamefont {Gomez~Alvarado}}, \bibinfo {author} {\bibfnamefont
  {H.}~\bibnamefont {Zhang}}, \bibinfo {author} {\bibfnamefont {D.~S.}\
  \bibnamefont {Parker}}, \emph {et~al.},\ }\bibfield  {title} {\bibinfo
  {title} {Stability frontiers in the {AM$_6$X$_6$} kagome metals: The
  {LnNb$_6$Sn$_6$ (Ln:Ce-Lu,Y)} family and density-wave transition in
  {LuNb$_6$Sn$_6$}},\ }\href@noop {} {\bibfield  {journal} {\bibinfo  {journal}
  {Journal of the American Chemical Society}\ } (\bibinfo {year}
  {2024})}\BibitemShut {NoStop}%
\bibitem [{\citenamefont {Asaba}\ \emph {et~al.}(2020)\citenamefont {Asaba},
  \citenamefont {Thomas}, \citenamefont {Curtis}, \citenamefont {Thompson},
  \citenamefont {Bauer},\ and\ \citenamefont {Ronning}}]{asaba2020anomalous}%
  \BibitemOpen
  \bibfield  {author} {\bibinfo {author} {\bibfnamefont {T.}~\bibnamefont
  {Asaba}}, \bibinfo {author} {\bibfnamefont {S.~M.}\ \bibnamefont {Thomas}},
  \bibinfo {author} {\bibfnamefont {M.}~\bibnamefont {Curtis}}, \bibinfo
  {author} {\bibfnamefont {J.~D.}\ \bibnamefont {Thompson}}, \bibinfo {author}
  {\bibfnamefont {E.~D.}\ \bibnamefont {Bauer}},\ and\ \bibinfo {author}
  {\bibfnamefont {F.}~\bibnamefont {Ronning}},\ }\bibfield  {title} {\bibinfo
  {title} {Anomalous {Hall} effect in the kagome ferrimagnet
  {GdMn$_6$Sn$_6$}},\ }\href@noop {} {\bibfield  {journal} {\bibinfo  {journal}
  {Physical Review B}\ }\textbf {\bibinfo {volume} {101}},\ \bibinfo {pages}
  {174415} (\bibinfo {year} {2020})}\BibitemShut {NoStop}%
\bibitem [{\citenamefont {Dhakal}\ \emph {et~al.}(2021)\citenamefont {Dhakal},
  \citenamefont {Cheenicode~Kabeer}, \citenamefont {Pathak}, \citenamefont
  {Kabir}, \citenamefont {Poudel}, \citenamefont {Filippone}, \citenamefont
  {Casey}, \citenamefont {Pradhan~Sakhya}, \citenamefont {Regmi}, \citenamefont
  {Sims} \emph {et~al.}}]{dhakal2021anisotropically}%
  \BibitemOpen
  \bibfield  {author} {\bibinfo {author} {\bibfnamefont {G.}~\bibnamefont
  {Dhakal}}, \bibinfo {author} {\bibfnamefont {F.}~\bibnamefont
  {Cheenicode~Kabeer}}, \bibinfo {author} {\bibfnamefont {A.~K.}\ \bibnamefont
  {Pathak}}, \bibinfo {author} {\bibfnamefont {F.}~\bibnamefont {Kabir}},
  \bibinfo {author} {\bibfnamefont {N.}~\bibnamefont {Poudel}}, \bibinfo
  {author} {\bibfnamefont {R.}~\bibnamefont {Filippone}}, \bibinfo {author}
  {\bibfnamefont {J.}~\bibnamefont {Casey}}, \bibinfo {author} {\bibfnamefont
  {A.}~\bibnamefont {Pradhan~Sakhya}}, \bibinfo {author} {\bibfnamefont
  {S.}~\bibnamefont {Regmi}}, \bibinfo {author} {\bibfnamefont
  {C.}~\bibnamefont {Sims}}, \emph {et~al.},\ }\bibfield  {title} {\bibinfo
  {title} {Anisotropically large anomalous and topological {Hall} effect in a
  kagome magnet},\ }\href@noop {} {\bibfield  {journal} {\bibinfo  {journal}
  {Physical Review B}\ }\textbf {\bibinfo {volume} {104}},\ \bibinfo {pages}
  {L161115} (\bibinfo {year} {2021})}\BibitemShut {NoStop}%
\bibitem [{\citenamefont {Yin}\ \emph {et~al.}(2020)\citenamefont {Yin},
  \citenamefont {Ma}, \citenamefont {Cochran}, \citenamefont {Xu},
  \citenamefont {Zhang}, \citenamefont {Tien}, \citenamefont {Shumiya},
  \citenamefont {Cheng}, \citenamefont {Jiang}, \citenamefont {Lian} \emph
  {et~al.}}]{yin2020quantum}%
  \BibitemOpen
  \bibfield  {author} {\bibinfo {author} {\bibfnamefont {J.-X.}\ \bibnamefont
  {Yin}}, \bibinfo {author} {\bibfnamefont {W.}~\bibnamefont {Ma}}, \bibinfo
  {author} {\bibfnamefont {T.~A.}\ \bibnamefont {Cochran}}, \bibinfo {author}
  {\bibfnamefont {X.}~\bibnamefont {Xu}}, \bibinfo {author} {\bibfnamefont
  {S.~S.}\ \bibnamefont {Zhang}}, \bibinfo {author} {\bibfnamefont {H.-J.}\
  \bibnamefont {Tien}}, \bibinfo {author} {\bibfnamefont {N.}~\bibnamefont
  {Shumiya}}, \bibinfo {author} {\bibfnamefont {G.}~\bibnamefont {Cheng}},
  \bibinfo {author} {\bibfnamefont {K.}~\bibnamefont {Jiang}}, \bibinfo
  {author} {\bibfnamefont {B.}~\bibnamefont {Lian}}, \emph {et~al.},\
  }\bibfield  {title} {\bibinfo {title} {Quantum-limit {Chern} topological
  magnetism in {TbMn$_6$Sn$_6$}},\ }\href@noop {} {\bibfield  {journal}
  {\bibinfo  {journal} {Nature}\ }\textbf {\bibinfo {volume} {583}},\ \bibinfo
  {pages} {533} (\bibinfo {year} {2020})}\BibitemShut {NoStop}%
\bibitem [{\citenamefont {Guo}\ \emph {et~al.}(2023)\citenamefont {Guo},
  \citenamefont {Ye}, \citenamefont {Guan},\ and\ \citenamefont
  {Jia}}]{guo2023triangular}%
  \BibitemOpen
  \bibfield  {author} {\bibinfo {author} {\bibfnamefont {K.}~\bibnamefont
  {Guo}}, \bibinfo {author} {\bibfnamefont {J.}~\bibnamefont {Ye}}, \bibinfo
  {author} {\bibfnamefont {S.}~\bibnamefont {Guan}},\ and\ \bibinfo {author}
  {\bibfnamefont {S.}~\bibnamefont {Jia}},\ }\bibfield  {title} {\bibinfo
  {title} {Triangular {Kondo} lattice in {YbV$_6$Sn$_6$} and its quantum
  critical behavior in a magnetic field},\ }\href@noop {} {\bibfield  {journal}
  {\bibinfo  {journal} {Physical Review B}\ }\textbf {\bibinfo {volume}
  {107}},\ \bibinfo {pages} {205151} (\bibinfo {year} {2023})}\BibitemShut
  {NoStop}%
\bibitem [{\citenamefont {Buchholz}\ and\ \citenamefont
  {Schuster}(1981)}]{buchholz1981intermetallische}%
  \BibitemOpen
  \bibfield  {author} {\bibinfo {author} {\bibfnamefont {W.}~\bibnamefont
  {Buchholz}}\ and\ \bibinfo {author} {\bibfnamefont {H.-U.}\ \bibnamefont
  {Schuster}},\ }\bibfield  {title} {\bibinfo {title} {Intermetallische phasen
  mit {B35-{\"U}berstruktur} und verwandtschaftsbeziehung zu
  {LiFe$_6$Ge$_6$}},\ }\href@noop {} {\bibfield  {journal} {\bibinfo  {journal}
  {Zeitschrift f{\"u}r anorganische und allgemeine Chemie}\ }\textbf {\bibinfo
  {volume} {482}},\ \bibinfo {pages} {40} (\bibinfo {year} {1981})}\BibitemShut
  {NoStop}%
\bibitem [{\citenamefont {Gon{\c{c}}alves}\ \emph {et~al.}(1994)\citenamefont
  {Gon{\c{c}}alves}, \citenamefont {Waerenborgh}, \citenamefont {Bonfait},
  \citenamefont {Amaro}, \citenamefont {Godinho}, \citenamefont {Almeida},\
  and\ \citenamefont {Spirlet}}]{gonccalves1994ufe6ge6}%
  \BibitemOpen
  \bibfield  {author} {\bibinfo {author} {\bibfnamefont {A.~P.}\ \bibnamefont
  {Gon{\c{c}}alves}}, \bibinfo {author} {\bibfnamefont {J.~C.}\ \bibnamefont
  {Waerenborgh}}, \bibinfo {author} {\bibfnamefont {G.}~\bibnamefont
  {Bonfait}}, \bibinfo {author} {\bibfnamefont {A.}~\bibnamefont {Amaro}},
  \bibinfo {author} {\bibfnamefont {M.~M.}\ \bibnamefont {Godinho}}, \bibinfo
  {author} {\bibfnamefont {M.}~\bibnamefont {Almeida}},\ and\ \bibinfo {author}
  {\bibfnamefont {J.~C.}\ \bibnamefont {Spirlet}},\ }\bibfield  {title}
  {\bibinfo {title} {{UFe$_6$Ge$_6$}: {A} new ternary magnetic compound},\
  }\href@noop {} {\bibfield  {journal} {\bibinfo  {journal} {Journal of Alloys
  and Compounds}\ }\textbf {\bibinfo {volume} {204}},\ \bibinfo {pages} {59}
  (\bibinfo {year} {1994})}\BibitemShut {NoStop}%
\bibitem [{\citenamefont {Thomas}\ \emph {et~al.}(2025)\citenamefont {Thomas},
  \citenamefont {Kengle}, \citenamefont {Simeth}, \citenamefont {Lim},
  \citenamefont {Riedel}, \citenamefont {Allen}, \citenamefont {Schmidt},
  \citenamefont {Ruf}, \citenamefont {Thompson}, \citenamefont {Ronning},
  \citenamefont {Scheie}, \citenamefont {Lane}, \citenamefont {Denlinger},
  \citenamefont {Blanco-Canosa}, \citenamefont {Zhu}, \citenamefont {Bauer},\
  and\ \citenamefont {Rosa}}]{thomas2025uv6sn6}%
  \BibitemOpen
  \bibfield  {author} {\bibinfo {author} {\bibfnamefont {S.~M.}\ \bibnamefont
  {Thomas}}, \bibinfo {author} {\bibfnamefont {C.~S.}\ \bibnamefont {Kengle}},
  \bibinfo {author} {\bibfnamefont {W.}~\bibnamefont {Simeth}}, \bibinfo
  {author} {\bibfnamefont {C.-y.}\ \bibnamefont {Lim}}, \bibinfo {author}
  {\bibfnamefont {Z.~W.}\ \bibnamefont {Riedel}}, \bibinfo {author}
  {\bibfnamefont {K.}~\bibnamefont {Allen}}, \bibinfo {author} {\bibfnamefont
  {A.}~\bibnamefont {Schmidt}}, \bibinfo {author} {\bibfnamefont
  {M.}~\bibnamefont {Ruf}}, \bibinfo {author} {\bibfnamefont {J.~D.}\
  \bibnamefont {Thompson}}, \bibinfo {author} {\bibfnamefont {F.}~\bibnamefont
  {Ronning}}, \bibinfo {author} {\bibfnamefont {A.~O.}\ \bibnamefont {Scheie}},
  \bibinfo {author} {\bibfnamefont {C.}~\bibnamefont {Lane}}, \bibinfo {author}
  {\bibfnamefont {J.}~\bibnamefont {Denlinger}}, \bibinfo {author}
  {\bibfnamefont {S.}~\bibnamefont {Blanco-Canosa}}, \bibinfo {author}
  {\bibfnamefont {J.-X.}\ \bibnamefont {Zhu}}, \bibinfo {author} {\bibfnamefont
  {E.~D.}\ \bibnamefont {Bauer}},\ and\ \bibinfo {author} {\bibfnamefont
  {P.~F.~S.}\ \bibnamefont {Rosa}},\ }\href {https://arxiv.org/abs/2503.13245}
  {\bibinfo {title} {{UV$_6$Sn$_6$}: a new kagome material with unusual 5$f$
  magnetism}} (\bibinfo {year} {2025}),\ \Eprint
  {https://arxiv.org/abs/2503.13245} {arXiv:2503.13245 [cond-mat.str-el]}
  \BibitemShut {NoStop}%
\bibitem [{\citenamefont {Patino}\ \emph {et~al.}(2025)\citenamefont {Patino},
  \citenamefont {Raymond}, \citenamefont {Knebel}, \citenamefont {Berre},
  \citenamefont {Savvin}, \citenamefont {Pachoud}, \citenamefont {Ressouche},
  \citenamefont {Fettinger}, \citenamefont {Leynaud}, \citenamefont {Pecaut},
  \citenamefont {Klavins}, \citenamefont {Hasselbach}, \citenamefont {Brison},
  \citenamefont {Lapertot},\ and\ \citenamefont {Taufour}}]{patino2025incom}%
  \BibitemOpen
  \bibfield  {author} {\bibinfo {author} {\bibfnamefont {M.~A.}\ \bibnamefont
  {Patino}}, \bibinfo {author} {\bibfnamefont {S.}~\bibnamefont {Raymond}},
  \bibinfo {author} {\bibfnamefont {G.}~\bibnamefont {Knebel}}, \bibinfo
  {author} {\bibfnamefont {P.~L.}\ \bibnamefont {Berre}}, \bibinfo {author}
  {\bibfnamefont {S.}~\bibnamefont {Savvin}}, \bibinfo {author} {\bibfnamefont
  {E.}~\bibnamefont {Pachoud}}, \bibinfo {author} {\bibfnamefont
  {E.}~\bibnamefont {Ressouche}}, \bibinfo {author} {\bibfnamefont {J.~C.}\
  \bibnamefont {Fettinger}}, \bibinfo {author} {\bibfnamefont {O.}~\bibnamefont
  {Leynaud}}, \bibinfo {author} {\bibfnamefont {J.}~\bibnamefont {Pecaut}},
  \bibinfo {author} {\bibfnamefont {P.}~\bibnamefont {Klavins}}, \bibinfo
  {author} {\bibfnamefont {K.}~\bibnamefont {Hasselbach}}, \bibinfo {author}
  {\bibfnamefont {J.-P.}\ \bibnamefont {Brison}}, \bibinfo {author}
  {\bibfnamefont {G.}~\bibnamefont {Lapertot}},\ and\ \bibinfo {author}
  {\bibfnamefont {V.}~\bibnamefont {Taufour}},\ }\href
  {https://arxiv.org/abs/2503.12922} {\bibinfo {title} {Incommensurate and
  commensurate antiferromagnetic orders in the kagome compound
  {UV$_{6}$Sn$_{6}$}}} (\bibinfo {year} {2025}),\ \Eprint
  {https://arxiv.org/abs/2503.12922} {arXiv:2503.12922 [cond-mat.str-el]}
  \BibitemShut {NoStop}%
\bibitem [{\citenamefont {Oshchapovsky}\ \emph {et~al.}(2010)\citenamefont
  {Oshchapovsky}, \citenamefont {Pavlyuk}, \citenamefont {F{\"a}ssler},\ and\
  \citenamefont {Hlukhyy}}]{oshchapovsky2010tbnb6sn6}%
  \BibitemOpen
  \bibfield  {author} {\bibinfo {author} {\bibfnamefont {I.}~\bibnamefont
  {Oshchapovsky}}, \bibinfo {author} {\bibfnamefont {V.}~\bibnamefont
  {Pavlyuk}}, \bibinfo {author} {\bibfnamefont {T.~F.}\ \bibnamefont
  {F{\"a}ssler}},\ and\ \bibinfo {author} {\bibfnamefont {V.}~\bibnamefont
  {Hlukhyy}},\ }\bibfield  {title} {\bibinfo {title} {{TbNb$_6$Sn$_6$}: {The}
  first ternary compound from the rare earth--niobium--tin system},\
  }\href@noop {} {\bibfield  {journal} {\bibinfo  {journal} {Acta
  Crystallographica Section E}\ }\textbf {\bibinfo {volume} {66}},\ \bibinfo
  {pages} {i82} (\bibinfo {year} {2010})}\BibitemShut {NoStop}%
\bibitem [{\citenamefont {Yue}\ and\ \citenamefont
  {Lei}(2012)}]{yue2012syntheses}%
  \BibitemOpen
  \bibfield  {author} {\bibinfo {author} {\bibfnamefont {C.-Y.}\ \bibnamefont
  {Yue}}\ and\ \bibinfo {author} {\bibfnamefont {X.-W.}\ \bibnamefont {Lei}},\
  }\bibfield  {title} {\bibinfo {title} {Syntheses and structures of
  {Sc$_2$Nb$_{4-x}$Sn$_5$}, {YNb$_6$Sn$_6$}, and {ErNb$_6$Sn$_5$}:
  {Exploratory} studies in ternary rare-earth niobium stannides},\ }\href@noop
  {} {\bibfield  {journal} {\bibinfo  {journal} {Inorganic Chemistry}\ }\textbf
  {\bibinfo {volume} {51}},\ \bibinfo {pages} {2461} (\bibinfo {year}
  {2012})}\BibitemShut {NoStop}%
\bibitem [{\citenamefont {Xiao}\ \emph {et~al.}(2025)\citenamefont {Xiao},
  \citenamefont {Duan}, \citenamefont {Li}, \citenamefont {Guo}, \citenamefont
  {Tan},\ and\ \citenamefont {Zhong}}]{xiao2025kagome}%
  \BibitemOpen
  \bibfield  {author} {\bibinfo {author} {\bibfnamefont {Y.}~\bibnamefont
  {Xiao}}, \bibinfo {author} {\bibfnamefont {Q.}~\bibnamefont {Duan}}, \bibinfo
  {author} {\bibfnamefont {Z.}~\bibnamefont {Li}}, \bibinfo {author}
  {\bibfnamefont {S.}~\bibnamefont {Guo}}, \bibinfo {author} {\bibfnamefont
  {H.}~\bibnamefont {Tan}},\ and\ \bibinfo {author} {\bibfnamefont
  {R.}~\bibnamefont {Zhong}},\ }\href {https://arxiv.org/abs/2501.00996}
  {\bibinfo {title} {Kagome metal {GdNb$_6$Sn$_6$}: {A} 4d playground for
  topological magnetism and electron correlations}} (\bibinfo {year} {2025}),\
  \Eprint {https://arxiv.org/abs/2501.00996} {arXiv:2501.00996
  [cond-mat.mtrl-sci]} \BibitemShut {NoStop}%
\bibitem [{\citenamefont {Pokharel}\ \emph {et~al.}(2021)\citenamefont
  {Pokharel}, \citenamefont {Teicher}, \citenamefont {Ortiz}, \citenamefont
  {Sarte}, \citenamefont {Wu}, \citenamefont {Peng}, \citenamefont {He},
  \citenamefont {Seshadri},\ and\ \citenamefont
  {Wilson}}]{pokharel2021electronic}%
  \BibitemOpen
  \bibfield  {author} {\bibinfo {author} {\bibfnamefont {G.}~\bibnamefont
  {Pokharel}}, \bibinfo {author} {\bibfnamefont {S.~M.~L.}\ \bibnamefont
  {Teicher}}, \bibinfo {author} {\bibfnamefont {B.~R.}\ \bibnamefont {Ortiz}},
  \bibinfo {author} {\bibfnamefont {P.~M.}\ \bibnamefont {Sarte}}, \bibinfo
  {author} {\bibfnamefont {G.}~\bibnamefont {Wu}}, \bibinfo {author}
  {\bibfnamefont {S.}~\bibnamefont {Peng}}, \bibinfo {author} {\bibfnamefont
  {J.}~\bibnamefont {He}}, \bibinfo {author} {\bibfnamefont {R.}~\bibnamefont
  {Seshadri}},\ and\ \bibinfo {author} {\bibfnamefont {S.~D.}\ \bibnamefont
  {Wilson}},\ }\bibfield  {title} {\bibinfo {title} {Electronic properties of
  the topological kagome metals {YV$_6$Sn$_6$} and {GdV$_6$Sn$_6$}},\
  }\href@noop {} {\bibfield  {journal} {\bibinfo  {journal} {Physical Review
  B}\ }\textbf {\bibinfo {volume} {104}},\ \bibinfo {pages} {235139} (\bibinfo
  {year} {2021})}\BibitemShut {NoStop}%
\bibitem [{\citenamefont {Zhang}\ \emph {et~al.}(2022)\citenamefont {Zhang},
  \citenamefont {Liu}, \citenamefont {Cui}, \citenamefont {Guo}, \citenamefont
  {Wang}, \citenamefont {Shi}, \citenamefont {Zhang}, \citenamefont {Wang},
  \citenamefont {Dong}, \citenamefont {Sun} \emph
  {et~al.}}]{zhang2022electronic}%
  \BibitemOpen
  \bibfield  {author} {\bibinfo {author} {\bibfnamefont {X.}~\bibnamefont
  {Zhang}}, \bibinfo {author} {\bibfnamefont {Z.}~\bibnamefont {Liu}}, \bibinfo
  {author} {\bibfnamefont {Q.}~\bibnamefont {Cui}}, \bibinfo {author}
  {\bibfnamefont {Q.}~\bibnamefont {Guo}}, \bibinfo {author} {\bibfnamefont
  {N.}~\bibnamefont {Wang}}, \bibinfo {author} {\bibfnamefont {L.}~\bibnamefont
  {Shi}}, \bibinfo {author} {\bibfnamefont {H.}~\bibnamefont {Zhang}}, \bibinfo
  {author} {\bibfnamefont {W.}~\bibnamefont {Wang}}, \bibinfo {author}
  {\bibfnamefont {X.}~\bibnamefont {Dong}}, \bibinfo {author} {\bibfnamefont
  {J.}~\bibnamefont {Sun}}, \emph {et~al.},\ }\bibfield  {title} {\bibinfo
  {title} {Electronic and magnetic properties of intermetallic kagome magnets
  {$R$V$_6$Sn$_6$ ($R$= Tb-Tm)}},\ }\href@noop {} {\bibfield  {journal}
  {\bibinfo  {journal} {Physical Review Materials}\ }\textbf {\bibinfo {volume}
  {6}},\ \bibinfo {pages} {105001} (\bibinfo {year} {2022})}\BibitemShut
  {NoStop}%
\bibitem [{\citenamefont {Zeng}\ \emph {et~al.}(2024)\citenamefont {Zeng},
  \citenamefont {Wang}, \citenamefont {Wang}, \citenamefont {Lin},
  \citenamefont {Gong}, \citenamefont {Ma}, \citenamefont {Han}, \citenamefont
  {Wang}, \citenamefont {Dai},\ and\ \citenamefont {Xia}}]{zeng2024magnetic}%
  \BibitemOpen
  \bibfield  {author} {\bibinfo {author} {\bibfnamefont {X.-Y.}\ \bibnamefont
  {Zeng}}, \bibinfo {author} {\bibfnamefont {H.}~\bibnamefont {Wang}}, \bibinfo
  {author} {\bibfnamefont {X.-Y.}\ \bibnamefont {Wang}}, \bibinfo {author}
  {\bibfnamefont {J.-F.}\ \bibnamefont {Lin}}, \bibinfo {author} {\bibfnamefont
  {J.}~\bibnamefont {Gong}}, \bibinfo {author} {\bibfnamefont {X.-P.}\
  \bibnamefont {Ma}}, \bibinfo {author} {\bibfnamefont {K.}~\bibnamefont
  {Han}}, \bibinfo {author} {\bibfnamefont {Y.-T.}\ \bibnamefont {Wang}},
  \bibinfo {author} {\bibfnamefont {Z.-Y.}\ \bibnamefont {Dai}},\ and\ \bibinfo
  {author} {\bibfnamefont {T.-L.}\ \bibnamefont {Xia}},\ }\bibfield  {title}
  {\bibinfo {title} {Magnetic and magnetotransport properties in the
  vanadium-based kagome metals {DyV$_6$Sn$_6$} and {HoV$_6$Sn$_6$}},\
  }\href@noop {} {\bibfield  {journal} {\bibinfo  {journal} {Physical Review
  B}\ }\textbf {\bibinfo {volume} {109}},\ \bibinfo {pages} {104412} (\bibinfo
  {year} {2024})}\BibitemShut {NoStop}%
\bibitem [{\citenamefont
  {Sheldrick}(2015{\natexlab{a}})}]{sheldrick2015shelxt}%
  \BibitemOpen
  \bibfield  {author} {\bibinfo {author} {\bibfnamefont {G.~M.}\ \bibnamefont
  {Sheldrick}},\ }\bibfield  {title} {\bibinfo {title} {{SHELXT}--integrated
  space-group and crystal-structure determination},\ }\href@noop {} {\bibfield
  {journal} {\bibinfo  {journal} {Acta Crystallographica Section A}\ }\textbf
  {\bibinfo {volume} {71}},\ \bibinfo {pages} {3} (\bibinfo {year}
  {2015}{\natexlab{a}})}\BibitemShut {NoStop}%
\bibitem [{\citenamefont {Sheldrick}(2008)}]{sheldrick2008short}%
  \BibitemOpen
  \bibfield  {author} {\bibinfo {author} {\bibfnamefont {G.~M.}\ \bibnamefont
  {Sheldrick}},\ }\bibfield  {title} {\bibinfo {title} {A short history of
  {SHELX}},\ }\href@noop {} {\bibfield  {journal} {\bibinfo  {journal} {Acta
  Crystallographica Section A}\ }\textbf {\bibinfo {volume} {64}},\ \bibinfo
  {pages} {112} (\bibinfo {year} {2008})}\BibitemShut {NoStop}%
\bibitem [{\citenamefont
  {Sheldrick}(2015{\natexlab{b}})}]{sheldrick2015crystal}%
  \BibitemOpen
  \bibfield  {author} {\bibinfo {author} {\bibfnamefont {G.~M.}\ \bibnamefont
  {Sheldrick}},\ }\bibfield  {title} {\bibinfo {title} {Crystal structure
  refinement with {SHELXL}},\ }\href@noop {} {\bibfield  {journal} {\bibinfo
  {journal} {Acta Crystallographica Section C}\ }\textbf {\bibinfo {volume}
  {71}},\ \bibinfo {pages} {3} (\bibinfo {year}
  {2015}{\natexlab{b}})}\BibitemShut {NoStop}%
\bibitem [{\citenamefont {Spek}(2003)}]{spek2003single}%
  \BibitemOpen
  \bibfield  {author} {\bibinfo {author} {\bibfnamefont {A.~L.}\ \bibnamefont
  {Spek}},\ }\bibfield  {title} {\bibinfo {title} {Single-crystal structure
  validation with the program {PLATON}},\ }\href@noop {} {\bibfield  {journal}
  {\bibinfo  {journal} {Journal of Applied Crystallography}\ }\textbf {\bibinfo
  {volume} {36}},\ \bibinfo {pages} {7} (\bibinfo {year} {2003})}\BibitemShut
  {NoStop}%
\bibitem [{\citenamefont {Dolomanov}\ \emph {et~al.}(2009)\citenamefont
  {Dolomanov}, \citenamefont {Bourhis}, \citenamefont {Gildea}, \citenamefont
  {Howard},\ and\ \citenamefont {Puschmann}}]{dolomanov2009olex2}%
  \BibitemOpen
  \bibfield  {author} {\bibinfo {author} {\bibfnamefont {O.~V.}\ \bibnamefont
  {Dolomanov}}, \bibinfo {author} {\bibfnamefont {L.~J.}\ \bibnamefont
  {Bourhis}}, \bibinfo {author} {\bibfnamefont {R.~J.}\ \bibnamefont {Gildea}},
  \bibinfo {author} {\bibfnamefont {J.~A.~K.}\ \bibnamefont {Howard}},\ and\
  \bibinfo {author} {\bibfnamefont {H.}~\bibnamefont {Puschmann}},\ }\bibfield
  {title} {\bibinfo {title} {{OLEX2}: {A} complete structure solution,
  refinement and analysis program},\ }\href@noop {} {\bibfield  {journal}
  {\bibinfo  {journal} {Journal of Applied Crystallography}\ }\textbf {\bibinfo
  {volume} {42}},\ \bibinfo {pages} {339} (\bibinfo {year} {2009})}\BibitemShut
  {NoStop}%
\bibitem [{\citenamefont {Momma}\ and\ \citenamefont
  {Izumi}(2011)}]{momma2011vesta}%
  \BibitemOpen
  \bibfield  {author} {\bibinfo {author} {\bibfnamefont {K.}~\bibnamefont
  {Momma}}\ and\ \bibinfo {author} {\bibfnamefont {F.}~\bibnamefont {Izumi}},\
  }\bibfield  {title} {\bibinfo {title} {{VESTA} 3 for three-dimensional
  visualization of crystal, volumetric and morphology data},\ }\href@noop {}
  {\bibfield  {journal} {\bibinfo  {journal} {Journal of Applied
  Crystallography}\ }\textbf {\bibinfo {volume} {44}},\ \bibinfo {pages} {1272}
  (\bibinfo {year} {2011})}\BibitemShut {NoStop}%
\bibitem [{\citenamefont {Fischer}\ \emph {et~al.}(2000)\citenamefont
  {Fischer}, \citenamefont {Frey}, \citenamefont {Koch}, \citenamefont
  {K{\"o}nnecke}, \citenamefont {Pomjakushin}, \citenamefont {Schefer},
  \citenamefont {Thut}, \citenamefont {Schlumpf}, \citenamefont {B{\"u}rge},
  \citenamefont {Greuter} \emph {et~al.}}]{fischer2000high}%
  \BibitemOpen
  \bibfield  {author} {\bibinfo {author} {\bibfnamefont {P.}~\bibnamefont
  {Fischer}}, \bibinfo {author} {\bibfnamefont {G.}~\bibnamefont {Frey}},
  \bibinfo {author} {\bibfnamefont {M.}~\bibnamefont {Koch}}, \bibinfo {author}
  {\bibfnamefont {M.}~\bibnamefont {K{\"o}nnecke}}, \bibinfo {author}
  {\bibfnamefont {V.}~\bibnamefont {Pomjakushin}}, \bibinfo {author}
  {\bibfnamefont {J.}~\bibnamefont {Schefer}}, \bibinfo {author} {\bibfnamefont
  {R.}~\bibnamefont {Thut}}, \bibinfo {author} {\bibfnamefont {N.}~\bibnamefont
  {Schlumpf}}, \bibinfo {author} {\bibfnamefont {R.}~\bibnamefont {B{\"u}rge}},
  \bibinfo {author} {\bibfnamefont {U.}~\bibnamefont {Greuter}}, \emph
  {et~al.},\ }\bibfield  {title} {\bibinfo {title} {High-resolution powder
  diffractometer {HRPT} for thermal neutrons at {SINQ}},\ }\href@noop {}
  {\bibfield  {journal} {\bibinfo  {journal} {Physica B: Condensed Matter}\
  }\textbf {\bibinfo {volume} {276}},\ \bibinfo {pages} {146} (\bibinfo {year}
  {2000})}\BibitemShut {NoStop}%
\bibitem [{\citenamefont
  {Rodr{\'\i}guez-Carvajal}(1993)}]{rodriguez1993recent}%
  \BibitemOpen
  \bibfield  {author} {\bibinfo {author} {\bibfnamefont {J.}~\bibnamefont
  {Rodr{\'\i}guez-Carvajal}},\ }\bibfield  {title} {\bibinfo {title} {Recent
  advances in magnetic structure determination by neutron powder diffraction},\
  }\href@noop {} {\bibfield  {journal} {\bibinfo  {journal} {Physica B:
  Condensed Matter}\ }\textbf {\bibinfo {volume} {192}},\ \bibinfo {pages} {55}
  (\bibinfo {year} {1993})}\BibitemShut {NoStop}%
\bibitem [{\citenamefont {Venturini}(2001)}]{venturini2001crystallographic}%
  \BibitemOpen
  \bibfield  {author} {\bibinfo {author} {\bibfnamefont {G.}~\bibnamefont
  {Venturini}},\ }\bibfield  {title} {\bibinfo {title} {Crystallographic and
  magnetic properties of {TbFe$_6$Ge$_{6-x}$Ga$_x$} compounds
  (0.5${\le}$x${\le}$3.5)},\ }\href@noop {} {\bibfield  {journal} {\bibinfo
  {journal} {Journal of Alloys and Compounds}\ }\textbf {\bibinfo {volume}
  {329}},\ \bibinfo {pages} {8} (\bibinfo {year} {2001})}\BibitemShut {NoStop}%
\bibitem [{\citenamefont {Schobinger-Papamantellos}\ \emph
  {et~al.}(1997{\natexlab{a}})\citenamefont {Schobinger-Papamantellos},
  \citenamefont {Rodr{\'\i}guez-Carvajal},\ and\ \citenamefont
  {Buschow}}]{schobinger1997atomic}%
  \BibitemOpen
  \bibfield  {author} {\bibinfo {author} {\bibfnamefont {P.}~\bibnamefont
  {Schobinger-Papamantellos}}, \bibinfo {author} {\bibfnamefont
  {J.}~\bibnamefont {Rodr{\'\i}guez-Carvajal}},\ and\ \bibinfo {author}
  {\bibfnamefont {K.~H.~J.}\ \bibnamefont {Buschow}},\ }\bibfield  {title}
  {\bibinfo {title} {Atomic disorder and canted ferrimagnetism in the
  {TbCr$_6$Ge$_6$} compound. {A} neutron study},\ }\href@noop {} {\bibfield
  {journal} {\bibinfo  {journal} {Journal of Alloys and Compounds}\ }\textbf
  {\bibinfo {volume} {255}},\ \bibinfo {pages} {67} (\bibinfo {year}
  {1997}{\natexlab{a}})}\BibitemShut {NoStop}%
\bibitem [{\citenamefont {Schobinger-Papamantellos}\ \emph
  {et~al.}(1997{\natexlab{b}})\citenamefont {Schobinger-Papamantellos},
  \citenamefont {Rodr{\'\i}guez-Carvajal},\ and\ \citenamefont
  {Buschow}}]{schobinger1997ferrimagnetism}%
  \BibitemOpen
  \bibfield  {author} {\bibinfo {author} {\bibfnamefont {P.}~\bibnamefont
  {Schobinger-Papamantellos}}, \bibinfo {author} {\bibfnamefont
  {J.}~\bibnamefont {Rodr{\'\i}guez-Carvajal}},\ and\ \bibinfo {author}
  {\bibfnamefont {K.~H.~J.}\ \bibnamefont {Buschow}},\ }\bibfield  {title}
  {\bibinfo {title} {Ferrimagnetism and disorder in the {RCr$_6$Ge$_6$}
  compounds ({R=Dy, Ho, Er, Y}): {A} neutron study},\ }\href@noop {} {\bibfield
   {journal} {\bibinfo  {journal} {Journal of Alloys and Compounds}\ }\textbf
  {\bibinfo {volume} {256}},\ \bibinfo {pages} {92} (\bibinfo {year}
  {1997}{\natexlab{b}})}\BibitemShut {NoStop}%
\bibitem [{\citenamefont {Konyk}\ \emph {et~al.}(2021)\citenamefont {Konyk},
  \citenamefont {Romaka}, \citenamefont {Stadnyk}, \citenamefont {Romaka},\
  and\ \citenamefont {Pashkevych}}]{konyk2021phase}%
  \BibitemOpen
  \bibfield  {author} {\bibinfo {author} {\bibfnamefont {M.}~\bibnamefont
  {Konyk}}, \bibinfo {author} {\bibfnamefont {L.}~\bibnamefont {Romaka}},
  \bibinfo {author} {\bibfnamefont {Y.}~\bibnamefont {Stadnyk}}, \bibinfo
  {author} {\bibfnamefont {V.~V.}\ \bibnamefont {Romaka}},\ and\ \bibinfo
  {author} {\bibfnamefont {V.}~\bibnamefont {Pashkevych}},\ }\bibfield  {title}
  {\bibinfo {title} {Phase equilibria in the {Gd--Cr--Ge} system at 1070~{K}},\
  }\href@noop {} {\bibfield  {journal} {\bibinfo  {journal} {Physics and
  Chemistry of Solid State}\ }\textbf {\bibinfo {volume} {22}},\ \bibinfo
  {pages} {248} (\bibinfo {year} {2021})}\BibitemShut {NoStop}%
\bibitem [{\citenamefont {Romaka}\ \emph {et~al.}(2022)\citenamefont {Romaka},
  \citenamefont {Stadnyk}, \citenamefont {Romaka},\ and\ \citenamefont
  {Konyk}}]{romaka2022interaction}%
  \BibitemOpen
  \bibfield  {author} {\bibinfo {author} {\bibfnamefont {L.}~\bibnamefont
  {Romaka}}, \bibinfo {author} {\bibfnamefont {Y.}~\bibnamefont {Stadnyk}},
  \bibinfo {author} {\bibfnamefont {V.~V.}\ \bibnamefont {Romaka}},\ and\
  \bibinfo {author} {\bibfnamefont {M.}~\bibnamefont {Konyk}},\ }\bibfield
  {title} {\bibinfo {title} {Interaction between the components in {Tm-Cr-Ge}
  system at 1070~{K}},\ }\href@noop {} {\bibfield  {journal} {\bibinfo
  {journal} {Physics and Chemistry of Solid State}\ }\textbf {\bibinfo {volume}
  {23}},\ \bibinfo {pages} {633} (\bibinfo {year} {2022})}\BibitemShut
  {NoStop}%
\bibitem [{\citenamefont {Romaka}\ \emph {et~al.}(2024)\citenamefont {Romaka},
  \citenamefont {Romaka}, \citenamefont {Konyk}, \citenamefont {Corredor},
  \citenamefont {Srowik}, \citenamefont {Kuzhel}, \citenamefont {Stadnyk},\
  and\ \citenamefont {Yatskiv}}]{romaka2024structure}%
  \BibitemOpen
  \bibfield  {author} {\bibinfo {author} {\bibfnamefont {V.~V.}\ \bibnamefont
  {Romaka}}, \bibinfo {author} {\bibfnamefont {L.}~\bibnamefont {Romaka}},
  \bibinfo {author} {\bibfnamefont {M.}~\bibnamefont {Konyk}}, \bibinfo
  {author} {\bibfnamefont {L.~T.}\ \bibnamefont {Corredor}}, \bibinfo {author}
  {\bibfnamefont {K.}~\bibnamefont {Srowik}}, \bibinfo {author} {\bibfnamefont
  {B.}~\bibnamefont {Kuzhel}}, \bibinfo {author} {\bibfnamefont
  {Y.}~\bibnamefont {Stadnyk}},\ and\ \bibinfo {author} {\bibfnamefont
  {Y.}~\bibnamefont {Yatskiv}},\ }\bibfield  {title} {\bibinfo {title}
  {Structure, bonding, and properties of {RCr$_6$Ge$_6$} intermetallics
  ({R=Gd--Lu})},\ }\href@noop {} {\bibfield  {journal} {\bibinfo  {journal}
  {Journal of Solid State Chemistry}\ }\textbf {\bibinfo {volume} {338}},\
  \bibinfo {pages} {124874} (\bibinfo {year} {2024})}\BibitemShut {NoStop}%
\bibitem [{\citenamefont {Yang}\ \emph {et~al.}(2024)\citenamefont {Yang},
  \citenamefont {Zeng}, \citenamefont {He}, \citenamefont {Xu}, \citenamefont
  {Du},\ and\ \citenamefont {Qu}}]{yang2024crystal}%
  \BibitemOpen
  \bibfield  {author} {\bibinfo {author} {\bibfnamefont {X.}~\bibnamefont
  {Yang}}, \bibinfo {author} {\bibfnamefont {Q.}~\bibnamefont {Zeng}}, \bibinfo
  {author} {\bibfnamefont {M.}~\bibnamefont {He}}, \bibinfo {author}
  {\bibfnamefont {X.}~\bibnamefont {Xu}}, \bibinfo {author} {\bibfnamefont
  {H.}~\bibnamefont {Du}},\ and\ \bibinfo {author} {\bibfnamefont
  {Z.}~\bibnamefont {Qu}},\ }\bibfield  {title} {\bibinfo {title} {Crystal
  growth, magnetic and electrical transport properties of the {Kagome} magnet
  {$R$Cr$_6$Ge$_6$} ({$R$=Gd--Tm})},\ }\href@noop {} {\bibfield  {journal}
  {\bibinfo  {journal} {Chinese Physics B}\ } (\bibinfo {year}
  {2024})}\BibitemShut {NoStop}%
\bibitem [{\citenamefont {Romaka}\ \emph {et~al.}(2011)\citenamefont {Romaka},
  \citenamefont {Stadnyk}, \citenamefont {Romaka}, \citenamefont {Demchenko},
  \citenamefont {Stadnyshyn},\ and\ \citenamefont
  {Konyk}}]{romaka2011peculiarities}%
  \BibitemOpen
  \bibfield  {author} {\bibinfo {author} {\bibfnamefont {L.}~\bibnamefont
  {Romaka}}, \bibinfo {author} {\bibfnamefont {Y.}~\bibnamefont {Stadnyk}},
  \bibinfo {author} {\bibfnamefont {V.~V.}\ \bibnamefont {Romaka}}, \bibinfo
  {author} {\bibfnamefont {P.}~\bibnamefont {Demchenko}}, \bibinfo {author}
  {\bibfnamefont {M.}~\bibnamefont {Stadnyshyn}},\ and\ \bibinfo {author}
  {\bibfnamefont {M.}~\bibnamefont {Konyk}},\ }\bibfield  {title} {\bibinfo
  {title} {Peculiarities of component interaction in {$\{$Gd, Er$\}$--V--Sn}
  ternary systems at 870~{K} and crystal structure of {RV$_6$Sn$_6$}
  stannides},\ }\href@noop {} {\bibfield  {journal} {\bibinfo  {journal}
  {Journal of Alloys and Compounds}\ }\textbf {\bibinfo {volume} {509}},\
  \bibinfo {pages} {8862} (\bibinfo {year} {2011})}\BibitemShut {NoStop}%
\bibitem [{\citenamefont {Romaka}\ \emph {et~al.}(2019)\citenamefont {Romaka},
  \citenamefont {Konyk}, \citenamefont {Stadnyk}, \citenamefont {Romaka},\ and\
  \citenamefont {Serkiz}}]{romaka2019lu}%
  \BibitemOpen
  \bibfield  {author} {\bibinfo {author} {\bibfnamefont {L.}~\bibnamefont
  {Romaka}}, \bibinfo {author} {\bibfnamefont {M.}~\bibnamefont {Konyk}},
  \bibinfo {author} {\bibfnamefont {Y.}~\bibnamefont {Stadnyk}}, \bibinfo
  {author} {\bibfnamefont {V.}~\bibnamefont {Romaka}},\ and\ \bibinfo {author}
  {\bibfnamefont {R.}~\bibnamefont {Serkiz}},\ }\bibfield  {title} {\bibinfo
  {title} {{Lu-V-$\{$Ge, Sn$\}$} ternary systems},\ }\href@noop {} {\bibfield
  {journal} {\bibinfo  {journal} {Physics and Chemistry of Solid State}\
  }\textbf {\bibinfo {volume} {20}},\ \bibinfo {pages} {69} (\bibinfo {year}
  {2019})}\BibitemShut {NoStop}%
\bibitem [{\citenamefont {Huang}\ \emph {et~al.}(2023)\citenamefont {Huang},
  \citenamefont {Cui}, \citenamefont {Huang}, \citenamefont {Huo},
  \citenamefont {Liu}, \citenamefont {Li}, \citenamefont {Liang}, \citenamefont
  {Chen}, \citenamefont {Sun}, \citenamefont {Shen}, \citenamefont {Zhang},\
  and\ \citenamefont {Wang}}]{huang2023anisotropic}%
  \BibitemOpen
  \bibfield  {author} {\bibinfo {author} {\bibfnamefont {X.}~\bibnamefont
  {Huang}}, \bibinfo {author} {\bibfnamefont {Z.}~\bibnamefont {Cui}}, \bibinfo
  {author} {\bibfnamefont {C.}~\bibnamefont {Huang}}, \bibinfo {author}
  {\bibfnamefont {M.}~\bibnamefont {Huo}}, \bibinfo {author} {\bibfnamefont
  {H.}~\bibnamefont {Liu}}, \bibinfo {author} {\bibfnamefont {J.}~\bibnamefont
  {Li}}, \bibinfo {author} {\bibfnamefont {F.}~\bibnamefont {Liang}}, \bibinfo
  {author} {\bibfnamefont {L.}~\bibnamefont {Chen}}, \bibinfo {author}
  {\bibfnamefont {H.}~\bibnamefont {Sun}}, \bibinfo {author} {\bibfnamefont
  {B.}~\bibnamefont {Shen}}, \bibinfo {author} {\bibfnamefont {Y.}~\bibnamefont
  {Zhang}},\ and\ \bibinfo {author} {\bibfnamefont {M.}~\bibnamefont {Wang}},\
  }\bibfield  {title} {\bibinfo {title} {Anisotropic magnetism and electronic
  properties of the kagome metal {SmV$_6$Sn$_6$}},\ }\href@noop {} {\bibfield
  {journal} {\bibinfo  {journal} {Physical Review Materials}\ }\textbf
  {\bibinfo {volume} {7}},\ \bibinfo {pages} {054403} (\bibinfo {year}
  {2023})}\BibitemShut {NoStop}%
\bibitem [{\citenamefont {Xia}\ and\ \citenamefont
  {Bobev}(2006)}]{xia2006ybmn6sn6}%
  \BibitemOpen
  \bibfield  {author} {\bibinfo {author} {\bibfnamefont {S.-Q.}\ \bibnamefont
  {Xia}}\ and\ \bibinfo {author} {\bibfnamefont {S.}~\bibnamefont {Bobev}},\
  }\bibfield  {title} {\bibinfo {title} {{YbMn$_6$Sn$_6$}},\ }\href@noop {}
  {\bibfield  {journal} {\bibinfo  {journal} {Acta Crystallographica Section
  E}\ }\textbf {\bibinfo {volume} {62}},\ \bibinfo {pages} {i7} (\bibinfo
  {year} {2006})}\BibitemShut {NoStop}%
\bibitem [{\citenamefont {Mazet}\ \emph {et~al.}(2002)\citenamefont {Mazet},
  \citenamefont {Isnard},\ and\ \citenamefont {Malaman}}]{mazet2002study}%
  \BibitemOpen
  \bibfield  {author} {\bibinfo {author} {\bibfnamefont {T.}~\bibnamefont
  {Mazet}}, \bibinfo {author} {\bibfnamefont {O.}~\bibnamefont {Isnard}},\ and\
  \bibinfo {author} {\bibfnamefont {B.}~\bibnamefont {Malaman}},\ }\bibfield
  {title} {\bibinfo {title} {A study of the new {Yb$_{{\sim}0.6}$Fe$_6$Sn$_6$}
  compound by neutron diffraction, $^{57}${Fe} and $^{119}${Sn} {M{\"o}ssbauer}
  spectroscopy experiments},\ }\href@noop {} {\bibfield  {journal} {\bibinfo
  {journal} {Journal of Magnetism and Magnetic Materials}\ }\textbf {\bibinfo
  {volume} {241}},\ \bibinfo {pages} {51} (\bibinfo {year} {2002})}\BibitemShut
  {NoStop}%
\bibitem [{\citenamefont {Schobinger-Papamantellos}\ \emph
  {et~al.}(1998)\citenamefont {Schobinger-Papamantellos}, \citenamefont
  {Buschow}, \citenamefont {De~Boer}, \citenamefont {Ritter}, \citenamefont
  {Isnard},\ and\ \citenamefont {Fauth}}]{schobinger1998fe}%
  \BibitemOpen
  \bibfield  {author} {\bibinfo {author} {\bibfnamefont {P.}~\bibnamefont
  {Schobinger-Papamantellos}}, \bibinfo {author} {\bibfnamefont {K.~H.~J.}\
  \bibnamefont {Buschow}}, \bibinfo {author} {\bibfnamefont {F.~R.}\
  \bibnamefont {De~Boer}}, \bibinfo {author} {\bibfnamefont {C.}~\bibnamefont
  {Ritter}}, \bibinfo {author} {\bibfnamefont {O.}~\bibnamefont {Isnard}},\
  and\ \bibinfo {author} {\bibfnamefont {F.}~\bibnamefont {Fauth}},\ }\bibfield
   {title} {\bibinfo {title} {The {Fe} ordering in {RFe$_6$Ge$_6$} compounds
  with non-magnetic {R (R=Y, Lu, Hf)} studied by neutron diffraction and
  magnetic measurements},\ }\href@noop {} {\bibfield  {journal} {\bibinfo
  {journal} {Journal of Alloys and Compounds}\ }\textbf {\bibinfo {volume}
  {267}},\ \bibinfo {pages} {59} (\bibinfo {year} {1998})}\BibitemShut
  {NoStop}%
\bibitem [{\citenamefont {Meier}\ \emph {et~al.}(2023)\citenamefont {Meier},
  \citenamefont {Madhogaria}, \citenamefont {Mozaffari}, \citenamefont
  {Marshall}, \citenamefont {Graf}, \citenamefont {McGuire}, \citenamefont
  {Arachchige}, \citenamefont {Allen}, \citenamefont {Driver}, \citenamefont
  {Cao} \emph {et~al.}}]{meier2023tiny}%
  \BibitemOpen
  \bibfield  {author} {\bibinfo {author} {\bibfnamefont {W.~R.}\ \bibnamefont
  {Meier}}, \bibinfo {author} {\bibfnamefont {R.~P.}\ \bibnamefont
  {Madhogaria}}, \bibinfo {author} {\bibfnamefont {S.}~\bibnamefont
  {Mozaffari}}, \bibinfo {author} {\bibfnamefont {M.}~\bibnamefont {Marshall}},
  \bibinfo {author} {\bibfnamefont {D.~E.}\ \bibnamefont {Graf}}, \bibinfo
  {author} {\bibfnamefont {M.~A.}\ \bibnamefont {McGuire}}, \bibinfo {author}
  {\bibfnamefont {H.~W.~S.}\ \bibnamefont {Arachchige}}, \bibinfo {author}
  {\bibfnamefont {C.~L.}\ \bibnamefont {Allen}}, \bibinfo {author}
  {\bibfnamefont {J.}~\bibnamefont {Driver}}, \bibinfo {author} {\bibfnamefont
  {H.}~\bibnamefont {Cao}}, \emph {et~al.},\ }\bibfield  {title} {\bibinfo
  {title} {Tiny {Sc} allows the chains to rattle: {Impact} of {Lu} and {Y}
  doping on the charge-density wave in {ScV$_6$Sn$_6$}},\ }\href@noop {}
  {\bibfield  {journal} {\bibinfo  {journal} {Journal of the American Chemical
  Society}\ }\textbf {\bibinfo {volume} {145}},\ \bibinfo {pages} {20943}
  (\bibinfo {year} {2023})}\BibitemShut {NoStop}%
\bibitem [{sup()}]{supplemental}%
  \BibitemOpen
  \href@noop {} {}\bibinfo {note} {See Supplemental Material at [URL will be
  inserted by publisher] for additional information and
  Refs.~\cite{stevens1952matrix,hutchings1964point}.}\BibitemShut {Stop}%
\bibitem [{\citenamefont {Scheie}(2021)}]{scheie2021pycrystalfield}%
  \BibitemOpen
  \bibfield  {author} {\bibinfo {author} {\bibfnamefont {A.}~\bibnamefont
  {Scheie}},\ }\bibfield  {title} {\bibinfo {title} {$pycrystalfield$: software
  for calculation, analysis and fitting of crystal electric field
  {Hamiltonians}},\ }\href@noop {} {\bibfield  {journal} {\bibinfo  {journal}
  {Applied Crystallography}\ }\textbf {\bibinfo {volume} {54}},\ \bibinfo
  {pages} {356} (\bibinfo {year} {2021})}\BibitemShut {NoStop}%
\bibitem [{\citenamefont {Xiao}\ \emph {et~al.}(2024)\citenamefont {Xiao},
  \citenamefont {Chen}, \citenamefont {Ni}, \citenamefont {Li}, \citenamefont
  {Wen}, \citenamefont {Cui}, \citenamefont {Zhang}, \citenamefont {Liu},
  \citenamefont {Wang}, \citenamefont {Zhong} \emph
  {et~al.}}]{xiao2024preparation}%
  \BibitemOpen
  \bibfield  {author} {\bibinfo {author} {\bibfnamefont {Y.}~\bibnamefont
  {Xiao}}, \bibinfo {author} {\bibfnamefont {Y.}~\bibnamefont {Chen}}, \bibinfo
  {author} {\bibfnamefont {H.}~\bibnamefont {Ni}}, \bibinfo {author}
  {\bibfnamefont {Y.}~\bibnamefont {Li}}, \bibinfo {author} {\bibfnamefont
  {Z.}~\bibnamefont {Wen}}, \bibinfo {author} {\bibfnamefont {Y.}~\bibnamefont
  {Cui}}, \bibinfo {author} {\bibfnamefont {Y.}~\bibnamefont {Zhang}}, \bibinfo
  {author} {\bibfnamefont {S.}~\bibnamefont {Liu}}, \bibinfo {author}
  {\bibfnamefont {C.}~\bibnamefont {Wang}}, \bibinfo {author} {\bibfnamefont
  {R.}~\bibnamefont {Zhong}}, \emph {et~al.},\ }\bibfield  {title} {\bibinfo
  {title} {Preparation, crystal structure, and properties of the kagome metal
  {ThV$_6$Sn$_6$}},\ }\href@noop {} {\bibfield  {journal} {\bibinfo  {journal}
  {Inorganic Chemistry}\ }\textbf {\bibinfo {volume} {63}},\ \bibinfo {pages}
  {23288} (\bibinfo {year} {2024})}\BibitemShut {NoStop}%
\bibitem [{\citenamefont {He}\ \emph {et~al.}(2024)\citenamefont {He},
  \citenamefont {Xu}, \citenamefont {Li}, \citenamefont {Zeng}, \citenamefont
  {Liu}, \citenamefont {Zhao}, \citenamefont {Zhou}, \citenamefont {Zhou},\
  and\ \citenamefont {Qu}}]{he2024quantum}%
  \BibitemOpen
  \bibfield  {author} {\bibinfo {author} {\bibfnamefont {M.}~\bibnamefont
  {He}}, \bibinfo {author} {\bibfnamefont {X.}~\bibnamefont {Xu}}, \bibinfo
  {author} {\bibfnamefont {D.}~\bibnamefont {Li}}, \bibinfo {author}
  {\bibfnamefont {Q.}~\bibnamefont {Zeng}}, \bibinfo {author} {\bibfnamefont
  {Y.}~\bibnamefont {Liu}}, \bibinfo {author} {\bibfnamefont {H.}~\bibnamefont
  {Zhao}}, \bibinfo {author} {\bibfnamefont {S.}~\bibnamefont {Zhou}}, \bibinfo
  {author} {\bibfnamefont {J.}~\bibnamefont {Zhou}},\ and\ \bibinfo {author}
  {\bibfnamefont {Z.}~\bibnamefont {Qu}},\ }\bibfield  {title} {\bibinfo
  {title} {Quantum oscillations in the kagome metals {(Ti, Zr, Hf)V$_6$Sn$_6$
  at Van Hove} filling},\ }\href@noop {} {\bibfield  {journal} {\bibinfo
  {journal} {Physical Review B}\ }\textbf {\bibinfo {volume} {109}},\ \bibinfo
  {pages} {155117} (\bibinfo {year} {2024})}\BibitemShut {NoStop}%
\bibitem [{\citenamefont {Boothroyd}(2020)}]{boothroyd2020principles}%
  \BibitemOpen
  \bibfield  {author} {\bibinfo {author} {\bibfnamefont {A.~T.}\ \bibnamefont
  {Boothroyd}},\ }\href@noop {} {\emph {\bibinfo {title} {Principles of neutron
  scattering from condensed matter}}}\ (\bibinfo  {publisher} {Oxford
  University Press},\ \bibinfo {year} {2020})\BibitemShut {NoStop}%
\bibitem [{\citenamefont {Rossat-Mignod}\ \emph {et~al.}(1977)\citenamefont
  {Rossat-Mignod}, \citenamefont {Burlet}, \citenamefont {Villain},
  \citenamefont {Bartholin}, \citenamefont {Tcheng-Si}, \citenamefont
  {Florence},\ and\ \citenamefont {Vogt}}]{rossat1977phase}%
  \BibitemOpen
  \bibfield  {author} {\bibinfo {author} {\bibfnamefont {J.}~\bibnamefont
  {Rossat-Mignod}}, \bibinfo {author} {\bibfnamefont {P.}~\bibnamefont
  {Burlet}}, \bibinfo {author} {\bibfnamefont {J.}~\bibnamefont {Villain}},
  \bibinfo {author} {\bibfnamefont {H.}~\bibnamefont {Bartholin}}, \bibinfo
  {author} {\bibfnamefont {W.}~\bibnamefont {Tcheng-Si}}, \bibinfo {author}
  {\bibfnamefont {D.}~\bibnamefont {Florence}},\ and\ \bibinfo {author}
  {\bibfnamefont {O.}~\bibnamefont {Vogt}},\ }\bibfield  {title} {\bibinfo
  {title} {Phase diagram and magnetic structures of {CeSb}},\ }\href@noop {}
  {\bibfield  {journal} {\bibinfo  {journal} {Physical Review B}\ }\textbf
  {\bibinfo {volume} {16}},\ \bibinfo {pages} {440} (\bibinfo {year}
  {1977})}\BibitemShut {NoStop}%
\bibitem [{\citenamefont {Kuroda}\ \emph {et~al.}(2020)\citenamefont {Kuroda},
  \citenamefont {Arai}, \citenamefont {Rezaei}, \citenamefont {Kunisada},
  \citenamefont {Sakuragi}, \citenamefont {Alaei}, \citenamefont {Kinoshita},
  \citenamefont {Bareille}, \citenamefont {Noguchi}, \citenamefont {Nakayama}
  \emph {et~al.}}]{kuroda2020devil}%
  \BibitemOpen
  \bibfield  {author} {\bibinfo {author} {\bibfnamefont {K.}~\bibnamefont
  {Kuroda}}, \bibinfo {author} {\bibfnamefont {Y.}~\bibnamefont {Arai}},
  \bibinfo {author} {\bibfnamefont {N.}~\bibnamefont {Rezaei}}, \bibinfo
  {author} {\bibfnamefont {S.}~\bibnamefont {Kunisada}}, \bibinfo {author}
  {\bibfnamefont {S.}~\bibnamefont {Sakuragi}}, \bibinfo {author}
  {\bibfnamefont {M.}~\bibnamefont {Alaei}}, \bibinfo {author} {\bibfnamefont
  {Y.}~\bibnamefont {Kinoshita}}, \bibinfo {author} {\bibfnamefont
  {C.}~\bibnamefont {Bareille}}, \bibinfo {author} {\bibfnamefont
  {R.}~\bibnamefont {Noguchi}}, \bibinfo {author} {\bibfnamefont
  {M.}~\bibnamefont {Nakayama}}, \emph {et~al.},\ }\bibfield  {title} {\bibinfo
  {title} {Devil's staircase transition of the electronic structures in
  {CeSb}},\ }\href@noop {} {\bibfield  {journal} {\bibinfo  {journal} {Nature
  Communications}\ }\textbf {\bibinfo {volume} {11}},\ \bibinfo {pages} {2888}
  (\bibinfo {year} {2020})}\BibitemShut {NoStop}%
\bibitem [{\citenamefont {Selke}(1988)}]{selke1988annni}%
  \BibitemOpen
  \bibfield  {author} {\bibinfo {author} {\bibfnamefont {W.}~\bibnamefont
  {Selke}},\ }\bibfield  {title} {\bibinfo {title} {The {ANNNI}
  model—theoretical analysis and experimental application},\ }\href@noop {}
  {\bibfield  {journal} {\bibinfo  {journal} {Physics Reports}\ }\textbf
  {\bibinfo {volume} {170}},\ \bibinfo {pages} {213} (\bibinfo {year}
  {1988})}\BibitemShut {NoStop}%
\bibitem [{\citenamefont {Ishii}\ \emph {et~al.}(2013)\citenamefont {Ishii},
  \citenamefont {Harima}, \citenamefont {Okamoto}, \citenamefont {Yamaura},\
  and\ \citenamefont {Hiroi}}]{ishii2013ycr6ge6}%
  \BibitemOpen
  \bibfield  {author} {\bibinfo {author} {\bibfnamefont {Y.}~\bibnamefont
  {Ishii}}, \bibinfo {author} {\bibfnamefont {H.}~\bibnamefont {Harima}},
  \bibinfo {author} {\bibfnamefont {Y.}~\bibnamefont {Okamoto}}, \bibinfo
  {author} {\bibfnamefont {J.-i.}\ \bibnamefont {Yamaura}},\ and\ \bibinfo
  {author} {\bibfnamefont {Z.}~\bibnamefont {Hiroi}},\ }\bibfield  {title}
  {\bibinfo {title} {{YCr$_6$Ge$_6$} as a candidate compound for a kagome
  metal},\ }\href@noop {} {\bibfield  {journal} {\bibinfo  {journal} {Journal
  of the Physical Society of Japan}\ }\textbf {\bibinfo {volume} {82}},\
  \bibinfo {pages} {023705} (\bibinfo {year} {2013})}\BibitemShut {NoStop}%
\bibitem [{\citenamefont {Pokharel}\ \emph {et~al.}(2022)\citenamefont
  {Pokharel}, \citenamefont {Ortiz}, \citenamefont {Chamorro}, \citenamefont
  {Sarte}, \citenamefont {Kautzsch}, \citenamefont {Wu}, \citenamefont {Ruff},\
  and\ \citenamefont {Wilson}}]{pokharel2022highly}%
  \BibitemOpen
  \bibfield  {author} {\bibinfo {author} {\bibfnamefont {G.}~\bibnamefont
  {Pokharel}}, \bibinfo {author} {\bibfnamefont {B.}~\bibnamefont {Ortiz}},
  \bibinfo {author} {\bibfnamefont {J.}~\bibnamefont {Chamorro}}, \bibinfo
  {author} {\bibfnamefont {P.}~\bibnamefont {Sarte}}, \bibinfo {author}
  {\bibfnamefont {L.}~\bibnamefont {Kautzsch}}, \bibinfo {author}
  {\bibfnamefont {G.}~\bibnamefont {Wu}}, \bibinfo {author} {\bibfnamefont
  {J.}~\bibnamefont {Ruff}},\ and\ \bibinfo {author} {\bibfnamefont {S.~D.}\
  \bibnamefont {Wilson}},\ }\bibfield  {title} {\bibinfo {title} {Highly
  anisotropic magnetism in the vanadium-based kagome metal {TbV$_6$Sn$_6$}},\
  }\href@noop {} {\bibfield  {journal} {\bibinfo  {journal} {Physical Review
  Materials}\ }\textbf {\bibinfo {volume} {6}},\ \bibinfo {pages} {104202}
  (\bibinfo {year} {2022})}\BibitemShut {NoStop}%
\bibitem [{\citenamefont {Lyu}\ \emph {et~al.}(2024)\citenamefont {Lyu},
  \citenamefont {Liu}, \citenamefont {Zhang}, \citenamefont {Liu},
  \citenamefont {Yang}, \citenamefont {Wang}, \citenamefont {Feng},
  \citenamefont {Dong}, \citenamefont {Wang}, \citenamefont {Wei} \emph
  {et~al.}}]{lyu2024anomalous}%
  \BibitemOpen
  \bibfield  {author} {\bibinfo {author} {\bibfnamefont {M.}~\bibnamefont
  {Lyu}}, \bibinfo {author} {\bibfnamefont {Y.}~\bibnamefont {Liu}}, \bibinfo
  {author} {\bibfnamefont {S.}~\bibnamefont {Zhang}}, \bibinfo {author}
  {\bibfnamefont {J.}~\bibnamefont {Liu}}, \bibinfo {author} {\bibfnamefont
  {J.}~\bibnamefont {Yang}}, \bibinfo {author} {\bibfnamefont {Y.}~\bibnamefont
  {Wang}}, \bibinfo {author} {\bibfnamefont {Y.}~\bibnamefont {Feng}}, \bibinfo
  {author} {\bibfnamefont {X.}~\bibnamefont {Dong}}, \bibinfo {author}
  {\bibfnamefont {B.}~\bibnamefont {Wang}}, \bibinfo {author} {\bibfnamefont
  {H.}~\bibnamefont {Wei}}, \emph {et~al.},\ }\bibfield  {title} {\bibinfo
  {title} {Anomalous {Hall} effect and electronic correlation in a
  spin-reoriented kagome antiferromagnet {LuFe$_6$Sn$_6$}},\ }\href@noop {}
  {\bibfield  {journal} {\bibinfo  {journal} {Chinese Physics B}\ }\textbf
  {\bibinfo {volume} {33}},\ \bibinfo {pages} {107507} (\bibinfo {year}
  {2024})}\BibitemShut {NoStop}%
\bibitem [{\citenamefont {Porter}\ \emph {et~al.}(2023)\citenamefont {Porter},
  \citenamefont {Pokharel}, \citenamefont {Kim}, \citenamefont {Ryan},\ and\
  \citenamefont {Wilson}}]{porter2023incommensurate}%
  \BibitemOpen
  \bibfield  {author} {\bibinfo {author} {\bibfnamefont {Z.}~\bibnamefont
  {Porter}}, \bibinfo {author} {\bibfnamefont {G.}~\bibnamefont {Pokharel}},
  \bibinfo {author} {\bibfnamefont {J.-W.}\ \bibnamefont {Kim}}, \bibinfo
  {author} {\bibfnamefont {P.~J.}\ \bibnamefont {Ryan}},\ and\ \bibinfo
  {author} {\bibfnamefont {S.~D.}\ \bibnamefont {Wilson}},\ }\bibfield  {title}
  {\bibinfo {title} {Incommensurate magnetic order in the {$\mathbb{Z}_2$}
  kagome metal {GdV$_6$Sn$_6$}},\ }\href@noop {} {\bibfield  {journal}
  {\bibinfo  {journal} {Physical Review B}\ }\textbf {\bibinfo {volume}
  {108}},\ \bibinfo {pages} {035134} (\bibinfo {year} {2023})}\BibitemShut
  {NoStop}%
\bibitem [{\citenamefont {Laughlin}(2019)}]{laughlin2019magnetic}%
  \BibitemOpen
  \bibfield  {author} {\bibinfo {author} {\bibfnamefont {D.~E.}\ \bibnamefont
  {Laughlin}},\ }\bibfield  {title} {\bibinfo {title} {Magnetic transformations
  and phase diagrams},\ }\href@noop {} {\bibfield  {journal} {\bibinfo
  {journal} {Metallurgical and Materials Transactions A}\ }\textbf {\bibinfo
  {volume} {50}},\ \bibinfo {pages} {2555} (\bibinfo {year}
  {2019})}\BibitemShut {NoStop}%
\bibitem [{\citenamefont {Nagaosa}\ \emph {et~al.}(2010)\citenamefont
  {Nagaosa}, \citenamefont {Sinova}, \citenamefont {Onoda}, \citenamefont
  {MacDonald},\ and\ \citenamefont {Ong}}]{nagaosa2010anomalous}%
  \BibitemOpen
  \bibfield  {author} {\bibinfo {author} {\bibfnamefont {N.}~\bibnamefont
  {Nagaosa}}, \bibinfo {author} {\bibfnamefont {J.}~\bibnamefont {Sinova}},
  \bibinfo {author} {\bibfnamefont {S.}~\bibnamefont {Onoda}}, \bibinfo
  {author} {\bibfnamefont {A.~H.}\ \bibnamefont {MacDonald}},\ and\ \bibinfo
  {author} {\bibfnamefont {N.~P.}\ \bibnamefont {Ong}},\ }\bibfield  {title}
  {\bibinfo {title} {Anomalous {Hall} effect},\ }\href@noop {} {\bibfield
  {journal} {\bibinfo  {journal} {Reviews of Modern Physics}\ }\textbf
  {\bibinfo {volume} {82}},\ \bibinfo {pages} {1539} (\bibinfo {year}
  {2010})}\BibitemShut {NoStop}%
\bibitem [{\citenamefont {Nakatsuji}\ \emph {et~al.}(2015)\citenamefont
  {Nakatsuji}, \citenamefont {Kiyohara},\ and\ \citenamefont
  {Higo}}]{nakatsuji2015large}%
  \BibitemOpen
  \bibfield  {author} {\bibinfo {author} {\bibfnamefont {S.}~\bibnamefont
  {Nakatsuji}}, \bibinfo {author} {\bibfnamefont {N.}~\bibnamefont
  {Kiyohara}},\ and\ \bibinfo {author} {\bibfnamefont {T.}~\bibnamefont
  {Higo}},\ }\bibfield  {title} {\bibinfo {title} {Large anomalous {Hall}
  effect in a non-collinear antiferromagnet at room temperature},\ }\href@noop
  {} {\bibfield  {journal} {\bibinfo  {journal} {Nature}\ }\textbf {\bibinfo
  {volume} {527}},\ \bibinfo {pages} {212} (\bibinfo {year}
  {2015})}\BibitemShut {NoStop}%
\bibitem [{\citenamefont {Wang}\ \emph {et~al.}(2018)\citenamefont {Wang},
  \citenamefont {Xu}, \citenamefont {Lou}, \citenamefont {Liu}, \citenamefont
  {Li}, \citenamefont {Huang}, \citenamefont {Shen}, \citenamefont {Weng},
  \citenamefont {Wang},\ and\ \citenamefont {Lei}}]{wang2018large}%
  \BibitemOpen
  \bibfield  {author} {\bibinfo {author} {\bibfnamefont {Q.}~\bibnamefont
  {Wang}}, \bibinfo {author} {\bibfnamefont {Y.}~\bibnamefont {Xu}}, \bibinfo
  {author} {\bibfnamefont {R.}~\bibnamefont {Lou}}, \bibinfo {author}
  {\bibfnamefont {Z.}~\bibnamefont {Liu}}, \bibinfo {author} {\bibfnamefont
  {M.}~\bibnamefont {Li}}, \bibinfo {author} {\bibfnamefont {Y.}~\bibnamefont
  {Huang}}, \bibinfo {author} {\bibfnamefont {D.}~\bibnamefont {Shen}},
  \bibinfo {author} {\bibfnamefont {H.}~\bibnamefont {Weng}}, \bibinfo {author}
  {\bibfnamefont {S.}~\bibnamefont {Wang}},\ and\ \bibinfo {author}
  {\bibfnamefont {H.}~\bibnamefont {Lei}},\ }\bibfield  {title} {\bibinfo
  {title} {Large intrinsic anomalous hall effect in half-metallic ferromagnet
  {Co$_3$Sn$_2$S$_2$} with magnetic {Weyl} fermions},\ }\href@noop {}
  {\bibfield  {journal} {\bibinfo  {journal} {Nature Communications}\ }\textbf
  {\bibinfo {volume} {9}},\ \bibinfo {pages} {1} (\bibinfo {year}
  {2018})}\BibitemShut {NoStop}%
\bibitem [{\citenamefont {Zeng}\ \emph {et~al.}(2022)\citenamefont {Zeng},
  \citenamefont {Yu}, \citenamefont {Luo}, \citenamefont {Chen}, \citenamefont
  {Fang}, \citenamefont {Ma}, \citenamefont {Mo}, \citenamefont {Shen},
  \citenamefont {Yuan},\ and\ \citenamefont {Zhong}}]{zeng2022large}%
  \BibitemOpen
  \bibfield  {author} {\bibinfo {author} {\bibfnamefont {H.}~\bibnamefont
  {Zeng}}, \bibinfo {author} {\bibfnamefont {G.}~\bibnamefont {Yu}}, \bibinfo
  {author} {\bibfnamefont {X.}~\bibnamefont {Luo}}, \bibinfo {author}
  {\bibfnamefont {C.}~\bibnamefont {Chen}}, \bibinfo {author} {\bibfnamefont
  {C.}~\bibnamefont {Fang}}, \bibinfo {author} {\bibfnamefont {S.}~\bibnamefont
  {Ma}}, \bibinfo {author} {\bibfnamefont {Z.}~\bibnamefont {Mo}}, \bibinfo
  {author} {\bibfnamefont {J.}~\bibnamefont {Shen}}, \bibinfo {author}
  {\bibfnamefont {M.}~\bibnamefont {Yuan}},\ and\ \bibinfo {author}
  {\bibfnamefont {Z.}~\bibnamefont {Zhong}},\ }\bibfield  {title} {\bibinfo
  {title} {Large anomalous {Hall} effect in kagom{\'e} ferrimagnetic
  {HoMn$_6$Sn$_6$} single crystal},\ }\href@noop {} {\bibfield  {journal}
  {\bibinfo  {journal} {Journal of Alloys and Compounds}\ }\textbf {\bibinfo
  {volume} {899}},\ \bibinfo {pages} {163356} (\bibinfo {year}
  {2022})}\BibitemShut {NoStop}%
\bibitem [{\citenamefont {Jones}\ \emph {et~al.}(2024)\citenamefont {Jones},
  \citenamefont {Das}, \citenamefont {Bhandari}, \citenamefont {Liu},
  \citenamefont {Siegfried}, \citenamefont {Ghimire}, \citenamefont {Tsirkin},
  \citenamefont {Mazin},\ and\ \citenamefont {Ghimire}}]{jones2024origin}%
  \BibitemOpen
  \bibfield  {author} {\bibinfo {author} {\bibfnamefont {D.~C.}\ \bibnamefont
  {Jones}}, \bibinfo {author} {\bibfnamefont {S.}~\bibnamefont {Das}}, \bibinfo
  {author} {\bibfnamefont {H.}~\bibnamefont {Bhandari}}, \bibinfo {author}
  {\bibfnamefont {X.}~\bibnamefont {Liu}}, \bibinfo {author} {\bibfnamefont
  {P.}~\bibnamefont {Siegfried}}, \bibinfo {author} {\bibfnamefont {M.~P.}\
  \bibnamefont {Ghimire}}, \bibinfo {author} {\bibfnamefont {S.~S.}\
  \bibnamefont {Tsirkin}}, \bibinfo {author} {\bibfnamefont {I.}~\bibnamefont
  {Mazin}},\ and\ \bibinfo {author} {\bibfnamefont {N.~J.}\ \bibnamefont
  {Ghimire}},\ }\bibfield  {title} {\bibinfo {title} {Origin of spin
  reorientation and intrinsic anomalous {Hall} effect in the kagome ferrimagnet
  {TbMn$_6$Sn$_6$}},\ }\href@noop {} {\bibfield  {journal} {\bibinfo  {journal}
  {Physical Review B}\ }\textbf {\bibinfo {volume} {110}},\ \bibinfo {pages}
  {115134} (\bibinfo {year} {2024})}\BibitemShut {NoStop}%
\bibitem [{\citenamefont {Chen}\ \emph {et~al.}(2014)\citenamefont {Chen},
  \citenamefont {Niu},\ and\ \citenamefont {MacDonald}}]{chen2014anomalous}%
  \BibitemOpen
  \bibfield  {author} {\bibinfo {author} {\bibfnamefont {H.}~\bibnamefont
  {Chen}}, \bibinfo {author} {\bibfnamefont {Q.}~\bibnamefont {Niu}},\ and\
  \bibinfo {author} {\bibfnamefont {A.~H.}\ \bibnamefont {MacDonald}},\
  }\bibfield  {title} {\bibinfo {title} {Anomalous {Hall} effect arising from
  noncollinear antiferromagnetism},\ }\href@noop {} {\bibfield  {journal}
  {\bibinfo  {journal} {Physical Review Letters}\ }\textbf {\bibinfo {volume}
  {112}},\ \bibinfo {pages} {017205} (\bibinfo {year} {2014})}\BibitemShut
  {NoStop}%
\bibitem [{\citenamefont {Thomas}\ \emph {et~al.}(2016)\citenamefont {Thomas},
  \citenamefont {Rosa}, \citenamefont {Lee}, \citenamefont {Parameswaran},
  \citenamefont {Fisk},\ and\ \citenamefont {Xia}}]{thomas2016hall}%
  \BibitemOpen
  \bibfield  {author} {\bibinfo {author} {\bibfnamefont {S.~M.}\ \bibnamefont
  {Thomas}}, \bibinfo {author} {\bibfnamefont {P.~F.~S.}\ \bibnamefont {Rosa}},
  \bibinfo {author} {\bibfnamefont {S.~B.}\ \bibnamefont {Lee}}, \bibinfo
  {author} {\bibfnamefont {S.~A.}\ \bibnamefont {Parameswaran}}, \bibinfo
  {author} {\bibfnamefont {Z.}~\bibnamefont {Fisk}},\ and\ \bibinfo {author}
  {\bibfnamefont {J.}~\bibnamefont {Xia}},\ }\bibfield  {title} {\bibinfo
  {title} {Hall effect anomaly and low-temperature metamagnetism in the {Kondo}
  compound {CeAgBi$_2$}},\ }\href@noop {} {\bibfield  {journal} {\bibinfo
  {journal} {Physical Review B}\ }\textbf {\bibinfo {volume} {93}},\ \bibinfo
  {pages} {075149} (\bibinfo {year} {2016})}\BibitemShut {NoStop}%
\bibitem [{\citenamefont {Chen}\ \emph {et~al.}(2021)\citenamefont {Chen},
  \citenamefont {Le}, \citenamefont {Fu}, \citenamefont {Lin}, \citenamefont
  {Schnelle}, \citenamefont {Sun},\ and\ \citenamefont
  {Felser}}]{chen2021large}%
  \BibitemOpen
  \bibfield  {author} {\bibinfo {author} {\bibfnamefont {D.}~\bibnamefont
  {Chen}}, \bibinfo {author} {\bibfnamefont {C.}~\bibnamefont {Le}}, \bibinfo
  {author} {\bibfnamefont {C.}~\bibnamefont {Fu}}, \bibinfo {author}
  {\bibfnamefont {H.}~\bibnamefont {Lin}}, \bibinfo {author} {\bibfnamefont
  {W.}~\bibnamefont {Schnelle}}, \bibinfo {author} {\bibfnamefont
  {Y.}~\bibnamefont {Sun}},\ and\ \bibinfo {author} {\bibfnamefont
  {C.}~\bibnamefont {Felser}},\ }\bibfield  {title} {\bibinfo {title} {Large
  anomalous {Hall} effect in the kagome ferromagnet {LiMn$_6$Sn$_6$}},\
  }\href@noop {} {\bibfield  {journal} {\bibinfo  {journal} {Physical Review
  B}\ }\textbf {\bibinfo {volume} {103}},\ \bibinfo {pages} {144410} (\bibinfo
  {year} {2021})}\BibitemShut {NoStop}%
\bibitem [{\citenamefont {Shang}\ \emph {et~al.}(2021)\citenamefont {Shang},
  \citenamefont {Xu}, \citenamefont {Gawryluk}, \citenamefont {Ma},
  \citenamefont {Shiroka}, \citenamefont {Shi},\ and\ \citenamefont
  {Pomjakushina}}]{shang2021anomalous}%
  \BibitemOpen
  \bibfield  {author} {\bibinfo {author} {\bibfnamefont {T.}~\bibnamefont
  {Shang}}, \bibinfo {author} {\bibfnamefont {Y.}~\bibnamefont {Xu}}, \bibinfo
  {author} {\bibfnamefont {D.~J.}\ \bibnamefont {Gawryluk}}, \bibinfo {author}
  {\bibfnamefont {J.~Z.}\ \bibnamefont {Ma}}, \bibinfo {author} {\bibfnamefont
  {T.}~\bibnamefont {Shiroka}}, \bibinfo {author} {\bibfnamefont
  {M.}~\bibnamefont {Shi}},\ and\ \bibinfo {author} {\bibfnamefont
  {E.}~\bibnamefont {Pomjakushina}},\ }\bibfield  {title} {\bibinfo {title}
  {Anomalous {Hall} resistivity and possible topological {Hall} effect in the
  {EuAl$_4$} antiferromagnet},\ }\href@noop {} {\bibfield  {journal} {\bibinfo
  {journal} {Physical Review B}\ }\textbf {\bibinfo {volume} {103}},\ \bibinfo
  {pages} {L020405} (\bibinfo {year} {2021})}\BibitemShut {NoStop}%
\bibitem [{\citenamefont {von Klitzing}(2017)}]{von2017quantum}%
  \BibitemOpen
  \bibfield  {author} {\bibinfo {author} {\bibfnamefont {K.}~\bibnamefont {von
  Klitzing}},\ }\bibfield  {title} {\bibinfo {title} {Quantum hall effect:
  Discovery and application},\ }\href@noop {} {\bibfield  {journal} {\bibinfo
  {journal} {Annual Review of Condensed Matter Physics}\ }\textbf {\bibinfo
  {volume} {8}},\ \bibinfo {pages} {13} (\bibinfo {year} {2017})}\BibitemShut
  {NoStop}%
\bibitem [{\citenamefont {Stevens}(1952)}]{stevens1952matrix}%
  \BibitemOpen
  \bibfield  {author} {\bibinfo {author} {\bibfnamefont {K.~W.~H.}\
  \bibnamefont {Stevens}},\ }\bibfield  {title} {\bibinfo {title} {Matrix
  elements and operator equivalents connected with the magnetic properties of
  rare earth ions},\ }\href@noop {} {\bibfield  {journal} {\bibinfo  {journal}
  {Proceedings of the Physical Society. Section A}\ }\textbf {\bibinfo {volume}
  {65}},\ \bibinfo {pages} {209} (\bibinfo {year} {1952})}\BibitemShut
  {NoStop}%
\bibitem [{\citenamefont {Hutchings}(1964)}]{hutchings1964point}%
  \BibitemOpen
  \bibfield  {author} {\bibinfo {author} {\bibfnamefont {M.~T.}\ \bibnamefont
  {Hutchings}},\ }\bibfield  {title} {\bibinfo {title} {Point-charge
  calculations of energy levels of magnetic ions in crystalline electric
  fields},\ }in\ \href@noop {} {\emph {\bibinfo {booktitle} {Solid State
  Physics}}},\ \bibinfo {series} {Solid State Physics}, Vol.~\bibinfo {volume}
  {16}\ (\bibinfo  {publisher} {Academic Press},\ \bibinfo {year} {1964})\ pp.\
  \bibinfo {pages} {227--273}\BibitemShut {NoStop}%
\end{thebibliography}%


\begin{thebibliography}{3}%
\makeatletter
\providecommand \@ifxundefined [1]{%
 \@ifx{#1\undefined}
}%
\providecommand \@ifnum [1]{%
 \ifnum #1\expandafter \@firstoftwo
 \else \expandafter \@secondoftwo
 \fi
}%
\providecommand \@ifx [1]{%
 \ifx #1\expandafter \@firstoftwo
 \else \expandafter \@secondoftwo
 \fi
}%
\providecommand \natexlab [1]{#1}%
\providecommand \enquote  [1]{``#1''}%
\providecommand \bibnamefont  [1]{#1}%
\providecommand \bibfnamefont [1]{#1}%
\providecommand \citenamefont [1]{#1}%
\providecommand \href@noop [0]{\@secondoftwo}%
\providecommand \href [0]{\begingroup \@sanitize@url \@href}%
\providecommand \@href[1]{\@@startlink{#1}\@@href}%
\providecommand \@@href[1]{\endgroup#1\@@endlink}%
\providecommand \@sanitize@url [0]{\catcode `\\12\catcode `\$12\catcode
  `\&12\catcode `\#12\catcode `\^12\catcode `\_12\catcode `\%12\relax}%
\providecommand \@@startlink[1]{}%
\providecommand \@@endlink[0]{}%
\providecommand \url  [0]{\begingroup\@sanitize@url \@url }%
\providecommand \@url [1]{\endgroup\@href {#1}{\urlprefix }}%
\providecommand \urlprefix  [0]{URL }%
\providecommand \Eprint [0]{\href }%
\providecommand \doibase [0]{https://doi.org/}%
\providecommand \selectlanguage [0]{\@gobble}%
\providecommand \bibinfo  [0]{\@secondoftwo}%
\providecommand \bibfield  [0]{\@secondoftwo}%
\providecommand \translation [1]{[#1]}%
\providecommand \BibitemOpen [0]{}%
\providecommand \bibitemStop [0]{}%
\providecommand \bibitemNoStop [0]{.\EOS\space}%
\providecommand \EOS [0]{\spacefactor3000\relax}%
\providecommand \BibitemShut  [1]{\csname bibitem#1\endcsname}%
\let\auto@bib@innerbib\@empty
\bibitem [{\citenamefont {Scheie}(2021)}]{scheie2021pycrystalfield}%
  \BibitemOpen
  \bibfield  {author} {\bibinfo {author} {\bibfnamefont {A.}~\bibnamefont
  {Scheie}},\ }\href@noop {} {\bibfield  {journal} {\bibinfo  {journal}
  {Applied Crystallography}\ }\textbf {\bibinfo {volume} {54}},\ \bibinfo
  {pages} {356} (\bibinfo {year} {2021})}\BibitemShut {NoStop}%
\bibitem [{\citenamefont {Stevens}(1952)}]{stevens1952matrix}%
  \BibitemOpen
  \bibfield  {author} {\bibinfo {author} {\bibfnamefont {K.~W.~H.}\
  \bibnamefont {Stevens}},\ }\href@noop {} {\bibfield  {journal} {\bibinfo
  {journal} {Proceedings of the Physical Society. Section A}\ }\textbf
  {\bibinfo {volume} {65}},\ \bibinfo {pages} {209} (\bibinfo {year}
  {1952})}\BibitemShut {NoStop}%
\bibitem [{\citenamefont {Hutchings}(1964)}]{hutchings1964point}%
  \BibitemOpen
  \bibfield  {author} {\bibinfo {author} {\bibfnamefont {M.~T.}\ \bibnamefont
  {Hutchings}},\ }in\ \href@noop {} {\emph {\bibinfo {booktitle} {Solid State
  Physics}}},\ \bibinfo {series} {Solid State Physics}, Vol.~\bibinfo {volume}
  {16}\ (\bibinfo  {publisher} {Academic Press},\ \bibinfo {year} {1964})\ pp.\
  \bibinfo {pages} {227--273}\BibitemShut {NoStop}%
\end{thebibliography}%

\end{document}


\begin{center}
\Large 
\textbf{Magnetic order and physical properties of the Kagome metal \UNS}\\
\vspace{1em}
Supplemental Material\\
\vspace{1em}
\normalsize
Z.~W.~Riedel, W.~Simeth, C.~S.~Kengle, S.~M.~Thomas, J.~D.~Thompson, A.~O.~Scheie, F.~Ronning, C.~Lane, Jian-Xin~Zhu, P.~F.~S.~Rosa, E.~D.~Bauer 
\end{center}

\vspace{0.5em}

\section{Single crystal x-ray diffraction data}
At room temperature, \UNS\ and \TNS\ crystallize in the disordered, SmMn$_6$Sn$_6$ structure type. The structure type contains site splitting between higher and lower occupancy U/Sn or Th/Sn site pairs. The 166 stoichiometry is maintained.

\begin{table}[h!]
\small
\centering 
\caption{\label{tab:scXRD_data_U} Room temperature refinement details for a \UNS\ single crystal} 
\begin{tabular}{l c}
\midrule
space group & $P$6/$mmm$ \\
$a$ & 5.7570(3)~\AA \\
$c$ & 9.5061(7)~\AA \\
$V$ & 272.85(3)~\AA \\
$Z$ & 1 \\
$\rho$ & 9.175~g~cm$^{-1}$ \\
$\mu$ & 34.209~mm$^{-1}$ \\
$F$(000) & 638 \\
$R_{\mathrm{1}}$ & 0.0210 \\
$wR_{\mathrm{2}}$ & 0.0568 \\
$R_{\mathrm{int}}$ & 0.0902 \\
\midrule
\end{tabular}
\vspace{3em}
\small
\centering 
\caption{\label{tab:scXRD_positions_U} The single crystal refinement data for \UNS\ includes disordered uranium (U1A, U1B) and tin (Sn1A, Sn1B) sites split 92.7(3)/7.3(3)\% (A/B). Anisotropic displacement parameters are in \AA$^2$.}
\resizebox{\columnwidth}{!}{
\begin{tabular}{c c c c c c c c c c c c}
\midrule Label & Site & x & y & z & occ. & U$_{11}$ & U$_{22}$ & U$_{33}$ & U$_{23}$ & U$_{13}$ & U$_{12}$ \\
\midrule
U1A & 1$a$ & 0 & 0 & 0 & 0.927(3) & 0.0063(3) & 0.0063(3) & 0.0082(5) & 0 & 0 & 0.0032(2) \\
U1B & 1$b$ & 0 & 0 & 0.5 & 0.073(3) & 0.0063(3) & 0.0063(3) & 0.0082(5) & 0 & 0 & 0.0032(2) \\
Nb1 & 6$i$ & 0.5 & 0 & 0.24843(9) & 1 & 0.0040(3) & 0.0037(4) & 0.0034(4) & 0 & 0 & 0.0018(2) \\
Sn1A & 2$e$ & 0 & 0 & 0.3385(2) & 0.927(3) & 0.0056(4) & 0.0056(4) & 0.0088(6) & 0 & 0 & 0.0028(2) \\
Sn1B & 2$e$ & 0 & 0 & 0.1578(2) & 0.073(3) & 0.0056(4) & 0.0056(4) & 0.0088(6) & 0 & 0 & 0.0028(2) \\
Sn2 & 2$d$ & 0.33333 & 0.66667 & 0.5 & 1 & 0.0075(4) & 0.0075(4) & 0.0035(5) & 0 & 0 & 0.0037(2) \\
Sn3 & 2$c$ & 0.33333 & 0.66667 & 0 & 1 & 0.0057(3) & 0.0057(3) & 0.0022(5) & 0 & 0 & 0.0028(2) \\
\end{tabular}
}
\end{table}

\begin{table}[!t]
\small
\centering 
\caption{\label{tab:scXRD_data_Th} Room temperature refinement details for a \TNS\ single crystal} 
\begin{tabular}{l c}
\midrule
space group & $P$6/$mmm$ \\
$a$ & 5.7925(4)~\AA \\
$c$ & 9.541(1)~\AA \\
$V$ & 277.23(4)~\AA \\
$Z$ & 1 \\
$\rho$ & 8.994~g~cm$^{-1}$ \\
$\mu$ & 32.478~mm$^{-1}$ \\
$F$(000) & 636 \\
$R_{\mathrm{1}}$ & 0.0260 \\
$wR_{\mathrm{2}}$ & 0.0645 \\
$R_{\mathrm{int}}$ & 0.0759 \\
\midrule
\end{tabular}
\vspace{3em}
\small
\centering 
\caption{\label{tab:scXRD_positions_Th} The single crystal refinement data for \TNS\ includes disordered thorium (Th1A, Th1B) and tin (Sn1A, Sn1B) sites split 94.0(2)/6.0(2)\% (A/B). Anisotropic displacement parameters are in \AA$^2$.}
\resizebox{\columnwidth}{!}{
\begin{tabular}{c c c c c c c c c c c c}
\midrule Label & Site & x & y & z & occ. & U$_{11}$ & U$_{22}$ & U$_{33}$ & U$_{23}$ & U$_{13}$ & U$_{12}$ \\
\midrule
Th1A & 1$a$ & 0 & 0 & 0 & 0.940(2) & 0.0046(2) & 0.0046(2) & 0.0054(2) & 0 & 0 & 0.00229(9) \\
Th1B & 1$b$ & 0 & 0 & 0.5 & 0.060(2) & 0.0046(2) & 0.0046(2) & 0.0054(2) & 0 & 0 & 0.00229(9) \\
Nb1 & 6$i$ & 0.5 & 0 & 0.24890(4) & 1 & 0.0045(2) & 0.0042(2) & 0.0014(3) & 0 & 0 & 0.0021(1) \\
Sn1A & 2$e$ & 0 & 0 & 0.34377(7) & 0.940(2) & 0.0053(2) & 0.0053(2) & 0.0050(3) & 0 & 0 & 0.0027(1) \\
Sn1B & 2$e$ & 0 & 0 & 0.1529(1) & 0.060(2) & 0.0053(2) & 0.0053(2) & 0.0050(3) & 0 & 0 & 0.0027(1) \\
Sn2 & 2$d$ & 0.33333 & 0.66667 & 0.5 & 1 & 0.0076(2) & 0.0076(2) & 0.0010(3) & 0 & 0 & 0.0038(1) \\
Sn3 & 2$c$ & 0.33333 & 0.66667 & 0 & 1 & 0.0057(2) & 0.0057(2) & 0.0013(3) & 0 & 0 & 0.0029(1) \\
\end{tabular}
}
\end{table}
\clearpage
\section{Crystalline electric field fits}
\begin{figure}[h!]
    \centering
    \includegraphics[width=0.6\columnwidth]{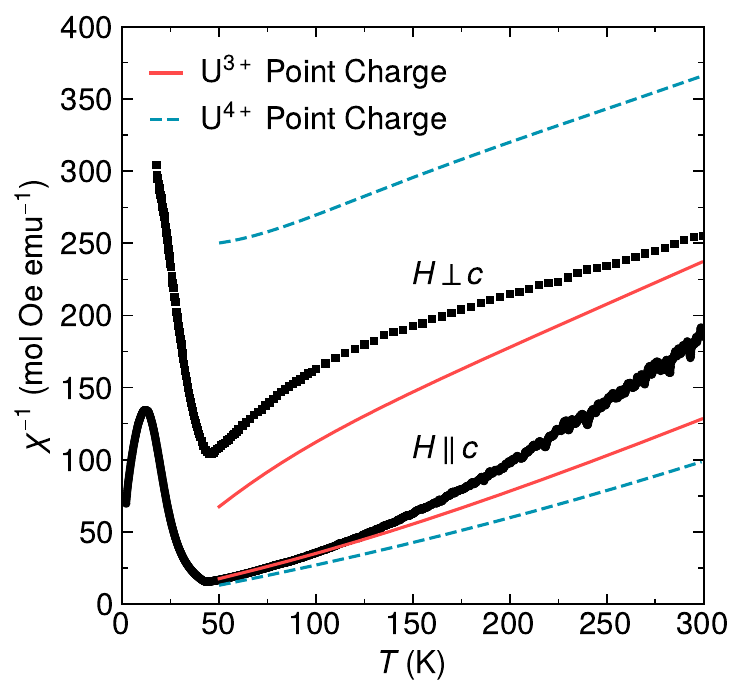}
    \caption{The inverse magnetic susceptibility of \UNS\ plotted with curves calculated from the point charge CEF model using trivalent or tetravalent uranium}
    \label{fig:ptcharge}
\end{figure}

The \UNS\ magnetic susceptibilities with $H{\parallel}c$ and $H{\perp}c$ were initially calculated with a point charge model using \textsc{PyCrystalField} \cite{scheie2021pycrystalfield}. An ideal \UNS\ structure was used with the primary U and Sn sites (U1A and Sn1A) fully occupied and the secondary sites unoccupied. The single-ion crystalline electric field (CEF) Hamiltonian followed the Stevens Operator formalism \cite{stevens1952matrix,hutchings1964point} with four symmetry-allowed parameters, giving $H_\mathrm{CEF}=B_2^0 O_2^0 + B_4^0 O_4^0 + B_6^0 O_6^0 + B_6^6 O_6^6$. $O_n^m$ are CEF Stevens Operators quantized along $c$, and $B_n^m$ are scalar fit parameters. The rudimentary point charge model results in a ground state doublet for U$^{3+}$ and U$^{4+}$ with the eigenstates shown in Tables~\ref{tab:ptcharge3} and \ref{tab:ptcharge4} and the inverse magnetic susceptibilities shown in Fig.~\ref{fig:ptcharge}. 

The \UNS\ anisotropic magnetic susceptibilities were then fit simultaneously ($T\geq50$~K) to three CEF models following Eq.~\ref{eq:CEF}. 
\begin{equation}
    \label{eq:CEF}
    \frac{1}{\chi} = \frac{1}{\chi_\mathrm{CEF}} - \lambda
\end{equation}
For case I, only $B_2^0$ and the molecular field term ($\lambda$) were fit, with the other crystal field terms coming from the point charge model. The fitting procedure led to a small, positive $B_2^0$ term ($\sim$10$^{-1}$~meV). Next, for case II, all four CEF parameters were fit without a molecular field term, producing larger, positive $B_2^0$ terms ($\sim$10$^{0}$~meV) for U$^{3+}$ and U$^{4+}$.
Finally, for case III, all four CEF parameters and the molecular field term were fit, again resulting in $B_2^{0}\sim10^{0}$~meV.
The interpretability of the fits is limited. Increasing the number of parameters necessarily improved the  fit quantitatively from case I to case III, but fitting multiple parameters to two datasets results in an underconstrained problem. All that can be reliably stated is that the U$^{3+}$ fits were closer to the data in each case and that each U$^{3+}$ fit had a ground state doublet primarily of $|\pm\frac{7}{2}\rangle$ character.

\begin{figure}[h!]
    \centering
    \includegraphics[width=0.6\columnwidth]{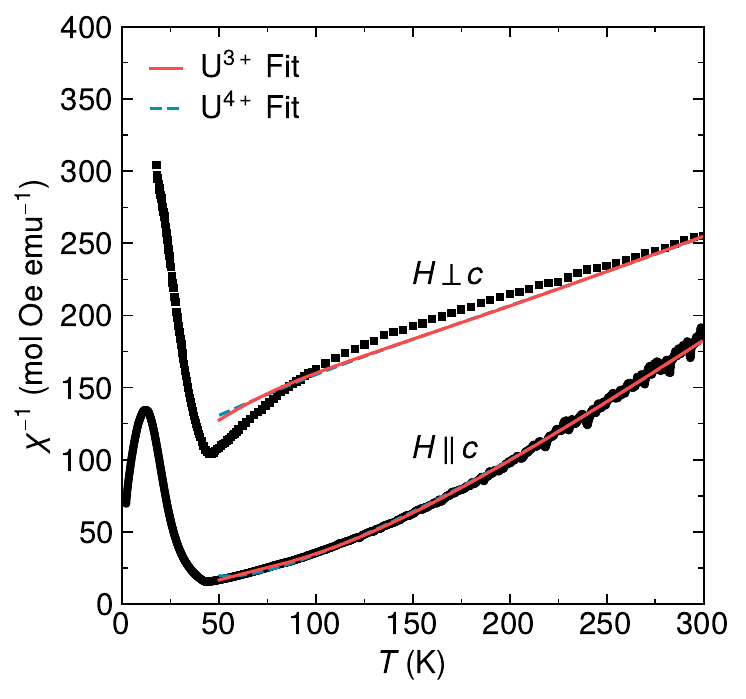}
    \caption{The inverse magnetic susceptibility of \UNS\ plotted with curves calculated from the case III CEF fits using trivalent and tetravalent uranium. Fits for U$^{3+}$ and U$^{4+}$ are nearly indistinguishable in the figure.}
    \label{fig:CEF_chi}
\end{figure}

As a reference, Tables~\ref{tab:U3plus_eigen} and \ref{tab:U4plus_eigen} list the eigenstates obtained from the case III fits to trivalent and tetravalent U, respectively; Table~\ref{tab:stevens} contains the fit $B_n^m$ parameters for the point charge model and case III; and Fig.~\ref{fig:CEF_chi} shows the case III fits to the magnetic susceptibility data. 

\begin{table}[h!]
\caption{Eigenstates of the point charge model for \UNS\ with U$^{3+}$}
\resizebox{\columnwidth}{!}{
\begin{tabular}{c|cccccccccc}
\midrule
E (meV) &$| -\frac{9}{2}\rangle$ & $| -\frac{7}{2}\rangle$ & $| -\frac{5}{2}\rangle$ & $| -\frac{3}{2}\rangle$ & $| -\frac{1}{2}\rangle$ & $| \frac{1}{2}\rangle$ & $| \frac{3}{2}\rangle$ & $| \frac{5}{2}\rangle$ & $| \frac{7}{2}\rangle$ & $| \frac{9}{2}\rangle$ \\
 \midrule
0.000 & 0.0 & 0.9785 & 0.0 & 0.0 & 0.0 & 0.0 & 0.0 & -0.2063 & 0.0 & 0.0 \tabularnewline
0.000 & 0.0 & 0.0 & -0.2063 & 0.0 & 0.0 & 0.0 & 0.0 & 0.0 & 0.9785 & 0.0 \tabularnewline
2.506 & -0.9973 & 0.0 & 0.0 & 0.0 & 0.0 & 0.0 & 0.0734 & 0.0 & 0.0 & 0.0 \tabularnewline
2.506 & 0.0 & 0.0 & 0.0 & 0.0734 & 0.0 & 0.0 & 0.0 & 0.0 & 0.0 & -0.9973 \tabularnewline
16.402 & 0.0 & 0.0 & -0.9785 & 0.0 & 0.0 & 0.0 & 0.0 & 0.0 & -0.2063 & 0.0 \tabularnewline
16.402 & 0.0 & 0.2063 & 0.0 & 0.0 & 0.0 & 0.0 & 0.0 & 0.9785 & 0.0 & 0.0 \tabularnewline
32.110 & -0.0734 & 0.0 & 0.0 & 0.0 & 0.0 & 0.0 & -0.9973 & 0.0 & 0.0 & 0.0 \tabularnewline
32.110 & 0.0 & 0.0 & 0.0 & 0.9973 & 0.0 & 0.0 & 0.0 & 0.0 & 0.0 & 0.0734 \tabularnewline
41.625 & 0.0 & 0.0 & 0.0 & 0.0 & 1.0 & 0.0 & 0.0 & 0.0 & 0.0 & 0.0 \tabularnewline
41.625 & 0.0 & 0.0 & 0.0 & 0.0 & 0.0 & 1.0 & 0.0 & 0.0 & 0.0 & 0.0 \tabularnewline
\end{tabular}
}
\label{tab:ptcharge3}
\end{table}

\begin{table*}
\caption{Eigenstates of the point charge model for \UNS\ with U$^{4+}$}
\resizebox{0.907\columnwidth}{!}{
\begin{tabular}{c|ccccccccc}
\midrule
E (meV) &$|-4\rangle$ & $|-3\rangle$ & $|-2\rangle$ & $|-1\rangle$ & $|0\rangle$ & $|1\rangle$ & $|2\rangle$ & $|3\rangle$ & $|4\rangle$ \\
 \midrule 
0.000 & -0.9998 & 0.0 & 0.0 & 0.0 & 0.0 & 0.0 & -0.0197 & 0.0 & 0.0 \tabularnewline
0.000 & 0.0 & 0.0 & -0.0197 & 0.0 & 0.0 & 0.0 & 0.0 & 0.0 & -0.9998 \tabularnewline
19.305 & 0.0 & -0.7071 & 0.0 & 0.0 & 0.0 & 0.0 & 0.0 & -0.7071 & 0.0 \tabularnewline
22.110 & 0.0 & -0.7071 & 0.0 & 0.0 & 0.0 & 0.0 & 0.0 & 0.7071 & 0.0 \tabularnewline
53.805 & -0.0197 & 0.0 & 0.0 & 0.0 & 0.0 & 0.0 & 0.9998 & 0.0 & 0.0 \tabularnewline
53.805 & 0.0 & 0.0 & -0.9998 & 0.0 & 0.0 & 0.0 & 0.0 & 0.0 & 0.0197 \tabularnewline
81.630 & 0.0 & 0.0 & 0.0 & -1.0 & 0.0 & 0.0 & 0.0 & 0.0 & 0.0 \tabularnewline
81.630 & 0.0 & 0.0 & 0.0 & 0.0 & 0.0 & -1.0 & 0.0 & 0.0 & 0.0 \tabularnewline
92.313 & 0.0 & 0.0 & 0.0 & 0.0 & -1.0 & 0.0 & 0.0 & 0.0 & 0.0 \tabularnewline
\end{tabular}
}
\label{tab:ptcharge4}
\end{table*}

\begin{table}[h!]
\caption{Eigenstates of the case III CEF fit with U$^{3+}$}
\resizebox{\columnwidth}{!}{
\begin{tabular}{c|cccccccccc}
\midrule
E (meV) &$| -\frac{9}{2}\rangle$ & $| -\frac{7}{2}\rangle$ & $| -\frac{5}{2}\rangle$ & $| -\frac{3}{2}\rangle$ & $| -\frac{1}{2}\rangle$ & $| \frac{1}{2}\rangle$ & $| \frac{3}{2}\rangle$ & $| \frac{5}{2}\rangle$ & $| \frac{7}{2}\rangle$ & $| \frac{9}{2}\rangle$ \\
\midrule
0.000 & 0.0 & 0.7313 & 0.0 & 0.0 & 0.0 & 0.0 & 0.0 & 0.6821 & 0.0 & 0.0 \tabularnewline
0.000 & 0.0 & 0.0 & -0.6821 & 0.0 & 0.0 & 0.0 & 0.0 & 0.0 & -0.7313 & 0.0 \tabularnewline
28.696 & 0.0 & 0.0 & 0.0 & 0.0 & 1.0 & 0.0 & 0.0 & 0.0 & 0.0 & 0.0 \tabularnewline
28.696 & 0.0 & 0.0 & 0.0 & 0.0 & 0.0 & 1.0 & 0.0 & 0.0 & 0.0 & 0.0 \tabularnewline
28.750 & 0.0832 & 0.0 & 0.0 & 0.0 & 0.0 & 0.0 & 0.9965 & 0.0 & 0.0 & 0.0 \tabularnewline
28.750 & 0.0 & 0.0 & 0.0 & 0.9965 & 0.0 & 0.0 & 0.0 & 0.0 & 0.0 & 0.0832 \tabularnewline
35.382 & 0.0 & 0.0 & -0.7313 & 0.0 & 0.0 & 0.0 & 0.0 & 0.0 & 0.6821 & 0.0 \tabularnewline
35.382 & 0.0 & 0.6821 & 0.0 & 0.0 & 0.0 & 0.0 & 0.0 & -0.7313 & 0.0 & 0.0 \tabularnewline
168.076 & -0.9965 & 0.0 & 0.0 & 0.0 & 0.0 & 0.0 & 0.0832 & 0.0 & 0.0 & 0.0 \tabularnewline
168.076 & 0.0 & 0.0 & 0.0 & -0.0832 & 0.0 & 0.0 & 0.0 & 0.0 & 0.0 & 0.9965 \tabularnewline
\end{tabular}
}
\label{tab:U3plus_eigen}
\end{table}

\begin{table}[h!]
\caption{Eigenstates of the case III CEF fit with U$^{4+}$}
\resizebox{0.907\columnwidth}{!}{
\begin{tabular}{c|ccccccccc}
\midrule
E (meV) &$|-4\rangle$ & $|-3\rangle$ & $|-2\rangle$ & $|-1\rangle$ & $|0\rangle$ & $|1\rangle$ & $|2\rangle$ & $|3\rangle$ & $|4\rangle$ \\
\midrule
0.000 & 0.0 & -0.7071 & 0.0 & 0.0 & 0.0 & 0.0 & 0.0 & 0.7071 & 0.0 \tabularnewline
10.489 & 0.0731 & 0.0 & 0.0 & 0.0 & 0.0 & 0.0 & -0.9973 & 0.0 & 0.0 \tabularnewline
10.489 & 0.0 & 0.0 & -0.9973 & 0.0 & 0.0 & 0.0 & 0.0 & 0.0 & 0.0731 \tabularnewline
43.137 & 0.0 & 0.0 & 0.0 & -1.0 & 0.0 & 0.0 & 0.0 & 0.0 & 0.0 \tabularnewline
43.137 & 0.0 & 0.0 & 0.0 & 0.0 & 0.0 & 1.0 & 0.0 & 0.0 & 0.0 \tabularnewline
47.594 & 0.0 & -0.7071 & 0.0 & 0.0 & 0.0 & 0.0 & 0.0 & -0.7071 & 0.0 \tabularnewline
57.703 & 0.0 & 0.0 & 0.0 & 0.0 & -1.0 & 0.0 & 0.0 & 0.0 & 0.0 \tabularnewline
257.146 & -0.9973 & 0.0 & 0.0 & 0.0 & 0.0 & 0.0 & -0.0731 & 0.0 & 0.0 \tabularnewline
257.146 & 0.0 & 0.0 & 0.0731 & 0.0 & 0.0 & 0.0 & 0.0 & 0.0 & 0.9973 \tabularnewline
\end{tabular}
}
\label{tab:U4plus_eigen}
\end{table}

\begin{table}[h!]
\centering
\caption{Point charge (PC) and case III (Fit) CEF parameters with $B_n^m$ parameters in units of meV and $\lambda$ parameters in units of mol Oe emu$^{-1}$}
\begin{tabular}{c|cccc}
\midrule
$B_n^m$ & U$^{3+}$ PC & U$^{4+}$ PC & U$^{3+}$ Fit & U$^{4+}$ Fit
\\
\midrule
$B_2^0$ & -6.582$\times$10$^{-1}$ & -1.873 & 2.052 & 4.275 \\
$B_4^0$ & 5.085$\times$10$^{-3}$ & 8.979$\times$10$^{-3}$& 2.439$\times$10$^{-2}$ & 6.426$\times$10$^{-2}$ \\
$B_6^0$ & 9.650$\times$10$^{-6}$ & -8.180$\times$10$^{-6}$& 2.755$\times$10$^{-4}$ & 2.763$\times$10$^{-4}$ \\
$B_6^6$ & 6.569$\times$10$^{-4}$ & -5.565$\times$10$^{-4}$& -3.502$\times$10$^{-3}$ & 9.443$\times$10$^{-3}$\\
$\lambda_\mathrm{c}$ & - & - & 77.36 & 86.27 \\
$\lambda_\mathrm{ab}$ & - & - & -75.72 & -55.08 \\
\end{tabular}
\label{tab:stevens}
\end{table}

\clearpage
\section{Additional magnetic property data}
\begin{figure}[h!]
    \centering
    \includegraphics[width=0.48\columnwidth]{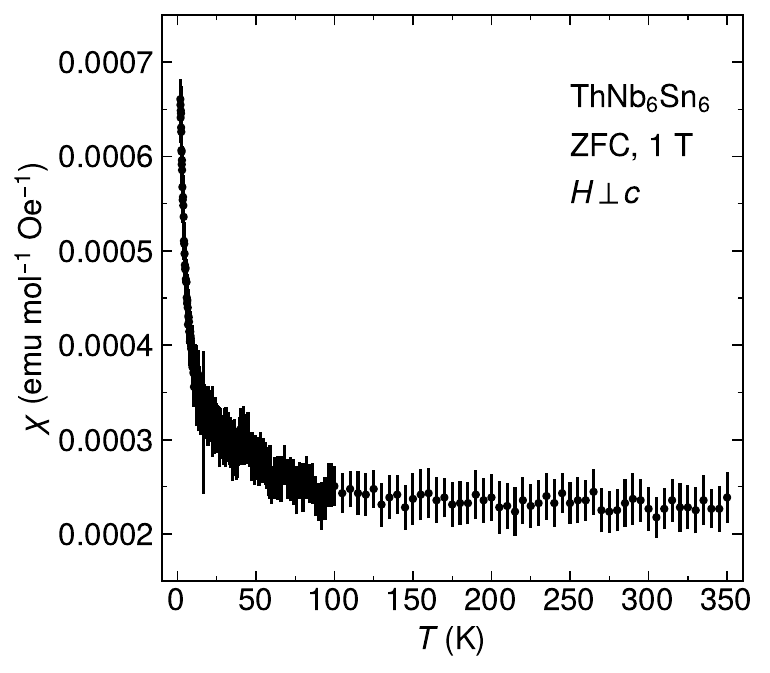}
    \caption{Magnetic susceptibility of \TNS}
    \label{fig:ThNb6Sn6_chi}
\end{figure}

The magnetic susceptibility of ThNb$_6$Sn$_6$ (Fig.~\ref{fig:ThNb6Sn6_chi}) shows Pauli paramagnetic behavior with $\chi_{\mathrm{P}}$~=~2.3(2)$\times$10$^{-4}$ emu~mol$^{-1}$~Oe$^{-1}$ and a small Curie tail (likely due to a surface impurity). The lack of a long-range ordering transition suggests that niobium does not magnetically order in \UNS.

The magnetization curves of \UNS\ collected from 0 to 16~T show a remnant magnetization at 2~K, 5~K, and 10~K. At 20~K, a small hysteresis persists, but the magnetization goes to zero at zero field during the magnetic field down sweep.

\begin{figure}[h!]
    \centering
    \includegraphics[width=0.48\columnwidth]{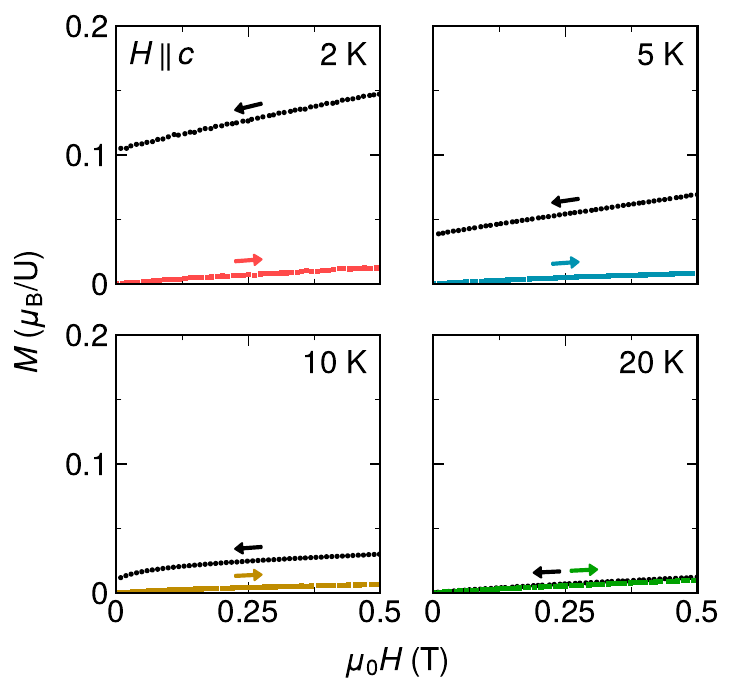}
    \caption{Low field magnetization curves of \UNS\ are shown with arrows indicating increasing (colorful) or decreasing (black) magnetic field.}
    \label{fig:remnant}
\end{figure}

\begin{figure}[h!]
    \centering
    \subfloat{\includegraphics[width=0.45\columnwidth]{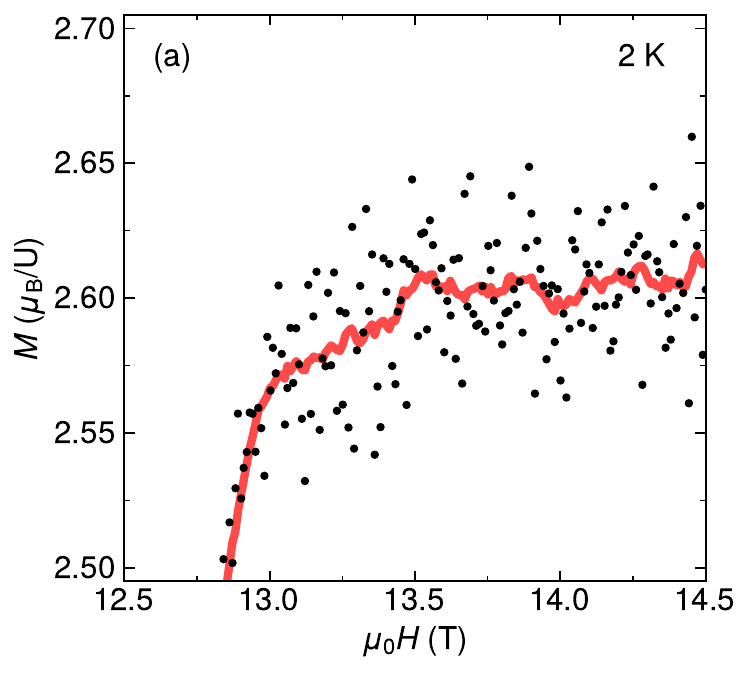}}
    \hspace{2em}
    \subfloat{\includegraphics[width=0.45\columnwidth]{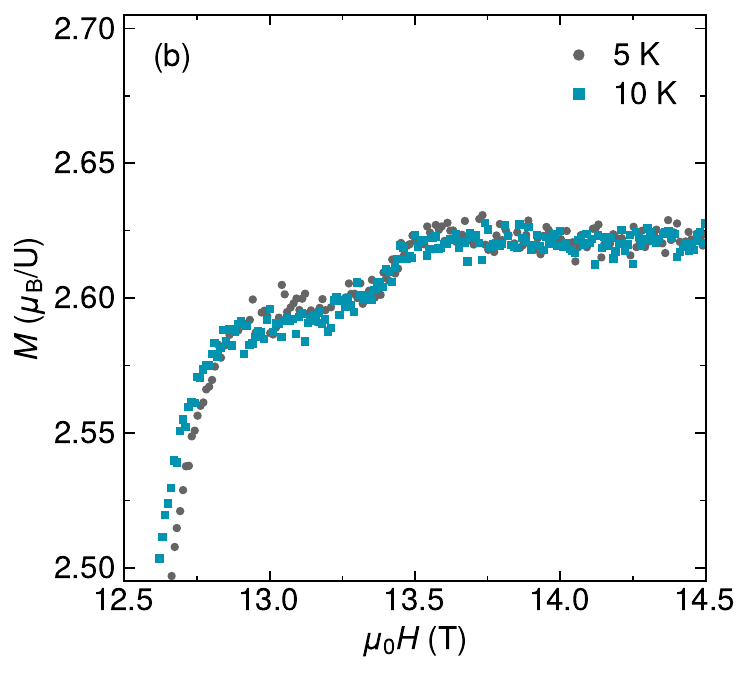}}
    \caption{(a) The high-field magnetization of \UNS\ at 2~K (black points) is shown with a 15-point window moving average of the data (red line) to highlight a feature similar to the P2-FP transition observed in the (b) 5~K and 10~K data.}
    \label{fig:MvH_P2_FF}
\end{figure}

\clearpage
\section{\UNS\ Sommerfeld coefficient fit}
A linear fit to the equation $C_{\mathrm{p}}$/$T$~=~${\beta}T^2$~+~${\gamma}$ for the low-temperature, zero-field heat capacity data provides the Sommerfeld coefficient ($\gamma$) for \UNS. The coefficient $\beta$ is also converted to the Debye temperature with Eq.~\ref{eq:debye}, where $R$ is the gas constant and $n$ is the number of atoms per formula unit ($n$=13).

\begin{equation}
    \label{eq:debye}
    {\theta}_{\mathrm{D}} = \left(\frac{12 {\pi}^4 n R}{5{\beta}}\right)^{1/3}
\end{equation}

\begin{figure}[h!]
    \centering
    \subfloat{\includegraphics[width=0.5\columnwidth]{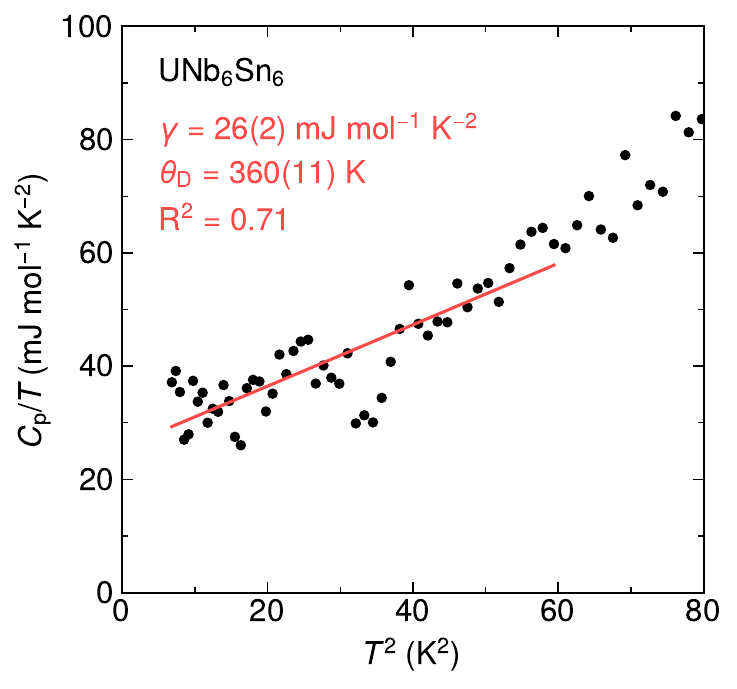}}
    \caption{Low temperature heat capacity data for \UNS\ is fit to the range $T^2{\le}60$~K$^2$.}
    \label{fig:Sommerfeld}
\end{figure}

\clearpage
\section{Additional resistivity data}
\begin{figure}[h!]
    \centering
    \includegraphics[width=0.5\columnwidth]{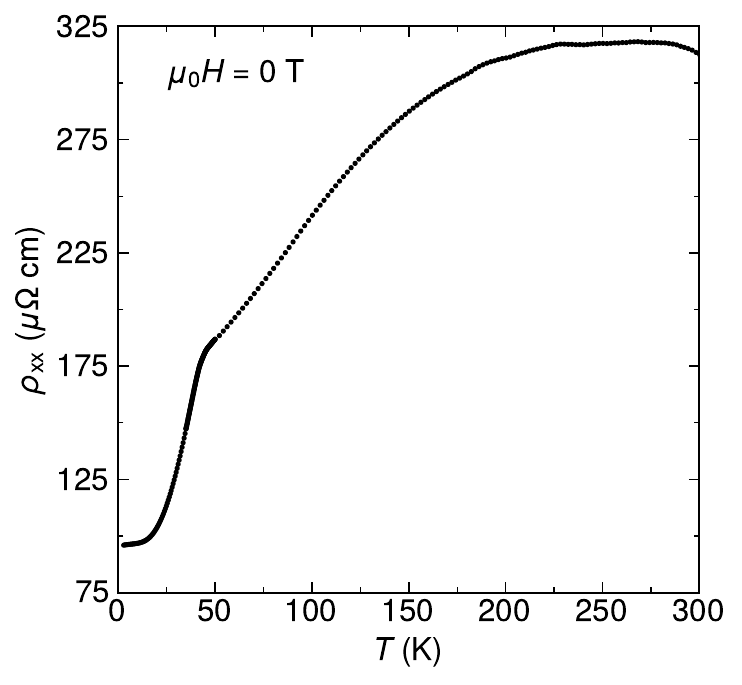}
    \caption{The longitudinal resistivity of \UNS\ at zero magnetic field ($RRR$~=~3.3)}
    \label{fig:rhoxx_0T_full}
\end{figure}

\begin{figure}[h!]
    \centering
    \subfloat{\includegraphics[width=0.45\columnwidth]{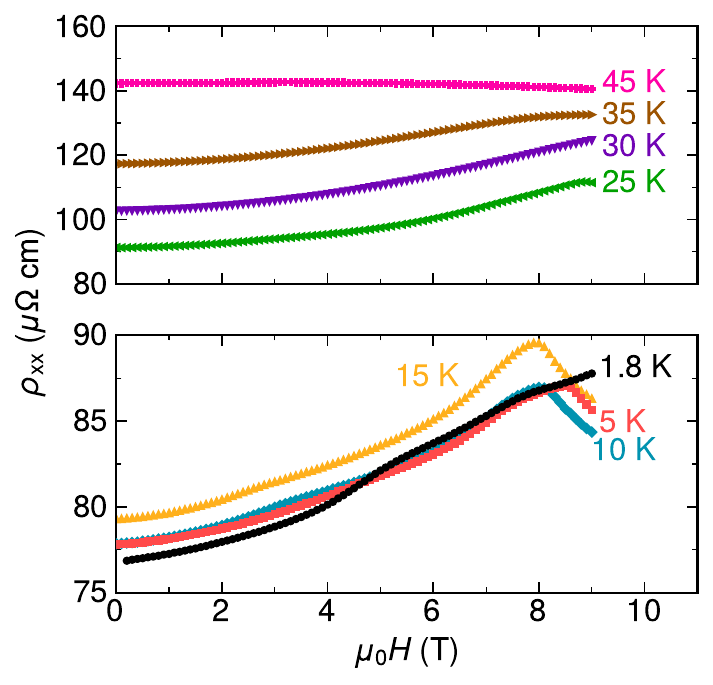}}
    \hspace{2em}
    \subfloat{\includegraphics[width=0.45\columnwidth]{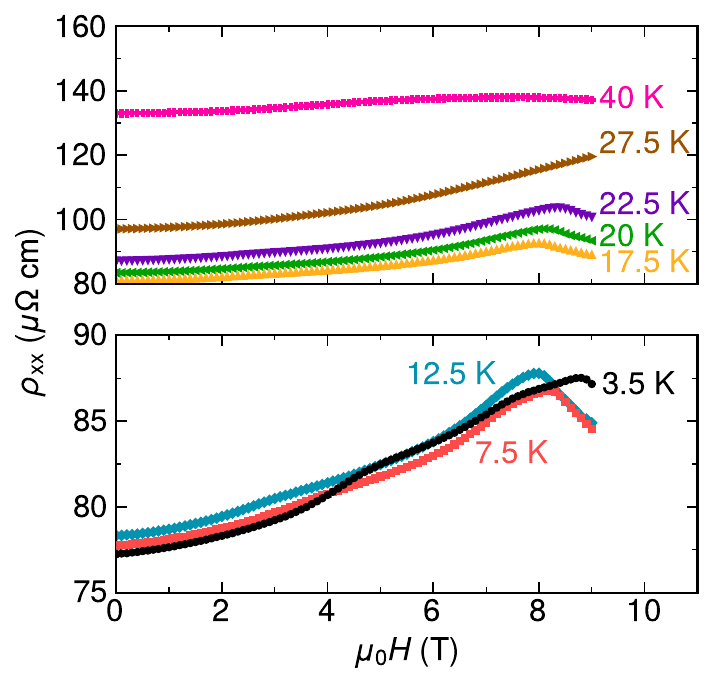}}
    
    \caption{Longitudinal resistivity data not shown in the main text that was used to map the magnetic phase regions of \UNS\ ($H{\parallel}c$). The phase transition points are labeled $\rho_\mathrm{xx,1}$($H$) in the main text's magnetic phase diagram.}
    \label{fig:xtra_rhoH}
\end{figure}

\begin{figure}
    \centering
    \includegraphics[width=0.5\columnwidth]{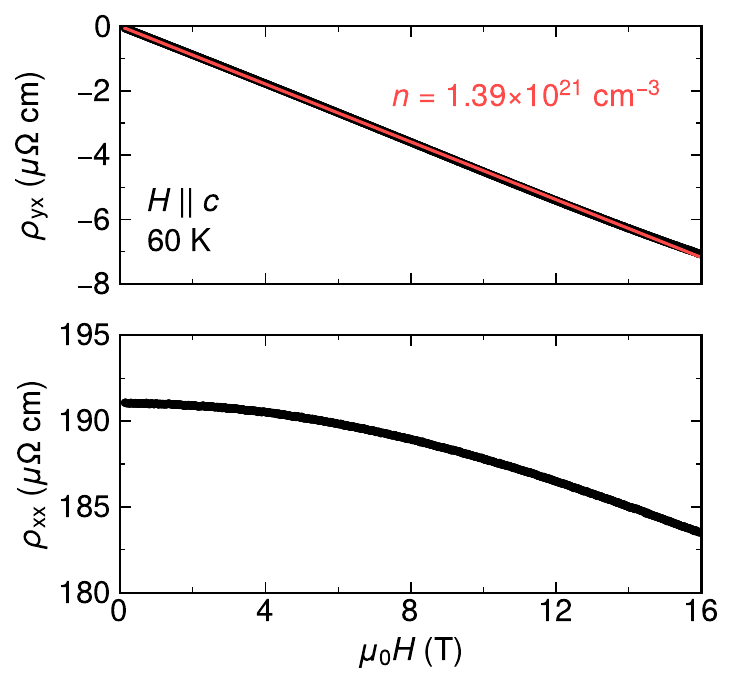}
    \caption{The longitudinal and transverse resistivity of \UNS\ in the paramagnetic state at 60~K show no features as a function of applied field. The transverse resistivity is fit to a single-carrier model with a carrier concentration of 1.39${\times}$10$^{21}$ cm$^{-3}$.}
    \label{fig:rhoxx_Hall_60K}
\end{figure}

\clearpage
\section{Comparison of Hall and magnetization features}
\begin{figure}[h!]
    \centering
    \subfloat{\includegraphics[width=0.45\columnwidth]{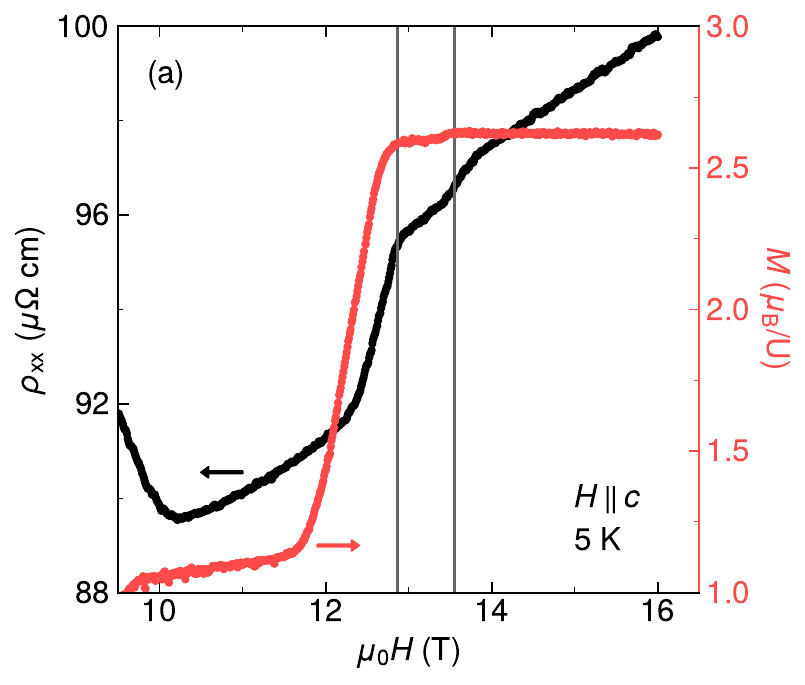}}
    \hspace{2em}
    \subfloat{\includegraphics[width=0.45\columnwidth]{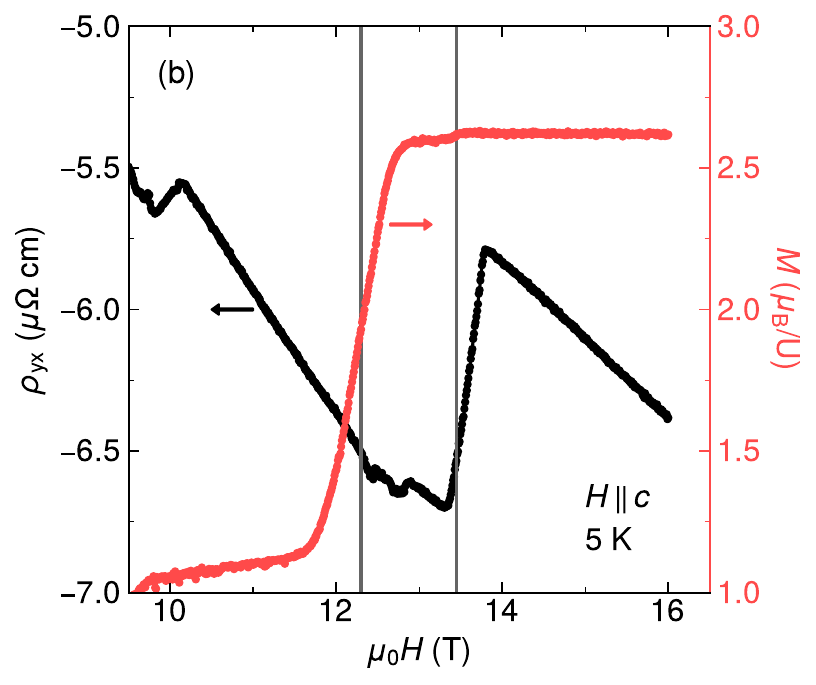}}
    \caption{(a) The 5~K longitudinal resistivity of \UNS\ is shown with the corresponding magnetization. Vertical lines are a guide to the eye to highlight the constant behavior of P2 in each dataset. (b) The 5~K Hall resistivity of \UNS\ is also shown with the magnetization. Vertical lines guide the eye to the large jump in magnetization from P6 to P2 and the comparatively minor shift in the Hall data.}
    \label{fig:rho_M_compare}
\end{figure}

Fig.~\ref{fig:rho_M_compare} compares the features of the resistivity and magnetization of \UNS\ near the P6-P2-FP phase boundaries at 5~K, where the P2 phase region is clear. The flat magnetization and constant slope $\rho_\mathrm{xx}$ regions align well for the P2 phase, reflecting good alignment between the two samples and a consistent data collection sweep rate. However, while a large magnetization shift occurs between the P6 and P2 phases in magnetization, only a small shift occurs in $\rho_\mathrm{yx}$.

\clearpage
\section{Alternative Hall analysis options}
\begin{figure}[h!]
    \centering
    \includegraphics[width=0.5\columnwidth]{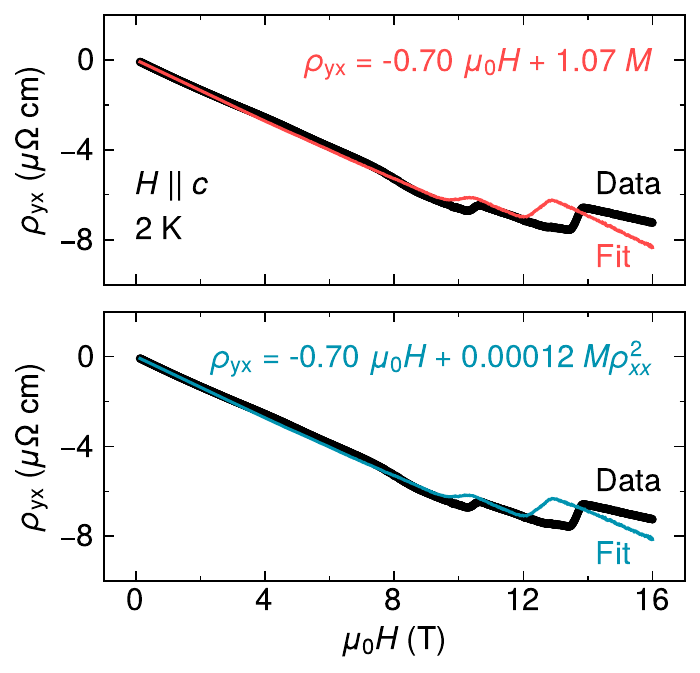}
    \caption{The [0,16]~T fit to the transverse resistivity used in the main text (top panel) is compared with a fit introducing a longitudinal resistivity dependence related to skew scattering (bottom panel).}
    \label{fig:Hall_2K_Mrhoxx2}
\end{figure}

For \UNS, Fig.~\ref{fig:Hall_2K_Mrhoxx2} compares the linear fit used in the main text with magnetic field ($\mu_\mathrm{0}H$) and magnetization ($M$) terms to a fit introducing a ${\rho}^{2}_{\mathrm{xx}}$ dependence related to skew scattering. The same magnetic field regions diverge between the fit and the data in each case. Therefore, the changes are not accounted for with a simple extrinsic contribution to the anomalous Hall effect and are likely intrinsic features.

\begin{figure}[h!]
    \centering
    \subfloat{\includegraphics[width=0.47\columnwidth]{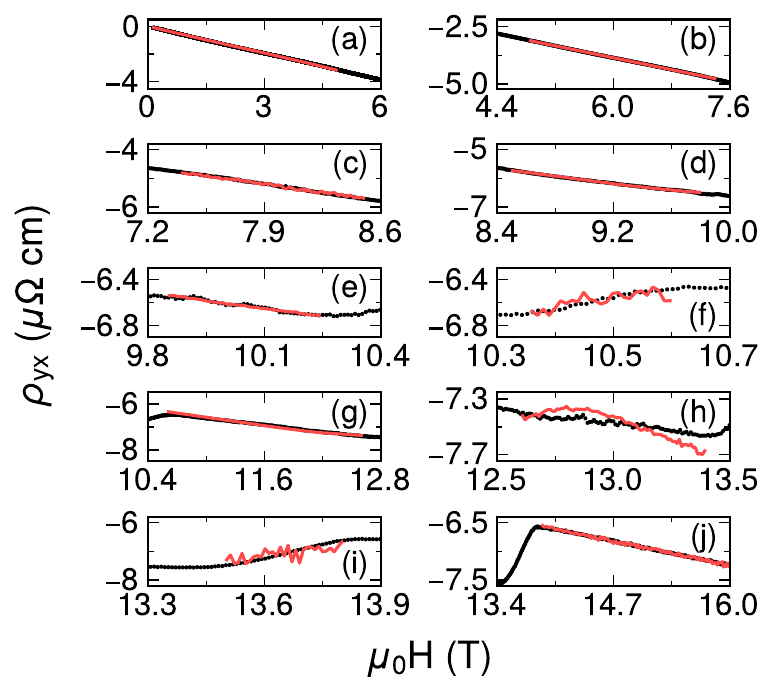}}
    \hspace{2em}
    \subfloat{\includegraphics[width=0.47\columnwidth]{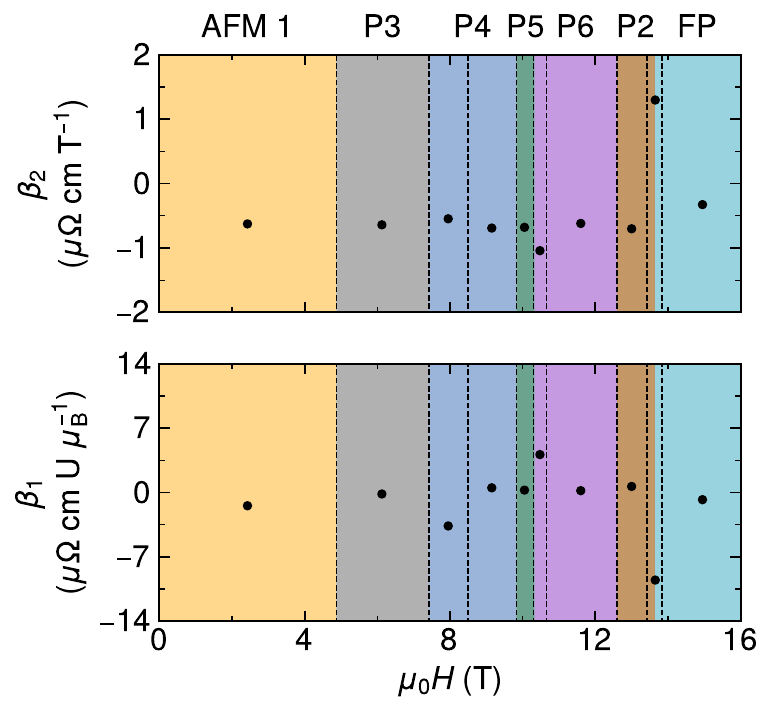}}
    \caption{Linear regions of the transverse resistivity are fit to ordinary and anomalous Hall terms without an intercept. The resulting coefficients are then plotted with dashed lines separating the fit ranges and colored regions representing the magnetic phase regions defined in the main text and listed above the color.}
    \label{fig:Hall_regions_without_intercept}
\end{figure}

\begin{figure}[h!]
    \centering
    \subfloat{\includegraphics[width=0.47\columnwidth]{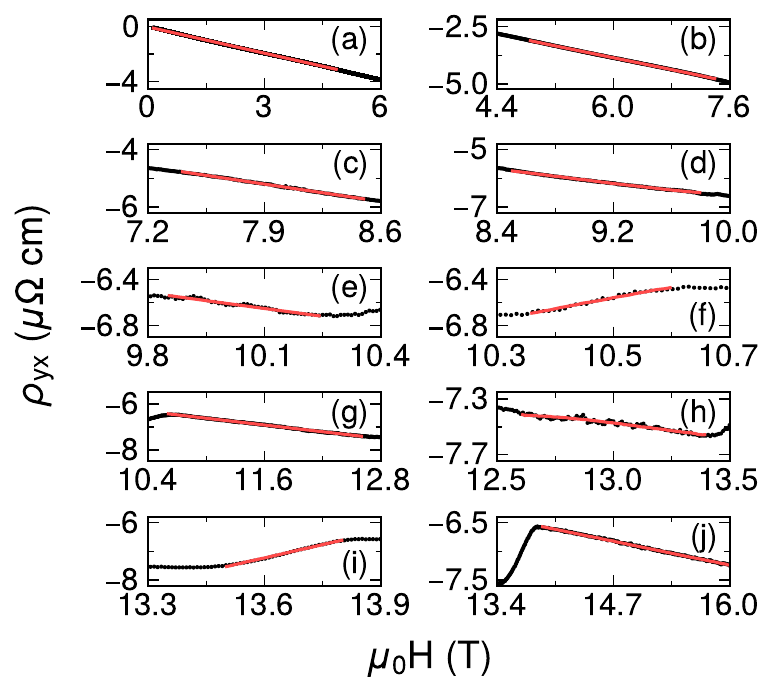}}
    \hspace{2em}
    \subfloat{\includegraphics[width=0.47\columnwidth]{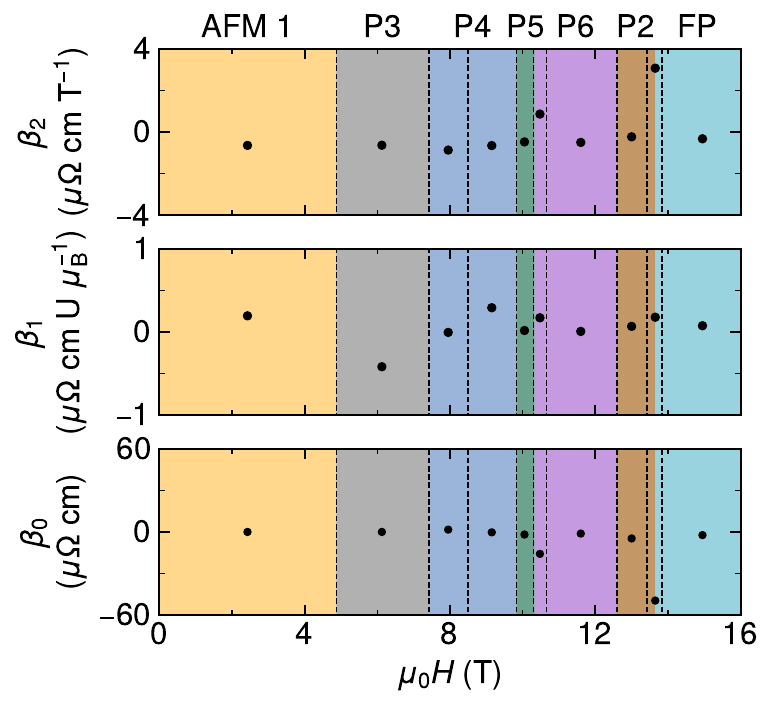}}
    \caption{Linear regions of the transverse resistivity are fit to ordinary and anomalous Hall terms as well as an intercept. The resulting coefficients are then plotted with dashed lines separating the fit ranges and colored regions representing the magnetic phase regions defined in the main text and listed above the color.}
    \label{fig:Hall_regions_with_intercept}
\end{figure}

In Fig.~\ref{fig:Hall_regions_without_intercept}, the 2~K, $H{\parallel}c$ transverse resistivity of \UNS\ is broken into linear regions to perform fits to the function ${\rho}_{\mathrm{yx}}$=${\beta}_{\mathrm{1}}(M)+{\beta}_{\mathrm{2}}({\mu}_{\mathrm{0}}H)$. In Fig.~\ref{fig:Hall_regions_with_intercept}, an additional intercept term, ${\beta}_{\mathrm{0}}$, is introduced to account for the offset from zero with increasing field. The resulting coefficients are tabulated in Tables~\ref{tab:region_coefs_without_intercept} and \ref{tab:region_coefs_with_intercept}. The regions in some cases do not align with the definitions of the magnetic phase regions used to build the main text's magnetic phase diagram. The discrepancy stems from the previously noted misalignment of magnetization shifts and transverse resistivity shifts. Qualitatively, the fit noticeably improves with the introduction of an intercept term in panels (f) and (i), where the resistivity has a positive slope. Those regions span the transitions into the P6 and field-polarized (FP) magnetic phases. For both the intercept and no-intercept cases, the largest coefficient shifts occur between P5 and P6 and between P2 and FP. The changes correspond to field regions in the main text where the difference in the data and [0,16]~T fit (${\Delta}{\sigma}_{\mathrm{yx}}$) hits a maximum before dropping to near zero. Therefore, each fitting method highlights significant changes in the ability of the magnetization to account for shifts in the transverse resistivity and suggests that an additional contribution, such as a large Berry curvature or significant Fermi surface changes, is relevant to the magnetic phases in those field regions. 

\begin{table}[h!]
\small
\centering 
\caption{\label{tab:region_coefs_without_intercept} The fit parameters are tabulated for the no-intercept case along with their standard errors. The carrier concentration is also calculated using the single-carrier model relationship $n$~=~($\beta_{\mathrm{2}}$$e$)$^{-1}$, where $e$ is the elementary charge.}
\begin{tabular}{c c c c c c}
\midrule Index & Phase(s) & Fit Range (T) & $\beta_\mathrm{1}$ ($\frac{{\mu}{\Omega}~\mathrm{cm}}{\mu_\mathrm{B}/\mathrm{U}}$) & $\beta_\mathrm{2}$ ($\frac{{\mu}{\Omega}~\mathrm{cm}}{\mathrm{T}}$) & $|n|$ ($\frac{1}{\mathrm{cm}^{3}}$) \\
\midrule
a & AFM 1 & 0--4.85 & -1.4(2) & -0.626(3) & 9.98$\times$10$^{20}$ \\
b & P3 & 4.85--7.4 & -0.16(2) & -0.6401(6) & 9.75$\times$10$^{20}$ \\
c & P4 & 7.4--8.5 & -3.6(2) & -0.546(6) & 1.14$\times$10$^{21}$ \\
d & P4 & 8.5--9.8 & 0.51(2) & -0.6905(8) & 9.04$\times$10$^{20}$ \\
e & P5 & 9.85--10.25 & 0.26(2) & -0.679(1) & 9.20$\times$10$^{20}$ \\
f & P6 & 10.35--10.6 & 4.1(4) & -1.04(4) & 5.99$\times$10$^{20}$ \\
g & P6 & 10.6--12.6 & 0.20(1) & -0.619(2) & 1.01$\times$10$^{21}$ \\
h & P2 & 12.6--13.4 & 0.66(8) & -0.70(1) & 8.89$\times$10$^{20}$ \\
i & P2/FP & 13.5--13.8 & -10(2) & 1.3(3) & 4.80$\times$10$^{20}$ \\
j & FP & 13.9--16 & -0.78(1) & -0.326(2) & 1.92$\times$10$^{21}$ \\
\end{tabular}
\end{table}

\begin{table}[h!]
\small
\centering 
\caption{\label{tab:region_coefs_with_intercept} The fit parameters are tabulated for the intercept case along with their standard errors. The carrier concentration is also calculated using the single-carrier model relationship $n$~=~($\beta_{\mathrm{2}}$$e$)$^{-1}$, where $e$ is the elementary charge.}
\begin{tabular}{c c c c c c c}
\midrule Index & Phase(s) & Fit Range (T) & $\beta_\mathrm{0}$ (${\mu}{\Omega}$~cm) & $\beta_\mathrm{1}$ ($\frac{{\mu}{\Omega}~\mathrm{cm}}{\mu_\mathrm{B}/\mathrm{U}}$) & $\beta_\mathrm{2}$ ($\frac{{\mu}{\Omega}~\mathrm{cm}}{\mathrm{T}}$) & $|n|$ ($\frac{1}{\mathrm{cm}^{3}}$) \\
\midrule
a & AFM 1 & 0--4.85 & -0.050(2) & 0.2(1) & -0.634(2) & 9.84$\times$10$^{20}$ \\
b & P3 & 4.85--7.4 & -0.037(8) & -0.42(6) & -0.628(3) & 9.94$\times$10$^{20}$ \\
c & P4 & 7.4--8.5 & 1.61(5) & 0.0(1) & -0.86(1) & 7.24$\times$10$^{20}$ \\
d & P4 & 8.5--9.8 & -0.34(4) & 0.292(5) & -0.64(3) & 9.68$\times$10$^{20}$ \\
e & P5 & 9.85--10.25 & -2(1) & 0.0(1) & -0.5(1) & 1.33$\times$10$^{21}$ \\
f & P6 & 10.35--10.6 & -15.9(7) & 0.2(2) & 0.87(8) & 7.19$\times$10$^{20}$ \\
g & P6 & 10.6--12.6 & -1.22(2) & 0.007(5) & -0.492(2) & 1.27$\times$10$^{21}$ \\
h & P2 & 12.6--13.4 & -4.73(9) & 0.07(2) & -0.224(9) & 2.78$\times$10$^{21}$ \\
i & P2/FP & 13.5--13.8 & -49.5(7) & 0.2(2) & 3.07(4) & 2.03$\times$10$^{20}$ \\
j & FP & 13.9--16 & -2.34(6) & 0.08(2) & -0.3187(8) & 1.96$\times$10$^{21}$ \\
\end{tabular}
\end{table}

\clearpage
\bibliography{UNb6Sn6.bib}